\documentclass{article} 
\usepackage[main, final]{neurips_2026}

\setcitestyle{numbers,square,sort&compress,citesep={,}}
\usepackage{colortbl}   
\usepackage{array}      

\newcolumntype{L}[1]{>{\raggedright\arraybackslash}p{#1}}
\usepackage{amsmath}

\usepackage{lineno}
\usepackage[utf8]{inputenc} 
\usepackage[T1]{fontenc}    
\usepackage{hyperref}       
\usepackage{url}            
\usepackage{float}
\usepackage{booktabs}       
\usepackage{amsfonts}       
\usepackage{nicefrac}       
\usepackage{microtype}      
\usepackage{todonotes}
\usepackage{subcaption}
\usepackage{enumitem}
\definecolor{darkblue}{rgb}{0, 0, 0.5}
\hypersetup{colorlinks=true, citecolor=darkblue, linkcolor=darkblue, urlcolor=darkblue}

\usepackage{multirow}
\usepackage{graphicx}
\usepackage{xspace}

\definecolor{forgetred}{RGB}{204,0,0}
\definecolor{forgetbg}{RGB}{255,230,230}
\definecolor{retainbg}{RGB}{217,255,217}

\usepackage{xcolor}
\definecolor{fig1colorblue}{HTML}{4C72B0}
\definecolor{fig1colorred}{HTML}{E74C3C}

\newcommand{\method}{TRACER\xspace}

\title{TRACER: Token ReAssignment for Concept ERasure in Generative Recommendation}

\author{
Ziheng Chen\\
Stony Brook University\\
\texttt{albertchen1993pokemon@gmail.com}\\
\And
Jiali Cheng\\
University of Massachusetts Lowell\\
\texttt{jiali\_cheng@uml.edu}\\
\And
Zezhong Fan\\
Columbia University\\
\texttt{zfan2274@gmail.com}\\
\And
Hadi Amiri\\
University of Massachusetts Lowell\\
\texttt{hadi\_amiri@uml.edu}\\
\And
Diyuan Wu\\
Institute of Science and Technology Austria\\
\texttt{diyuan.wu@ist.ac.at}\\
\And
Gabriele Tolomei\\
Sapienza University of Rome\\
\texttt{tolomei@di.uniroma1.it}\\
\And
Yang Zhang\\
National University of Singapore\\
\texttt{zhangy@nus.edu.sg}
}

\usepackage{amsmath,amsfonts,bm}

\newtheorem{theorem}{Theorem}
\newtheorem{lemma}[theorem]{Lemma}
\newtheorem{proposition}[theorem]{Proposition}

\newtheorem{assumption}[theorem]{Assumption}

\def\eqref#1{equation~\ref{#1}}

\def\1{\bm{1}}

\DeclareMathAlphabet{\mathsfit}{\encodingdefault}{\sfdefault}{m}{sl}
\SetMathAlphabet{\mathsfit}{bold}{\encodingdefault}{\sfdefault}{bx}{n}

\newcommand{\forget}{\mathcal{D}_f}
\newcommand{\remain}{\mathcal{D}_r}
\newcommand{\all}{\mathcal{D}}
\newcommand{\LLMori}{\mathcal{M}_{\theta}}
\newcommand{\un}{\mathcal{M}_{un}}
\newcommand{\optimal}{\mathcal{\mathcal{M}}_{\theta^{*}}}

\newcommand{\T}{\mathcal{T}}

\begin{document}

\maketitle

\begin{abstract}
Generative recommendation formulates next-item prediction as autoregressive generation over semantic ID (SID) sequences derived from users' historical interactions, making modern recommender systems structurally similar to large language models (LLMs). As privacy and safety concerns grow, these systems increasingly require concept unlearning to remove sensitive or harmful concepts associated with items. However, existing LLM unlearning methods cannot be directly applied to generative recommendation. Unlike word tokens with explicit semantics, SIDs are abstract identifiers that are often shared by both forget and retain items, leading to severe conflicts between concept removal and recommendation utility preservation.

To address this challenge, we propose TRACER, an end-to-end concept unlearning framework based on token reassignment. Rather than directly suppressing shared SIDs, TRACER reassigns concept-related items to alternative tokens that better facilitate forgetting while minimizing side effects on retained items. We further introduce a coherence regularizer to preserve semantic consistency among retain items during unlearning. Experiments on real-world recommendation datasets demonstrate that TRACER effectively removes target concepts while substantially better preserving recommendation utility than existing unlearning baselines.

\end{abstract}


\section{Introduction}
\label{sec:intro} 
Generative recommendation (GenRec) \cite{wang2024learnable,han2025mtgr,liu2025onerec,shen2026fine} has emerged as a promising paradigm for sequential recommendation, demonstrating strong empirical performance across various domains \cite{deng2025onerec,kong2025minionerec,rajput2023recommender,liu2026palm}. Unlike conventional recommendation systems, GenRec formulates recommendation as a sequence-to-sequence generation task and typically consists of two key components: item tokenization and autoregressive generation. Specifically, an item tokenizer assigns each item a semantic ID (SID), a sequence of discrete semantic tokens for indexing, while the generative model autoregressively predicts the next item from the SID sequence of purchased items. This design enables recommender systems to exploit the rich textual information associated with items and to inherit the representational and generative capabilities of Large Language Models (LLMs), showing strong empirical performance across diverse domains \cite{zhou2025onerec,kong2025minionerec,wang2024learnable}.


However, as GenRec becomes increasingly adopted, new concerns arise from sensitive information involved in model training, such as malicious descriptions~\citep{fan2024salun,gandikota2023erasing,liu2024rethinking}, outdated information~\citep{cheng2023multimodal,xu2024knowledge,cheng2026tmc}, and copyrighted text~\citep{eldan2023whos,shi2025muse,Cheng2026}. In this setting, concept unlearning in GenRec aims to remove the model’s ability to recommend items associated with a target concept while preserving its recommendation utility on unrelated items.
For example, removing a counterfeit sports brand requires the model to stop recommending items associated with that brand, while still preserving users’ general sports preferences.
Simply filtering concept-related items at generation time is insufficient, as it only modifies the final outputs while leaving the internal concept representation unchanged. Consequently, concept items in user histories can still influence predictions, and blocked items may be replaced with poorly matched alternatives, degrading recommendation utility.


Although concept unlearning remains underexplored in GenRec, it has been widely studied in LLMs~\cite{gandikota2024erasing,gur2025precise,cheng2025tool,cheng2026toward,zhu2025pathology}.
Existing methods typically construct the forget set from data containing concept-related token sequences and suppress their likelihood during unlearning. However, this strategy cannot be directly applied to GenRec. Unlike language tokens, which are often interpretable lexical units and can help localize target concepts, SID tokens in GenRec are abstract \textit{semantic identifiers (SIDs)}  shared by items across both forget and retain sets. Therefore, a target concept cannot be reliably localized to a set of SID tokens.  
Besides, directly suppressing such shared tokens affects both the forget and the retain set, leading to severe gradient conflict during unlearning.

\definecolor{fig1colorred}{HTML}{DC2626}
\definecolor{fig1colorblue}{HTML}{2B5EA7}

\definecolor{fig1colorred}{HTML}{DC2626}
\definecolor{fig1colorblue}{HTML}{2B5EA7}

\begin{figure*}[t]
    \centering
    \vspace{-30pt}
    \includegraphics[width=\textwidth,height=0.35\textwidth]{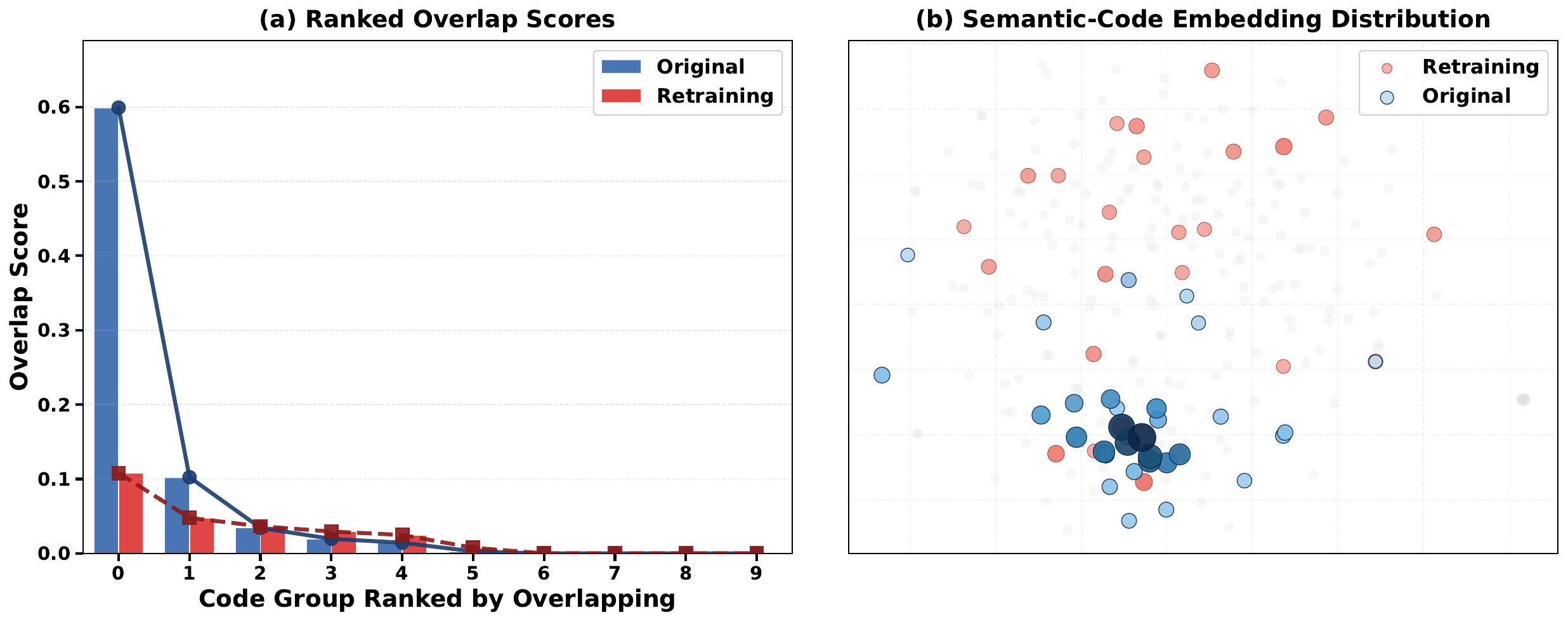}
    \caption{\textcolor{fig1colorred}{Retraining} vs. \textcolor{fig1colorblue}{Original} GenRec model.
    (a): \textcolor{fig1colorblue}{Original} has highly overlapped semantic tokens, while \textcolor{fig1colorred}{Retraining} mitigates overlap score significantly.
    (b): \textcolor{fig1colorblue}{Original} produces highly clustered embeddings, while \textcolor{fig1colorred}{Retraining} produces dispersed embeddings.
    The overlap score is computed as the product of token frequencies in the forget and retain sets.}
    \vspace{-15pt}
    \label{fig:motivation}
\end{figure*}

To characterize successful concept unlearning, we study the SID assignment of the \emph{retrained} model with concept-free data in Figure~\ref{fig:motivation}. Specifically, we randomly select several brands as target concepts, remove their brand information from item descriptions, and filter out the corresponding brand-related interactions.
Comparing the SID assignments of the original and retrained models reveals two key observations. First, SID overlap between forget and retain items drops sharply after retraining. Second, overlapped token embeddings become more dispersed after retraining than in the original model, suggesting that retraining weakens the concept-specific token structure. In contrast, unrelated tokens remain relatively stable. These observations motivate us to reshape SID assignments to reduce forget-retain overlap, rather than merely suppressing the likelihood of forget items.

To this end, we propose TRACER, a framework that achieves concept unlearning through differentiable token reassignment: rather than blindly suppressing forget items' likelihood, TRACER directly reshapes the SID structure to disentangle forget items from retain items. Specifically, TRACER relaxes the hard codebook lookup at each level into a soft distribution over candidate tokens, and introduces a lightweight learnable perturbation on top of the pretrained assignment. This design preserves the original tokenizer structure while allowing the unlearning objective to selectively redirect concept-related items toward tokens that retain items rarely use. Moreover, unlearning can disrupt recommendation coherence, especially when concept-related items appear as target items in the training data. TRACER preserves recommendation coherence by encouraging the recommender to generate semantically similar alternatives that do not contain the target concept. Although TRACER is primarily designed for concept unlearning, it can also be adapted to item-level and interaction-level unlearning.\looseness-1


The main contributions of this work are summarized as follows:
\begin{itemize}[leftmargin=*,topsep=2pt, itemsep=1pt]
    \item We present the first study of concept unlearning in GenRec and reveal the limitations of directly applying existing LLM concept unlearning methods to this setting.
    \item We propose TRACER, an end-to-end GenRec unlearning framework that jointly optimizes concept-related token reassignment to satisfy the unlearning objective. We further introduce a coherence loss to preserve recommendation consistency after unlearning and provide theoretical analysis to justify the effectiveness of our algorithm.
    \item We conduct extensive experiments on real-world datasets and show the advantage of TRACER over existing methods. 
\end{itemize}


\section{Background and Preliminaries}
\label{sec:prelim}

\subsection{Generative Recommendation}
\label{sec:GR}

Let $\mathcal{U}$ and $\mathcal{I}$ denote the user and item sets. Each user $u$ has an interaction sequence $\mathcal{H}_u=\{i_1,\ldots,i_T\}$ with $i_j\in\mathcal{I}$. The goal is to predict the next item $i_{T+1}$ from $\mathcal{H}_u$. 
To leverage LLMs for generative recommendation, GenRec typically follows a two-stage paradigm. 

\paragraph{Stage 1: Semantic Item Tokenization}
For each item $i \in \mathcal{I}$, a tokenizer $\mathcal{T}$ maps its textual description into a \emph{Semantic Identifier (SID)}, denoted by
\begin{equation}
    \boldsymbol{s}_i = \mathcal{T}(\boldsymbol{z}_i) = [s_i^1,\ldots,s_i^L],
\end{equation}
where $\boldsymbol{z}_i$ is the item embedding produced by a frozen text encoder, and each $s_i^l$ is a discrete semantic token selected from the $l$-th codebook. The codebooks are organized hierarchically: earlier levels capture coarse semantic information, while later levels encode increasingly fine-grained residual information. 
Concretely, residual quantization assigns one token at each level. Let $\boldsymbol{r}_i^1=\boldsymbol{z}_i$ be the initial residual, and let $\mathcal{C}^l=\{\boldsymbol{c}_1^l,\ldots,\boldsymbol{c}_K^l\}$ denote the codebook at level $l$. The token assignment and residual update are given by
\begin{equation}
\label{eq:RQ}
    s_i^l = \arg\min_{k} \left\|\boldsymbol{r}_i^l-\boldsymbol{c}_k^l\right\|^2,
    \qquad
    \boldsymbol{r}_i^{l+1} = \boldsymbol{r}_i^l-\boldsymbol{c}_{s_i^l}^l .
\end{equation}
Repeating this procedure for $L$ levels yields the final SID $\boldsymbol{s}_i=[s_i^1,\ldots,s_i^L]$.
The codebooks can be constructed in different ways depending on the tokenizer design. 
More generally, generative recommendation methods differ in whether SID tokenization is performed as a separate preprocessing step or learned jointly with the recommendation model.



\paragraph{Stage 2: Recommendation via Autoregressive Generation.}
For each user $u \in \mathcal{U}$, we observe a chronological interaction history $\mathcal{H}_u = (i_1,i_2,\ldots,i_{T-1}), i_t \in \mathcal{I}$. The goal of sequential recommendation is to predict the next item $i_T$ conditioned on this history. In GenRec, each item is represented by its SID described in the prior section. Therefore the interaction history becomes a sequence of semantic token sequences $\boldsymbol{s}_{\mathcal{H}_u} =     [\boldsymbol{s}_{i_1},\boldsymbol{s}_{i_2},\ldots,\boldsymbol{s}_{i_{T-1}}]$.
%
%
The GenRec model parameterized by $\theta$ predicts the next-item SID autoregressively by maximizing its log-likelihood:
\begin{equation}
\label{eq:autoregressive_sid_log}
    \log p_\theta(\boldsymbol{s}_{i_T}\mid \boldsymbol{x}_u)
    =
    \sum_{j=1}^{L}
    \log p_\theta(s_{i_T}^j \mid \boldsymbol{s}_{\mathcal{H}_u}, s_{i_T}^{<j}),
\end{equation}
where $s_{i_T}^{<j}=(s_{i_T}^1,\ldots,s_{i_T}^{j-1})$ denotes the previously generated prefix of the target SID.

At inference time, the model generates the next SID, which realizes the next token prediction process, and serves as the basis for next-item recommendation. In particular, the generated SID can be one-to-one mapped back to an item via the codebook.



\subsection{GenRec Concept Unlearning}
\label{sec:prelim_unlearning}
Machine unlearning aims to remove the influence of a designated training subset from a trained model without retraining from scratch~\citep{jia2024soul,cheng2024mubench,fan2025towards,cheng25d_interspeech,fu2025learning,chen2026open}. Given a dataset $\all$ and a forget set $\forget$, the retained data is $\remain=\all\setminus\forget$. The goal is to obtain an unlearned model $\un$ that eliminates the influence of $\forget$ while preserving performance on $\remain$~\citep{fan2024salun,chen2025frog,chen2025future}. Since retraining on $\remain$ to obtain $\optimal$ is costly, unlearning seeks to approximate this retrained model by updating the original model $\LLMori$:
\begin{equation*}
    \LLMori \xrightarrow{\forget} \un \approx \optimal.
\end{equation*}

\paragraph{Definition (GenRec Concept Unlearning)} Given a sensitive concept $c$, GenRec concept unlearning aims to: 1) prevent the model from recommending concept-related items, and 2) ensure that the presence of such items in a user's history $\mathcal{H}_u$ does not influence the prediction of future items.

Based on this definition, for a sensitive concept $c$, we define the forget set as
$\mathcal{D}_f^c = \{(\mathcal{H}_u, i_t) \in \mathcal{D} : i_t \in \mathcal{I}_c\}$,
where $\mathcal{I}_c \subseteq \mathcal{I}$ contains the items associated with $c$. The overall forget set is defined as the union over all target concepts $\forget=\bigcup\limits_{c}\mathcal{D}_f^c $, and retain set is given by $\remain = \mathcal{D} \setminus \forget$.
For interactions in $D_r$, concept-related textual information should ideally be removed before \textit{semantic item tokenization}, as if such information had never been present in the item descriptions. 
Following the approximate unlearning, we formulate unlearning as a trade-off between forgetting unwanted knowledge and preserving retain utility. Specifically, the unlearning objective combines a forget loss $\mathcal{L}_{\mathrm{F}}$ and a retain loss $\mathcal{L}_{\mathrm{R}}$:
\begin{equation}
\min_{\theta} \; \mathcal{L}_{\mathrm{unlearn}}(\theta)
=
\mathcal{L}_{\mathrm{R}}(\theta)
+
\lambda \mathcal{L}_{\mathrm{F}}(\theta),
\end{equation}
where $\lambda$ controls the strength of forgetting. We introduce the details of these two losses in Section~\ref{sec:loss}.

\paragraph{Challenges of GenRec Concept Unlearning}
Compared with traditional LLM concept unlearning~\citep{gandikota2024erasing,cao2024rwku,cheng2025tool,cheng2026toward}, GenRec concept unlearning faces two unique challenges. 
\emph{First}, forget and retain items can substantially overlap in the SID space, so suppressing forget-item SIDs may also harm retained items sharing the same tokens. Since concept information cannot be cleanly removed once encoded into SIDs, we propose token reassignment to reduce forget--retain conflict while preserving retain utility.
\emph{Second}, unlearning may disrupt recommendation coherence when concept-related items appear as training targets~\citep{cheng2023gnndelete,chen2026cure}. This motivates a coherence loss that encourages semantically consistent alternatives while suppressing the target concept.
Given the above unique challenges, existing LLM unlearning method may fall short in GenRec concept unlearning. 

\newtheorem{definition}{Definition}

\begin{figure*}[t]
    \centering
    \vspace{-25pt}
    \includegraphics[width=\linewidth]{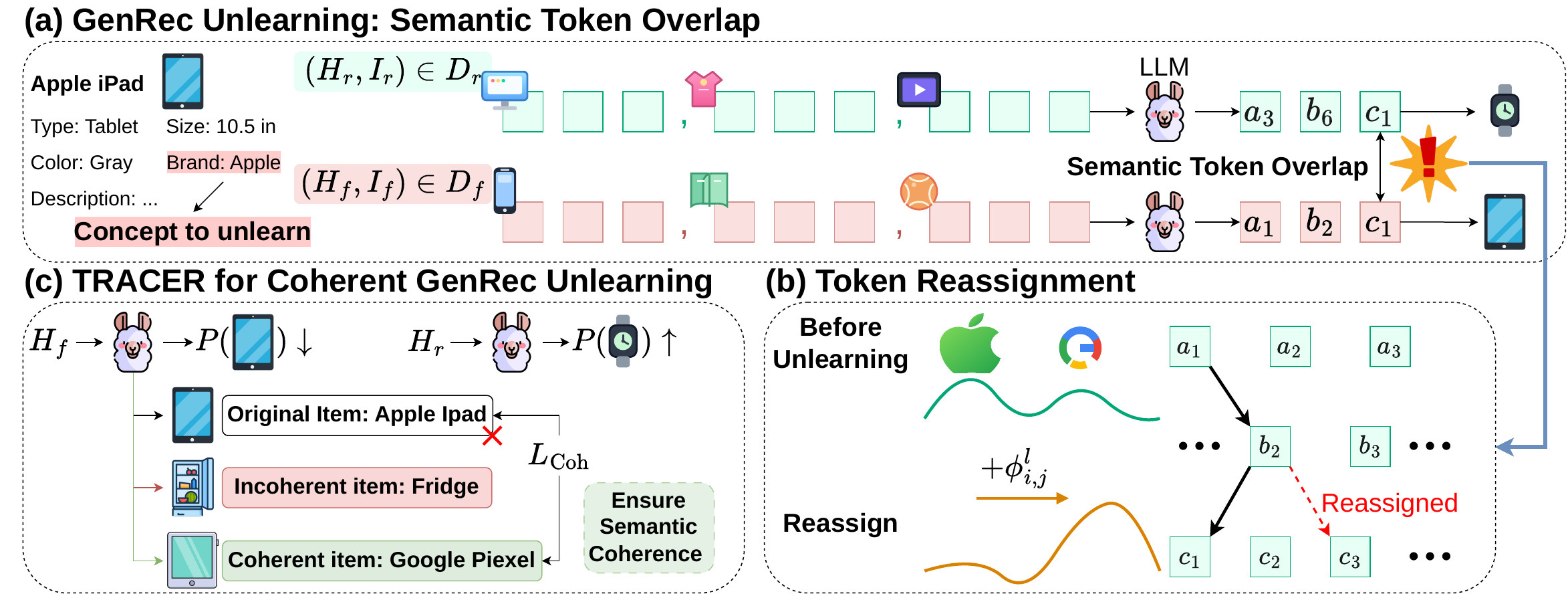}
    \caption{Overview of TRACER.
    \textbf{a) Semantic token overlap in GenRec unlearning:} removing conceptual knowledge from GenRec models is challenging, since many non-interpretable semantic tokens are shared by both forget and retain sets. 
    \textbf{b) Token reassignment:} TRACER reassigns the semantic token of unlearning-affected items to the closest item of the same category, which preserves semantic coherence to the maximum extent.
    \textbf{c) Coherenct unlearning:} TRACER unlearn the target concept without affecting the recommendation ability and semantic coherence of the GenRec model.
    }
    \label{fig:pipeline}
\end{figure*}

\section{Methodology}
Our method, TRACER, consists of two modules: (1) semantic token reassignment, which alleviates overlap in semantic tokens (Figure~\ref{fig:pipeline} (b), \S~\ref{sec:aprobook}), and (2) coherence loss, which mitigates the degradation of semantic coherence (Figure~\ref{fig:pipeline} (c), \S~\ref{sec:coherence}).

\paragraph{Overview and Motivation}
To better understand semantic token overlap--the major issue in GenRec unlearning, we investigate the oracle model retrained on concept-free data.
We first filter out concept-related information in item descriptions. Then we retrain LETTER~\cite{wang2024learnable}, a state-of-the-art GenRec model that jointly learns the tokenizer and the LLM recommender. Figure~\ref{fig:motivation} shows that the resulting SID assignment exhibits lower forget-retain overlap and a weaker concept-related token structure. This observation suggests an intuitive strategy: reassign semantic tokens for concept-related items to satisfy the unlearning objective, while preserving the tokenization of the remaining items. Inspired by this finding, we first introduce a differentiable token reassignment mechanism in Section~\ref{sec:aprobook}, and then present the unlearning objective in Section~\ref{sec:loss}.

\subsection{Flexible and Optimizable Token Reassignment}
\label{sec:aprobook}


We construct a differentiable surrogate for the original tokenizer $\mathcal{T}$ to fulfill the motivation, which preserves the core assignment mechanism of $\mathcal{T}$ while making the token selection process trainable.
Specifically, $\mathcal{T}$ can be viewed as a hierarchical token selector that maps each item to the nearest semantic token at each codebook level. We therefore introduce a soft tokenizer $\mathcal{T}_{\phi}$ to approximate $\mathcal{T}$ by replacing the discrete assignment in Equation~\ref{eq:RQ} with a differentiable relaxation.
Inspired by~\cite{lucic2022focus}, the hard assignment operation that selects the nearest codeword can be relaxed into a temperature-controlled softmax over all codewords. Specifically, for item $i$ at level $l$, the probability of assigning token $k$ is defined as
\begin{equation}
\label{eq:softmax}
q(s_i^l = k)
=
\frac{
\exp\left(-\|\boldsymbol{r}_i^l-\boldsymbol{c}_k^l\|^2/\tau\right)
}{
\sum_{j=1}^{K}
\exp\left(-\|\boldsymbol{r}_i^l-\boldsymbol{c}_j^l\|^2/\tau\right)
}.
\end{equation}
In this way, the token assignment is relaxed from a one-hot selection into a probability distribution over all codewords at each level. This approximation preserves the semantic structure of the original tokenizer by using the negative squared distance between the residual $\boldsymbol{r}_i^l$ and each codeword $\boldsymbol{c}_k^l$ as the assignment logit. More importantly, the soft tokenizer makes token assignment differentiable, allowing it to adapt to the unlearning objective through gradient-based optimization.



To enable end-to-end token reassignment while preserving the semantic hierarchy of the original tokenizer $\T$, we further introduce a learnable perturbation parameter $\phi$ over the codeword logit. 
Specifically, for the forget item $i$, the assignment probability at level $l$ is defined by applying a softmax over the perturbed logits $-\|\boldsymbol{r}_i^{l}-\boldsymbol{c}^{l}_{k}\|^{2}+\phi_{i,k}^{l}$, where $\phi_{i,k}^{l}$ captures the adjustment to the assignment preference for codeword $\boldsymbol{c}^{l}_{k}$.
To avoid excessive deviation from the original tokenizer, we impose an $L_1$ regularization term on $\phi$. In this way, the new tokenizer $\T_{\phi}$ inherits the semantic structure from $\T$, while allowing the learning objective to make small and targeted token reassignment:
\begin{equation}
\label{eq:reassign}
\begin{aligned}
q_{\phi}(s_{i}^{l}=k)
=
\frac{
\exp\left((-\|\boldsymbol{r}_i^{l}-\boldsymbol{c}^{l}_{k}\|^{2}+\phi_{i,k}^{l})/\tau\right)
}{
\sum_{j=1}^{K}
\exp\left((-\|\boldsymbol{r}_i^{l}-\boldsymbol{c}^{l}_{j}\|^{2}+\phi_{i,j}^{l})/\tau\right)
},
\qquad
\mathcal{L}_{\mathrm{reg}}
=
\sum_{i,l,k}\left|\phi_{i,k}^{l}\right|.
\end{aligned}
\end{equation}
 The soft assignment is integrated into the LLM through a soft token embedding $\tilde{\mathbf{e}}_{i}^{l}$, defined as the expectation of token embeddings under $q_{\phi}$:
$\tilde{\mathbf{e}}_{i}^{l}
=
\sum_{k=1}^{K}
q_{\phi}(s_i^l=k)\mathbf{e}^{l}_{k}$,
where $\mathbf{e}^{l}_{k}$ is the LLM token embedding of codeword $c_{k}^{l}$. We apply reassignment only to target-concept items $i\in\mathcal{I}_c$, while preserving the original SIDs of all other items by fixing $\phi_{i,k}^{l}=0$.

\subsection{Concept Unlearning with Semantic Coherence}
\label{sec:coherence}

\label{sec:loss}
To preserve the model's behavior on data unrelated to the erased concept, we further optimize the standard generation objective on retain interactions $(\mathcal{H}_r,i_r)\in \remain$. 
This retain loss encourages the model to maintain its original recommendation ability for retained items and mitigates unintended degradation caused by concept unlearning. 
Let $\mathcal{T}(i_r)=[s_{i_r}^{1},\cdots,s_{i_r}^{L}]$ denote the SID sequence of the retain target item $i_r$. 
We define
\begin{equation}
\mathcal{L}_{R}
=
-
\sum_{(\mathcal{H}_r,i_r)\in\remain}
\sum_{\ell=1}^{L}
\log
p_\theta\!\left(
s_{i_r}^{\ell}
\mid
\mathcal{T}(\mathcal{H}_r),\, s_{i_r}^{<\ell}
\right).
\end{equation}

Following the forgetting objective, we erase the target concept by reducing the likelihood of recommending concept-related items. For each forget interaction $(\mathcal{H}_f,i)\in\forget$, where $i\in\mathcal{I}_c$, we suppress the generation of its original SID sequence $\mathcal{T}(i)=[s_i^{1},\cdots,s_i^{L}]$ via a token-level objective:
\begin{equation}
\mathcal{L}_{F}
=
\sum_{(\mathcal{H}_f,i)\in\forget}
\sum_{\ell=1}^{L}
\log
p_\theta\!\left(
s_i^{\ell}
\mid
\mathcal{T}(\mathcal{H}_f),\, s_i^{<\ell}
\right).
\end{equation}
Minimizing this term discourages the model from assigning high probability to the original SID tokens of concept-related target items, thereby reducing their autoregressive generation likelihood.

However, directly suppressing the generation probability of a concept-related item $i$ may also penalize semantic tokens that are not specific to the target concept. This can disrupt the learned semantic structure and produce recommendations that are poorly aligned with the user history. To preserve recommendation coherence, we further encourage the model to generate retain items that are semantically similar to $i$ under the same history $\mathcal{H}_u$. Specifically, for each forget item $i$, we construct a positive set $\mathcal{P}(i)$ from the retain set based on item-description similarity. We compute the similarity using the frozen encoder representation $\bm{z}_i$ before token assignment, as defined in Section~\ref{sec:GR}. We select the top-$K$ most similar retain items, so $|\mathcal{P}(i)|=K$. Given item $i_p\in \mathcal{P}(i_T)$ with SIDs 
$\mathcal{T}(i_p)=[s_{p}^{1},\cdots,s_{p}^{L}]$, we define the coherence loss as
\begin{equation}
\label{eq:coh}
\mathcal{L}_{\mathrm{Coh}}
=
-
\frac{1}{K}
\sum_{(\mathcal{H}_f,i)\in\forget}
\sum_{i_p\in \mathcal{P}(i_T)}
\sum_{\ell=1}^{L}
\log
p_\theta\!\left(
s_{p}^{\ell}
\mid
\mathcal{T}(\mathcal{H}_f),\, s_{p}^{<\ell}
\right).
\end{equation}

The overall loss for the TRACER objective is a weighted combination of these components:
\begin{equation}
\label{eq:joint}
\mathcal{L}
=\mathcal{L}_{\mathrm{R}}+\lambda_1\mathcal{L}_{\mathrm{F}}+\lambda_2\mathcal{L}_{\mathrm{Coh}}+\lambda_3\mathcal{L}_{\mathrm{reg}}
\end{equation}

\subsection{Selective Updating of Perturbation Parameters}
Since the reassignment parameter \(\phi\) controls the token distribution of forget items, updating all entries of \(\phi\) may unnecessarily perturb concept-irrelevant token assignments and degrade recommendation utility. To make the reassignment more targeted, we selectively update \(\phi\). In particular, given item $i\in\mathcal{I}_c$ and token $t,$ we define the mask 
\begin{equation}
\label{eq:mask}
\begin{split}
    M_{i,t}^{\ell} &=
\mathbf{1}
\left[
\rho_{\ell}(t) > \bar{\rho}_{i,\ell}
\right]
\cdot
\mathbf{1}
\left[
\nabla_{\phi_{i,b}^{\ell}} \mathcal{L}_F > 0
 \right], 
\quad 
     \rho_{\ell}(t)  := \frac{1}{|\remain|}
\sum_{j\in\remain}
\mathbf{1}\left[t_j^\ell=k\right],\quad\bar{\rho}_{i,\ell}
=
\mathbb{E}_{q_\phi}\rho_\ell(k)
\end{split}
\end{equation} Here, $\rho_{\ell}(t)$ measures how frequently token $t$ appears in the retain set, while $\bar{\rho}_{i,\ell}$ denotes the expected retain-token frequency under the current assignment distribution $q_{\phi}$. The key intuition is that tokens heavily shared by forget and retain items are more likely to induce gradient conflict and concept leakage. Therefore, we restrict the update of $\phi$ to entries associated with high-overlap tokens according to Eq.~\ref{eq:mask}, while leaving irrelevant token assignments unchanged.

\subsection{Theoretical Analysis}
\label{sec:theory}

We define \(\Omega\) as the average token overlap between a forget item
\(i\in\mathcal F\) and a retain item \(j\in\mathcal R\), averaged over all
codebook levels. Specifically,
\begin{equation}
\Omega
=
\mathbb{E}_{i\sim\mathcal F,\;j\sim\mathcal R}
\left[
\omega(i,j)
\right],
\qquad
\omega(i,j)
=
\frac{1}{L}
\sum_{\ell=1}^{L}
\sum_{k\in\mathcal T^\ell}
\mathbf{1}[s_i^\ell=k]\mathbf{1}[s_j^\ell=k].
\end{equation}
Here, \(\mathcal T^\ell\) denotes the token set at codebook level \(\ell\),
and \(s_i^\ell\) is the token assigned to item \(i\) at level \(\ell\).

\begin{proposition}
\label{prop:ulb-retain}
Suppose Assumption~\ref{asm:loss} holds. Let
\[
\theta_+
:=
\theta
-
\eta\nabla_\theta \mathcal L_{\mathrm{unlearn}}(\theta)
\]
be the model parameters after one gradient descent step on the unlearning
objective. Then the retain-loss change satisfies
\begin{equation}
\label{eq:retain-change-upper-simplified}
\mathcal L_R(\theta_+) - \mathcal L_R(\theta)
\le
\eta\alpha\frac{\xi^2}{L}\Omega
+
C_U .
\end{equation}
Furthermore, if for any fixed token \(t\) and any conditioning sequences
\(\mathcal T,\mathcal T'\), the token-level gradients satisfy\looseness-1
\[
\left\langle
\nabla_\theta \ell(\theta;t\mid\mathcal T),
\nabla_\theta \ell(\theta;t\mid\mathcal T')
\right\rangle
\ge
\gamma
>
0,
\]
then
\begin{equation}
\label{eq:retain-change-lower-simplified}
\mathcal L_R(\theta_+) - \mathcal L_R(\theta)
\ge
\eta\alpha\frac{\gamma}{L}\Omega
+
C_L .
\end{equation}
Here, \(C_U\) and \(C_L\) collect residual terms and
second-order optimization remainders.
\end{proposition}


We defer the proof and further discussion of
 to Appendix~\ref{apx:theory}.

\begin{lemma}
\label{lem:reduce_con}
Under Assumption~\ref{asm:overlap}(Appendix), fix the model parameter \(\theta\) and
let
\[
    \phi_+
    =
    \phi
    -
    \eta_\phi
    M\odot
    \nabla_{\phi}\mathcal L_{\mathrm{unlearn}}(\phi),
\]
where \(M\) is defined in Eq.~\ref{eq:mask} and \(\odot\) denotes
element-wise product. For sufficiently small \(\eta_\phi\), we have
\begin{equation*}
    \Omega(\phi_+) - \Omega(\phi) \leq 0,
    \qquad
    \mathcal L_F(\phi_+) - \mathcal L_F(\phi) \leq 0.
\end{equation*}
\end{lemma}

This suggests that token reassignment can help mitigate retain-loss
degradation by suppressing overlap-mediated forget-retain interference
during unlearning.

\section{Experiments}
\paragraph{Datasets and preprocessing.}
We evaluate our TRACER on three real-world Amazon subsets, namely
\textit{Industrial\_and\_Scientific}, \textit{Sports\_and\_Outdoors}, and
\textit{Toys\_and\_Games}, which cover diverse recommendation domains.
Following \cite{kong2025minionerec}, we remove users and items with fewer than five interactions, and then split each dataset chronologically into train ($80\%$), validate (10\%), and test (10\%).\looseness-1

\paragraph{Concept unlearning setup.}
In the main text, we use brand as the target concept to erase. Brand information is explicitly provided in the metadata and naturally shared across multiple items, making it well-suited for concept-level forgetting. For each dataset, we randomly select 5\% of brands as target concepts and construct the forget set from interactions whose target items are associated with these brands, which account for approximately 10\% of the full dataset. To further evaluate the versatility of our framework beyond a single metadata concept, we additionally report appendix experiments on price band, where valid item prices are discretized into several coarse-grained groups, and the same unlearning protocol is applied.

\paragraph{Models.}
We evaluate our framework on a diverse collection of generative recommendation models
spanning both separate-training and joint-training paradigms.
The separate-training models include \textit{MiniOneRec}~\cite{kong2025minionerec}, \textit{Tiger}~\cite{rajput2023recommender},
\textit{P5-SID}~\cite{wang2024learnable} while the joint-training models include
\textit{LETTER}~\cite{wang2024learnable} and \textit{ETEGRec}~\cite{liu2025generative}.
These models differ substantially in architecture and the way
recommendation signals are integrated with language modeling.

\paragraph{Backbones.}
MiniOneRec is built on Qwen2.5-7B~\cite{yang2025qwen3}, whereas Tiger, P5-SID, LETTER, and ETEGRec are all based on the T5~\cite{raffel2020exploring}.
This backbone diversity—spanning different model scales, generation paradigms, and training strategies—allows us to evaluate whether TRACER generalizes across fundamentally different architectures.
\paragraph{Baselines.}
Since GenRec models rely on LLMs to generate semantic tokens, we compare TRACER with representative LLM concept unlearning methods, including ELM~\cite{gandikota2024erasing}, RMU~\cite{li2024wmdp}, RepNoise~\cite{rosati2024representation}, and FLAT~\cite{wang2024llm}. In addition, we include recent unlearning methods designed to mitigate retain-forget conflicts, such as BLUR~\cite{reisizadeh2026blur} and PISCES~\cite{gur2025precise}.




\paragraph{Evaluation metrics.}
We evaluate methods from three complementary dimensions: \emph{utility},
\emph{forgetting effectiveness}, and \emph{recommendation coherence}. Implementation
details are provided in the appendix.\looseness-1

\textbf{1. Utility.}
We measure recommendation quality on the retain set using standard ranking metrics,
including \textit{NDCG}, \textit{MRR}, and \textit{HR}. These metrics assess
whether the model preserves its ranking ability on non-forgotten data after
unlearning.

\textbf{2. Unlearning Effectiveness.}
We evaluate forgetting on the forget set using the same ranking metrics,
\textit{NDCG}, \textit{MRR}. Lower scores on the forget set
indicate more effective removal of the target information. 

\textbf{3. Recommendation Coherence.}
To evaluate whether the model remains semantically coherent after unlearning, we report the semantic similarity between the generated items and the original target items. The similarity is computed using item descriptions encoded by the frozen encoder described in Section~\ref{sec:GR}. This metric captures whether the model can still produce conceptually consistent recommendations while suppressing the target concept.

\section{Results}
We evaluate TRACER from four perspectives: 1) unlearning effectiveness; 2) flexibility with different GenRec backbones; 3) robustness to attacks; and 4) the contributions of each component of TRACER.


\paragraph{TRACER achieves effective and targeted unlearning}

Across multiple GenRec models, TRACER consistently achieves the strongest unlearning effectiveness while preserving recommendation utility (see Table~\ref{tab:main_results}). 
Specifically on MOR, TRACER obtains the best HR@5, HR@10, NDCG@5, NDCG@10, MRR scores, showing that it most effectively removes the target concept. On Forget HR@5, TRACER outperforms PISCES, the best performing baseline, more than 20\%. On Retain HR@5, TRACER outperforms all unlearning methods. We observe similar patterns on other metrics and other GenRec models.
%
Among the baselines, BLUR and FLAT explicitly consider gradient conflict in their formulation. We discovery that BLUR, a method optimized for retain-forget conflict, has poor unlearning effectiveness. This is because BLUR explicitly prioritize retain performance over forget effectiveness.

\providecommand{\ms} [2]{$#1 {\scriptstyle \pm #2}$}
\providecommand{\msb}[2]{$\mathbf{#1} {\scriptstyle \pm #2}$}
\providecommand{\msu}[2]{$\underline{#1} {\scriptstyle \pm #2}$}

\begin{table}[]
\caption{Unlearning results on \textbf{Industrial \& Scientific} and \textbf{Sports \& Outdoors} using MOR and LETTER as backbones. Bold = best among unlearning methods; underline = second best. Retain metrics ($\uparrow$): higher is better; Forget metrics ($\downarrow$): lower is better; Sim.~($\uparrow$): higher is better.}
\label{tab:main_results}
\resizebox{1.0\textwidth}{!}{
\begin{tabular}{lccccc|ccccc}
 \toprule
\multirow{2}{*}{\textbf{Method}} & \multicolumn{2}{c}{\textbf{Forget} $\downarrow$} & \multicolumn{2}{c}{\textbf{Retain} $\uparrow$} & \textbf{Sem.} $\uparrow$ & \multicolumn{2}{c}{\textbf{Forget} $\downarrow$} & \multicolumn{2}{c}{\textbf{Retain} $\uparrow$} & \textbf{Sem.} $\uparrow$ \\ 
\cmidrule(lr){2-3}\cmidrule(lr){4-5}\cmidrule(l){6-6}\cmidrule(l){7-8} \cmidrule(lr){9-10}\cmidrule(lr){11-11}
 & HR@5 & NDC@5 & HR@5 & NDC@5 & Sim. & HR@5 & NDC@5 & HR@5 & NDC@5 & Sim. \\ 
 \cmidrule{1-11}
 \cmidrule{1-11}
& \multicolumn{5}{c|}{\textbf{Industrial \& Scientific, MOR}} & \multicolumn{5}{c}{\textbf{Sports \& Outdoors, MOR}} \\ \hline
Original & \ms{0.170}{.003} & \ms{0.110}{.002} & \ms{0.126}{.003} & \ms{0.100}{.002} & \ms{0.851}{.005} & \ms{0.090}{.002} & \ms{0.063}{.002} & \ms{0.078}{.002} & \ms{0.067}{.002} & \ms{0.840}{.005} \\ \cmidrule{2-11}
ELM & \ms{0.143}{.004} & \ms{0.091}{.003} & \ms{0.084}{.004} & \ms{0.068}{.003} & \msu{0.824}{.006} & \ms{0.074}{.003} & \ms{0.055}{.002} & \ms{0.051}{.004} & \ms{0.044}{.002} & \msu{0.812}{.006} \\
RMU & \ms{0.151}{.005} & \ms{0.094}{.003} & \ms{0.061}{.005} & \ms{0.051}{.004} & \ms{0.730}{.008} & \ms{0.078}{.004} & \ms{0.054}{.003} & \ms{0.037}{.005} & \ms{0.033}{.003} & \ms{0.720}{.007} \\
REPNoise & \ms{0.131}{.004} & \ms{0.081}{.003} & \ms{0.068}{.004} & \ms{0.056}{.003} & \ms{0.630}{.008} & \ms{0.069}{.003} & \ms{0.044}{.002} & \ms{0.058}{.003} & \ms{0.037}{.002} & \ms{0.626}{.007} \\
FLAT & \ms{0.119}{.003} & \ms{0.074}{.002} & \ms{0.119}{.003} & \ms{0.091}{.003} & \ms{0.710}{.007} & \ms{0.059}{.002} & \ms{0.039}{.002} & \msu{0.074}{.003} & \ms{0.062}{.002} & \ms{0.693}{.006} \\
BLUR & \ms{0.139}{.004} & \ms{0.087}{.003} & \msu{0.120}{.003} & \msu{0.095}{.003} & \ms{0.680}{.006} & \ms{0.073}{.003} & \ms{0.047}{.002} & \ms{0.072}{.003} & \msu{0.065}{.002} & \ms{0.666}{.006} \\
PISCES & \msu{0.101}{.003} & \msu{0.066}{.002} & \ms{0.115}{.003} & \ms{0.090}{.002} & \ms{0.790}{.005} & \msu{0.052}{.002} & \msu{0.038}{.001} & \ms{0.072}{.002} & \ms{0.063}{.002} & \ms{0.773}{.005} \\ \cmidrule{2-11}
\textbf{TRACER} & \msb{0.080}{.002} & \msb{0.058}{.001} & \msb{0.123}{.002} & \msb{0.097}{.002} & \msb{0.828}{.004} & \msb{0.044}{.001} & \msb{0.034}{.001} & \msb{0.078}{.002} & \msb{0.066}{.002} & \msb{0.825}{.004} \\ 
\cmidrule{1-11}
\cmidrule{1-11}
& \multicolumn{5}{c|}{\textbf{Industrial \& Scientific, LETTER}} & \multicolumn{5}{c}{\textbf{Sports \& Outdoors, LETTER}} \\ \cmidrule{1-11}
Original & \ms{0.192}{.004} & \ms{0.121}{.003} & \ms{0.128}{.003} & \ms{0.103}{.002} & \ms{0.832}{.005} & \ms{0.103}{.002} & \ms{0.071}{.002} & \ms{0.081}{.002} & \ms{0.069}{.002} & \ms{0.821}{.005} \\
ELM & \ms{0.167}{.004} & \ms{0.104}{.003} & \ms{0.086}{.004} & \ms{0.070}{.003} & \msu{0.821}{.006} & \ms{0.082}{.003} & \ms{0.057}{.002} & \ms{0.053}{.004} & \ms{0.057}{.002} & \msu{0.802}{.006} \\
RMU & \ms{0.178}{.005} & \ms{0.108}{.004} & \ms{0.059}{.005} & \ms{0.049}{.004} & \ms{0.730}{.007} & \ms{0.085}{.004} & \ms{0.057}{.003} & \ms{0.061}{.004} & \ms{0.031}{.003} & \ms{0.725}{.007} \\
REPNoise & \ms{0.145}{.004} & \ms{0.090}{.003} & \ms{0.070}{.004} & \ms{0.057}{.003} & \ms{0.690}{.008} & \ms{0.070}{.003} & \ms{0.055}{.002} & \ms{0.062}{.004} & \ms{0.050}{.002} & \ms{0.681}{.007} \\
FLAT & \ms{0.148}{.004} & \ms{0.091}{.003} & \ms{0.113}{.003} & \ms{0.089}{.003} & \ms{0.710}{.007} & \ms{0.078}{.003} & \ms{0.054}{.002} & \ms{0.071}{.003} & \msu{0.066}{.002} & \ms{0.686}{.006} \\
BLUR & \ms{0.164}{.004} & \ms{0.101}{.003} & \msu{0.123}{.003} & \msu{0.096}{.003} & \ms{0.720}{.006} & \ms{0.079}{.003} & \ms{0.056}{.002} & \ms{0.076}{.003} & \ms{0.065}{.002} & \ms{0.713}{.006} \\
PISCES & \msu{0.125}{.003} & \msu{0.077}{.002} & \ms{0.117}{.003} & \ms{0.092}{.002} & \ms{0.807}{.005} & \msu{0.061}{.002} & \msu{0.045}{.002} & \msu{0.077}{.002} & \ms{0.063}{.002} & \ms{0.782}{.005} \\ \cmidrule{2-11}
\textbf{TRACER} & \msb{0.102}{.002} & \msb{0.068}{.002} & \msb{0.124}{.002} & \msb{0.098}{.002} & \msb{0.822}{.004} & \msb{0.054}{.001} & \msb{0.041}{.001} & \msb{0.078}{.002} & \msb{0.067}{.002} & \msb{0.821}{.003} \\ \bottomrule
\end{tabular}
}
\end{table}

\paragraph{TRACER preserves recommendation ability and coherence.}
Meanwhile, TRACER maintains nearly the same utility as the original model on the retain split. On average, it preserves 97.1\% of the original model performance across HR, NDCG, and MRR. In particular, its HR@10 reaches 0.159 compared with 0.162 for the original model, retaining approximately 98.0\% of the original HR@10. TRACER is also the best or tied-best unlearning method in 24 out of 25 retain-side comparisons, suggesting that it selectively suppresses forgotten behavior rather than broadly damaging the recommender.
The coherence results further show that TRACER does not achieve forgetting by simply disrupting generation quality. TRACER reaches an average semantic similarity of 0.834, close to the original model's 0.852 and higher than all evaluated unlearning baselines. Its relative coherence drop is only about 2.0\%, indicating that TRACER preserves the model's semantic structure while weakening the information targeted for removal.
In summary, these results demonstrate that TRACER can unlearn conceptual knowledge from GenRec models in a targeted and effective manner.  We show the item-level unlearning in Appendix~\ref{app:item_level}.



\paragraph{TRACER Enables Backbone-Agnostic GenRec Unlearning}
We conduct experiments across different backbones, including both hard-tokenization and end-to-end GenRec models. On TIGER and MOR, TRACER demonstrates superior performance, achieving $16.5\%$ lower targeting scores than PISCES while preserving $96.8\%$ of the original model utility on unrelated samples across HR, NDCG, and MRR. For end-to-end backbones such as LETTER and ETEGRec, TRACER also consistently achieves the best performance in both concept erasing and utility preservation. 

\paragraph{TRACER is Efficient}
Moreover, TRACER runs $33.5\%$ and $36.7\%$ faster than PISCES and REPNoise, respectively. TRACER also requires the least time to achieve effective unlearning compared with existing methods. Additional experimental results and efficiency analysis are provided in Appendix~\ref{sec:additional_results}.



\begin{figure*}[t]
    \centering
    \vspace{-10pt}
    \includegraphics[width=0.85\linewidth]{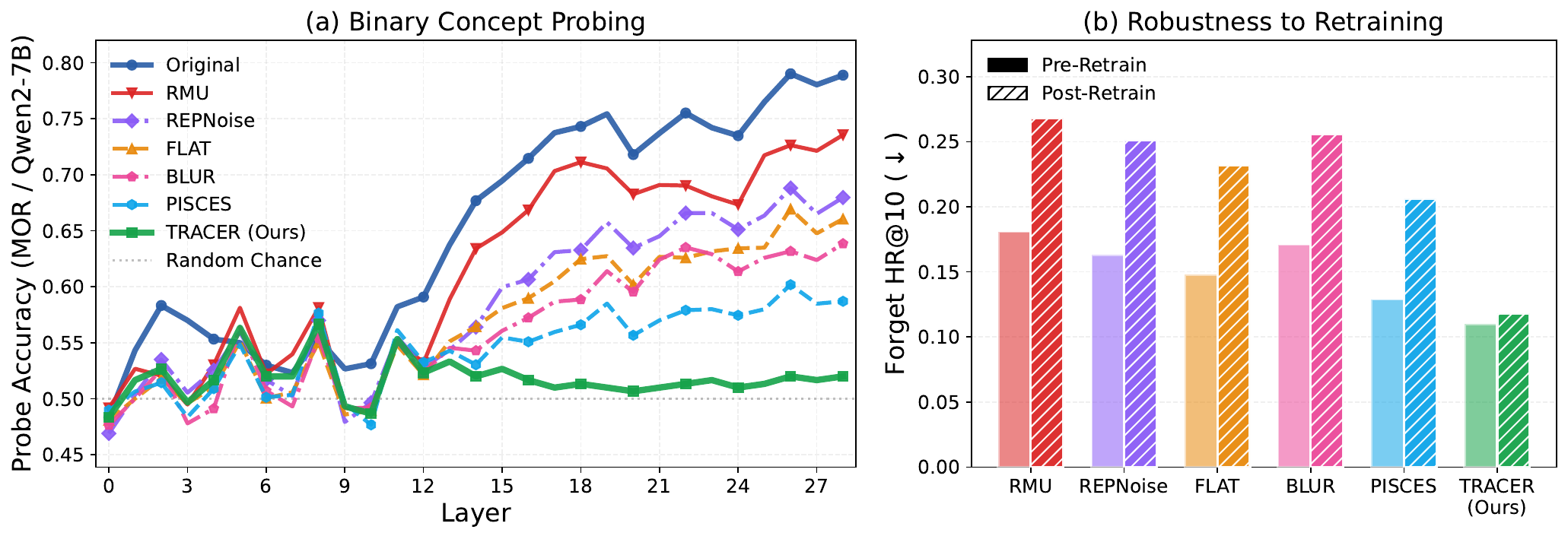}
    \caption{Probing Results on MOR.}
    \label{fig:probing}
\end{figure*}

\paragraph{TRACER is Robustness to Attacks}
This setting tests whether the erased concept can be relearned from concept-related interactions under the post-unlearning tokenizer. Following~\cite{li2024wmdp}, we evaluate the robustness of TRACER against fine-tuning attacks. We construct a stronger fine-tuning set in which $20\%$ of the target items are associated with target concept. For these items, we use their post-reassignment SIDs, corresponding to the deployed SID space after unlearning. As shown in Figure~\ref{fig:probing} (right), TRACER remains robust against fine-tuning attacks even after 2000 fine-tuning steps.\looseness-1

\paragraph{Concept Probing Analysis}
To estimate whether the erased concept remains in the model’s internal representations, we conduct a probing analysis following~\cite{gandikota2024erasing}. This analysis examines whether the model still encodes the target concept even if it no longer explicitly generates or recommends it. Specifically, we treat interactions in $\forget$ as positive samples, and randomly sample the same number of interactions from $D_r^c$ as negative samples. We then train a linear probe on the hidden representations from each model layer and report the probing accuracy across layers. As shown in Figure~\ref{fig:probing}(left), TRACER and PISCES consistently achieve low probing accuracies, only slightly above random guessing, across all layers for brand-related items, indicating that the target concept is effectively removed from the internal representations.



\begin{figure*}[t]
    \centering
    \includegraphics[width=0.85\linewidth]{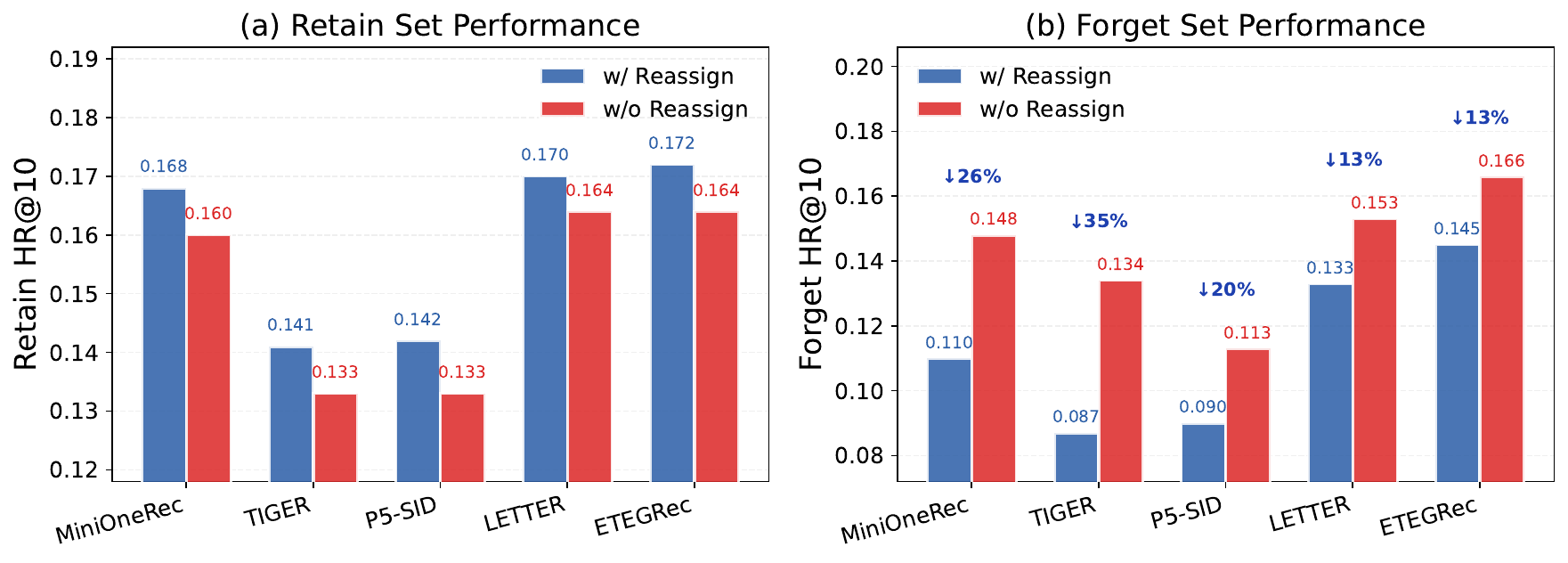}
    \caption{Effect of Reassignment.}
    \label{fig:ablation_1}
\end{figure*}

\paragraph{Ablation Study}
We conduct ablation studies to examine the contribution of each component of TRACER. Figure~\ref{fig:ablation_1} shows the impact of removing the reassignment module across different GenRec backbones, demonstrating that it substantially degrades both model utility and unlearning effectiveness. Figure~\ref{fig:ablation_lambda}(Appendix) studies the effect of the hyperparameter $\lambda$ in Eq.~\ref{eq:joint}. The results show that $\mathcal{L}_{\mathrm{Coh}}$ is crucial for maintaining recommendation coherence, while the regularizer is important for preserving model utility. More fine-grained ablations and hyperparameter analyses are provided in Appendix~\ref{sec:ablation_K_tau}.


\paragraph{Limitations.}
TRACER is designed for concept unlearning in generative recommendation and therefore assumes that sensitive information is associated with semantic concepts rather than directly encoded in item identifiers (SIDs). Extending the framework to settings where personal information is embedded within SIDs or other latent representations remains an important direction for future work. In addition, while TRACER reassigns target concept tokens to tokens that are infrequently used by retain items, the potential interaction between this reassignment strategy and the popularity distribution of items has not been systematically investigated. In particular, low-popularity retain items whose tokens overlap with reassignment targets may be disproportionately affected. Although our experiments demonstrate strong overall recommendation performance, a more detailed analysis of fairness and popularity-related effects would provide a more comprehensive understanding of the proposed approach.
\section{Conclusion}

We introduce the first concept unlearning task for generative recommendation (GenRec), where the goal is to remove target concept information while preserving recommendation utility and semantic coherence. To address this task, we propose TRACER, a token reassignment framework that learns to reassign GenRec semantic tokens for effective unlearning, while preserving recommendation coherence. 
Extensive experiments demonstrate the effectiveness~\cite{zhu2026medeyes}, flexibility, and robustness of TRACER, which outperforms existing unlearning baselines.


\bibliographystyle{ACM-Reference-Format}
\bibliography{reference}


\begin{thebibliography}{62}


\ifx \showCODEN    \undefined \def \showCODEN     #1{\unskip}     \fi
\ifx \showISBNx    \undefined \def \showISBNx     #1{\unskip}     \fi
\ifx \showISBNxiii \undefined \def \showISBNxiii  #1{\unskip}     \fi
\ifx \showISSN     \undefined \def \showISSN      #1{\unskip}     \fi
\ifx \showLCCN     \undefined \def \showLCCN      #1{\unskip}     \fi
\ifx \shownote     \undefined \def \shownote      #1{#1}          \fi
\ifx \showarticletitle \undefined \def \showarticletitle #1{#1}   \fi
\ifx \showURL      \undefined \def \showURL       {\relax}        \fi
\providecommand\bibfield[2]{#2}
\providecommand\bibinfo[2]{#2}
\providecommand\natexlab[1]{#1}
\providecommand\showeprint[2][]{arXiv:#2}

\bibitem[Bourtoule et~al\mbox{.}(2021)]%
        {bourtoule2021machine}
\bibfield{author}{\bibinfo{person}{Lucas Bourtoule}, \bibinfo{person}{Varun Chandrasekaran}, \bibinfo{person}{Christopher~A Choquette-Choo}, \bibinfo{person}{Hengrui Jia}, \bibinfo{person}{Adelin Travers}, \bibinfo{person}{Baiwu Zhang}, \bibinfo{person}{David Lie}, {and} \bibinfo{person}{Nicolas Papernot}.} \bibinfo{year}{2021}\natexlab{}.
\newblock \showarticletitle{Machine unlearning}. In \bibinfo{booktitle}{\emph{IEEE Symposium on Security and Privacy (SP)}}.
\newblock


\bibitem[Cao et~al\mbox{.}(2024)]%
        {cao2024rwku}
\bibfield{author}{\bibinfo{person}{Pengfei Cao}, \bibinfo{person}{Chenhao Wang}, \bibinfo{person}{Zhitao He}, \bibinfo{person}{Hongbang Yuan}, \bibinfo{person}{Jiachun Li}, \bibinfo{person}{Yubo Chen}, \bibinfo{person}{Kang Liu}, \bibinfo{person}{Jun Zhao}, {et~al\mbox{.}}} \bibinfo{year}{2024}\natexlab{}.
\newblock \showarticletitle{Rwku: Benchmarking real-world knowledge unlearning for large language models}.
\newblock \bibinfo{journal}{\emph{Advances in Neural Information Processing Systems}}  \bibinfo{volume}{37} (\bibinfo{year}{2024}), \bibinfo{pages}{98213--98263}.
\newblock


\bibitem[Cao and Yang(2015)]%
        {cao2015towards}
\bibfield{author}{\bibinfo{person}{Yinzhi Cao} {and} \bibinfo{person}{Junfeng Yang}.} \bibinfo{year}{2015}\natexlab{}.
\newblock \showarticletitle{Towards making systems forget with machine unlearning}. In \bibinfo{booktitle}{\emph{Proceedings of the IEEE Symposium on Security and Privacy}}.
\newblock


\bibitem[Chen et~al\mbox{.}(2022)]%
        {chen2022recommendation}
\bibfield{author}{\bibinfo{person}{Chong Chen}, \bibinfo{person}{Fei Sun}, \bibinfo{person}{Min Zhang}, {and} \bibinfo{person}{Bolin Ding}.} \bibinfo{year}{2022}\natexlab{}.
\newblock \showarticletitle{Recommendation unlearning}. In \bibinfo{booktitle}{\emph{Proceedings of the ACM web conference 2022}}. \bibinfo{pages}{2768--2777}.
\newblock


\bibitem[Chen et~al\mbox{.}(2026b)]%
        {chen2026open}
\bibfield{author}{\bibinfo{person}{Junkai Chen}, \bibinfo{person}{Yuhao He}, \bibinfo{person}{Qianshan Wei}, \bibinfo{person}{Junxiang You}, \bibinfo{person}{Jingwen Shao}, \bibinfo{person}{Junkai Lin}, \bibinfo{person}{Zhongkai Yue}, \bibinfo{person}{Xiaotian Ye}, \bibinfo{person}{Zhengbo Jiao}, \bibinfo{person}{Jiali Cheng}, {et~al\mbox{.}}} \bibinfo{year}{2026}\natexlab{b}.
\newblock \showarticletitle{Open-MMUnlearning: Unifying Methods and Evaluation for MLLM Unlearning}.
\newblock \bibinfo{journal}{\emph{arXiv preprint arXiv:2610.10358}} (\bibinfo{year}{2026}).
\newblock


\bibitem[Chen et~al\mbox{.}(2025a)]%
        {chen2025frog}
\bibfield{author}{\bibinfo{person}{Ziheng Chen}, \bibinfo{person}{Jiali Cheng}, \bibinfo{person}{Hadi Amiri}, \bibinfo{person}{Kaushiki Nag}, \bibinfo{person}{Lu Lin}, \bibinfo{person}{Sijia Liu}, \bibinfo{person}{Gabriele Tolomei}, {and} \bibinfo{person}{Xiangguo Sun}.} \bibinfo{year}{2025}\natexlab{a}.
\newblock \showarticletitle{FROG: Fair Removal on Graph}. In \bibinfo{booktitle}{\emph{Proceedings of the 34th ACM International Conference on Information and Knowledge Management}}. \bibinfo{pages}{415--424}.
\newblock


\bibitem[Chen et~al\mbox{.}(2026a)]%
        {chen2026cure}
\bibfield{author}{\bibinfo{person}{Ziheng Chen}, \bibinfo{person}{Jiali Cheng}, \bibinfo{person}{Zezhong Fan}, \bibinfo{person}{Hadi Amiri}, \bibinfo{person}{Yunzhi Yao}, \bibinfo{person}{Xiangguo Sun}, {and} \bibinfo{person}{Yang Zhang}.} \bibinfo{year}{2026}\natexlab{a}.
\newblock \showarticletitle{CURE: Circuit-Aware Unlearning for LLM-based Recommendation}.
\newblock \bibinfo{journal}{\emph{arXiv preprint arXiv:2604.04982}} (\bibinfo{year}{2026}).
\newblock


\bibitem[Chen et~al\mbox{.}(2025b)]%
        {chen2025future}
\bibfield{author}{\bibinfo{person}{Ziheng Chen}, \bibinfo{person}{Jin Huang}, \bibinfo{person}{Jiali Cheng}, \bibinfo{person}{Yuchan Guo}, \bibinfo{person}{Mengjie Wang}, \bibinfo{person}{Lalitesh Morishetti}, \bibinfo{person}{Kaushiki Nag}, {and} \bibinfo{person}{Hadi Amiri}.} \bibinfo{year}{2025}\natexlab{b}.
\newblock \showarticletitle{FUTURE: Flexible Unlearning for Tree Ensemble}. In \bibinfo{booktitle}{\emph{Proceedings of the 34th ACM International Conference on Information and Knowledge Management}}. \bibinfo{pages}{4680--4684}.
\newblock


\bibitem[Cheng and Amiri(2024)]%
        {cheng2024mubench}
\bibfield{author}{\bibinfo{person}{Jiali Cheng} {and} \bibinfo{person}{Hadi Amiri}.} \bibinfo{year}{2024}\natexlab{}.
\newblock \showarticletitle{MU-bench: A multitask multimodal benchmark for machine unlearning}.
\newblock \bibinfo{journal}{\emph{arXiv preprint arXiv:2406.14796}} (\bibinfo{year}{2024}).
\newblock


\bibitem[Cheng and Amiri(2025a)]%
        {cheng2023multimodal}
\bibfield{author}{\bibinfo{person}{Jiali Cheng} {and} \bibinfo{person}{Hadi Amiri}.} \bibinfo{year}{2025}\natexlab{a}.
\newblock \showarticletitle{MultiDelete for Multimodal Machine Unlearning}. In \bibinfo{booktitle}{\emph{Computer Vision -- ECCV 2024}}, \bibfield{editor}{\bibinfo{person}{Ale{\v{s}} Leonardis}, \bibinfo{person}{Elisa Ricci}, \bibinfo{person}{Stefan Roth}, \bibinfo{person}{Olga Russakovsky}, \bibinfo{person}{Torsten Sattler}, {and} \bibinfo{person}{G{\"u}l Varol}} (Eds.). \bibinfo{publisher}{Springer Nature Switzerland}, \bibinfo{address}{Cham}, \bibinfo{pages}{165--184}.
\newblock
\showISBNx{978-3-031-72940-9}


\bibitem[Cheng and Amiri(2025b)]%
        {cheng25d_interspeech}
\bibfield{author}{\bibinfo{person}{Jiali Cheng} {and} \bibinfo{person}{Hadi Amiri}.} \bibinfo{year}{2025}\natexlab{b}.
\newblock \showarticletitle{{Speech Unlearning}}. In \bibinfo{booktitle}{\emph{{Interspeech 2025}}}. \bibinfo{pages}{3209--3213}.
\newblock
\showISSN{2958-1796}
\href{https://doi.org/10.21437/Interspeech.2025-2412}{doi:\nolinkurl{10.21437/Interspeech.2025-2412}}


\bibitem[Cheng and Amiri(2025c)]%
        {cheng2025tool}
\bibfield{author}{\bibinfo{person}{Jiali Cheng} {and} \bibinfo{person}{Hadi Amiri}.} \bibinfo{year}{2025}\natexlab{c}.
\newblock \showarticletitle{Tool Unlearning for Tool-Augmented {LLM}s}. In \bibinfo{booktitle}{\emph{Forty-second International Conference on Machine Learning}}.
\newblock


\bibitem[Cheng and Amiri(2026)]%
        {Cheng2026}
\bibfield{author}{\bibinfo{person}{Jiali Cheng} {and} \bibinfo{person}{Hadi Amiri}.} \bibinfo{year}{2026}\natexlab{}.
\newblock \bibinfo{booktitle}{\emph{Machine Unlearning Across Tasks and Modalities}}.
\newblock \bibinfo{publisher}{Springer Nature Switzerland}, \bibinfo{address}{Cham}, \bibinfo{pages}{227--240}.
\newblock
\showISBNx{978-3-032-17282-2}
\href{https://doi.org/10.1007/978-3-032-17282-2_15}{doi:\nolinkurl{10.1007/978-3-032-17282-2_15}}


\bibitem[Cheng et~al\mbox{.}(2026a)]%
        {cheng2026toward}
\bibfield{author}{\bibinfo{person}{Jiali Cheng}, \bibinfo{person}{Ziheng Chen}, \bibinfo{person}{Chirag Agarwal}, {and} \bibinfo{person}{Hadi Amiri}.} \bibinfo{year}{2026}\natexlab{a}.
\newblock \showarticletitle{Toward Understanding Unlearning Difficulty: A Mechanistic Perspective and Circuit-Guided Difficulty Metric}.
\newblock \bibinfo{journal}{\emph{arXiv preprint arXiv:2601.09624}} (\bibinfo{year}{2026}).
\newblock


\bibitem[Cheng et~al\mbox{.}(2023)]%
        {cheng2023gnndelete}
\bibfield{author}{\bibinfo{person}{Jiali Cheng}, \bibinfo{person}{George Dasoulas}, \bibinfo{person}{Huan He}, \bibinfo{person}{Chirag Agarwal}, {and} \bibinfo{person}{Marinka Zitnik}.} \bibinfo{year}{2023}\natexlab{}.
\newblock \showarticletitle{{GNND}elete: A General Strategy for Unlearning in Graph Neural Networks}. In \bibinfo{booktitle}{\emph{The Eleventh International Conference on Learning Representations}}.
\newblock
\urldef\tempurl%
\url{https://openreview.net/forum?id=X9yCkmT5Qrl}
\showURL{%
\tempurl}


\bibitem[Cheng et~al\mbox{.}(2026b)]%
        {cheng2026tmc}
\bibfield{author}{\bibinfo{person}{Jiali Cheng}, \bibinfo{person}{Rui Pan}, {and} \bibinfo{person}{Hadi Amiri}.} \bibinfo{year}{2026}\natexlab{b}.
\newblock \showarticletitle{Investigating Tool-Memory Conflicts in Tool-Augmented LLMs}.
\newblock \bibinfo{journal}{\emph{arXiv preprint arXiv:2601.09760}} (\bibinfo{year}{2026}).
\newblock


\bibitem[Deng et~al\mbox{.}(2025)]%
        {deng2025onerec}
\bibfield{author}{\bibinfo{person}{Jiaxin Deng}, \bibinfo{person}{Shiyao Wang}, \bibinfo{person}{Kuo Cai}, \bibinfo{person}{Lejian Ren}, \bibinfo{person}{Qigen Hu}, \bibinfo{person}{Weifeng Ding}, \bibinfo{person}{Qiang Luo}, {and} \bibinfo{person}{Guorui Zhou}.} \bibinfo{year}{2025}\natexlab{}.
\newblock \showarticletitle{Onerec: Unifying retrieve and rank with generative recommender and iterative preference alignment}.
\newblock \bibinfo{journal}{\emph{arXiv preprint arXiv:2502.18965}} (\bibinfo{year}{2025}).
\newblock


\bibitem[Eldan and Russinovich(2023)]%
        {eldan2023whos}
\bibfield{author}{\bibinfo{person}{Ronen Eldan} {and} \bibinfo{person}{Mark Russinovich}.} \bibinfo{year}{2023}\natexlab{}.
\newblock \showarticletitle{Who's harry potter? approximate unlearning in llms}.
\newblock \bibinfo{journal}{\emph{arXiv preprint arXiv:2310.02238}} (\bibinfo{year}{2023}).
\newblock


\bibitem[Fan et~al\mbox{.}(2025)]%
        {fan2025towards}
\bibfield{author}{\bibinfo{person}{Chongyu Fan}, \bibinfo{person}{Jinghan Jia}, \bibinfo{person}{Yihua Zhang}, \bibinfo{person}{Anil Ramakrishna}, \bibinfo{person}{Mingyi Hong}, {and} \bibinfo{person}{Sijia Liu}.} \bibinfo{year}{2025}\natexlab{}.
\newblock \showarticletitle{Towards llm unlearning resilient to relearning attacks: A sharpness-aware minimization perspective and beyond}.
\newblock \bibinfo{journal}{\emph{arXiv preprint arXiv:2502.05374}} (\bibinfo{year}{2025}).
\newblock


\bibitem[Fan et~al\mbox{.}(2024)]%
        {fan2024salun}
\bibfield{author}{\bibinfo{person}{Chongyu Fan}, \bibinfo{person}{Jiancheng Liu}, \bibinfo{person}{Yihua Zhang}, \bibinfo{person}{Eric Wong}, \bibinfo{person}{Dennis Wei}, {and} \bibinfo{person}{Sijia Liu}.} \bibinfo{year}{2024}\natexlab{}.
\newblock \showarticletitle{SalUn: Empowering Machine Unlearning via Gradient-based Weight Saliency in Both Image Classification and Generation}. In \bibinfo{booktitle}{\emph{The Twelfth International Conference on Learning Representations}}.
\newblock
\urldef\tempurl%
\url{https://openreview.net/forum?id=gn0mIhQGNM}
\showURL{%
\tempurl}


\bibitem[Fu et~al\mbox{.}(2025)]%
        {fu2025learning}
\bibfield{author}{\bibinfo{person}{Xingyu Fu}, \bibinfo{person}{Siyi Liu}, \bibinfo{person}{Yinuo Xu}, \bibinfo{person}{Pan Lu}, \bibinfo{person}{Guangqiuse Hu}, \bibinfo{person}{Tianbo Yang}, \bibinfo{person}{Taran Anantasagar}, \bibinfo{person}{Christopher Shen}, \bibinfo{person}{Yikai Mao}, \bibinfo{person}{Yuanzhe Liu}, {et~al\mbox{.}}} \bibinfo{year}{2025}\natexlab{}.
\newblock \showarticletitle{Learning human-perceived fakeness in ai-generated videos via multimodal llms}.
\newblock \bibinfo{journal}{\emph{arXiv preprint arXiv:2509.22646}} (\bibinfo{year}{2025}).
\newblock


\bibitem[Gandikota et~al\mbox{.}(2024)]%
        {gandikota2024erasing}
\bibfield{author}{\bibinfo{person}{Rohit Gandikota}, \bibinfo{person}{Sheridan Feucht}, \bibinfo{person}{Samuel Marks}, {and} \bibinfo{person}{David Bau}.} \bibinfo{year}{2024}\natexlab{}.
\newblock \showarticletitle{Erasing conceptual knowledge from language models}.
\newblock \bibinfo{journal}{\emph{arXiv preprint arXiv:2410.02760}} (\bibinfo{year}{2024}).
\newblock


\bibitem[Gandikota et~al\mbox{.}(2023)]%
        {gandikota2023erasing}
\bibfield{author}{\bibinfo{person}{Rohit Gandikota}, \bibinfo{person}{Joanna Materzy\'nska}, \bibinfo{person}{Jaden Fiotto-Kaufman}, {and} \bibinfo{person}{David Bau}.} \bibinfo{year}{2023}\natexlab{}.
\newblock \showarticletitle{Erasing Concepts from Diffusion Models}. In \bibinfo{booktitle}{\emph{Proceedings of the 2023 IEEE International Conference on Computer Vision}}.
\newblock


\bibitem[Golatkar et~al\mbox{.}(2020)]%
        {golatkar2020eternal}
\bibfield{author}{\bibinfo{person}{Aditya Golatkar}, \bibinfo{person}{Alessandro Achille}, {and} \bibinfo{person}{Stefano Soatto}.} \bibinfo{year}{2020}\natexlab{}.
\newblock \showarticletitle{Eternal sunshine of the spotless net: Selective forgetting in deep networks}. In \bibinfo{booktitle}{\emph{Proceedings of the IEEE/CVF conference on computer vision and pattern recognition}}. \bibinfo{pages}{9304--9312}.
\newblock


\bibitem[Guo et~al\mbox{.}(2019)]%
        {guo2019certified}
\bibfield{author}{\bibinfo{person}{Chuan Guo}, \bibinfo{person}{Tom Goldstein}, \bibinfo{person}{Awni Hannun}, {and} \bibinfo{person}{Laurens Van Der~Maaten}.} \bibinfo{year}{2019}\natexlab{}.
\newblock \showarticletitle{Certified data removal from machine learning models}.
\newblock \bibinfo{journal}{\emph{arXiv preprint arXiv:1911.03030}} (\bibinfo{year}{2019}).
\newblock


\bibitem[Gupta et~al\mbox{.}(2025)]%
        {gupta2025dameflame}
\bibfield{author}{\bibinfo{person}{Neha~R. Gupta}, \bibinfo{person}{Vittorio Orlandi}, \bibinfo{person}{Chia-Rui Chang}, \bibinfo{person}{Tianyu Wang}, \bibinfo{person}{Marco Morucci}, \bibinfo{person}{Pritam Dey}, \bibinfo{person}{Thomas~J. Howell}, \bibinfo{person}{Xian Sun}, \bibinfo{person}{Angikar Ghosal}, \bibinfo{person}{Sudeepa Roy}, \bibinfo{person}{Cynthia Rudin}, {and} \bibinfo{person}{Alexander Volfovsky}.} \bibinfo{year}{2025}\natexlab{}.
\newblock \showarticletitle{dame-flame: A Python Package Providing Fast Interpretable Matching for Causal Inference}.
\newblock \bibinfo{journal}{\emph{Journal of Statistical Software}} \bibinfo{volume}{113}, \bibinfo{number}{2} (\bibinfo{year}{2025}), \bibinfo{pages}{1--26}.
\newblock
\href{https://doi.org/10.18637/jss.v113.i02}{doi:\nolinkurl{10.18637/jss.v113.i02}}


\bibitem[Gur-Arieh et~al\mbox{.}(2025)]%
        {gur2025precise}
\bibfield{author}{\bibinfo{person}{Yoav Gur-Arieh}, \bibinfo{person}{Clara~Haya Suslik}, \bibinfo{person}{Yihuai Hong}, \bibinfo{person}{Fazl Barez}, {and} \bibinfo{person}{Mor Geva}.} \bibinfo{year}{2025}\natexlab{}.
\newblock \showarticletitle{Precise in-parameter concept erasure in large language models}. In \bibinfo{booktitle}{\emph{Proceedings of the 2025 Conference on Empirical Methods in Natural Language Processing}}. \bibinfo{pages}{18997--19017}.
\newblock


\bibitem[Han et~al\mbox{.}(2025)]%
        {han2025mtgr}
\bibfield{author}{\bibinfo{person}{Ruidong Han}, \bibinfo{person}{Bin Yin}, \bibinfo{person}{Shangyu Chen}, \bibinfo{person}{He Jiang}, \bibinfo{person}{Fei Jiang}, \bibinfo{person}{Xiang Li}, \bibinfo{person}{Chi Ma}, \bibinfo{person}{Mincong Huang}, \bibinfo{person}{Xiaoguang Li}, \bibinfo{person}{Chunzhen Jing}, {et~al\mbox{.}}} \bibinfo{year}{2025}\natexlab{}.
\newblock \showarticletitle{Mtgr: Industrial-scale generative recommendation framework in meituan}. In \bibinfo{booktitle}{\emph{Proceedings of the 34th ACM International Conference on Information and Knowledge Management}}. \bibinfo{pages}{5731--5738}.
\newblock


\bibitem[Jia et~al\mbox{.}(2024)]%
        {jia2024soul}
\bibfield{author}{\bibinfo{person}{Jinghan Jia}, \bibinfo{person}{Yihua Zhang}, \bibinfo{person}{Yimeng Zhang}, \bibinfo{person}{Jiancheng Liu}, \bibinfo{person}{Bharat Runwal}, \bibinfo{person}{James Diffenderfer}, \bibinfo{person}{Bhavya Kailkhura}, {and} \bibinfo{person}{Sijia Liu}.} \bibinfo{year}{2024}\natexlab{}.
\newblock \showarticletitle{SOUL: Unlocking the Power of Second-Order Optimization for LLM Unlearning}.
\newblock \bibinfo{journal}{\emph{arXiv preprint arXiv:2404.18239}} (\bibinfo{year}{2024}).
\newblock


\bibitem[Kong et~al\mbox{.}(2025)]%
        {kong2025minionerec}
\bibfield{author}{\bibinfo{person}{Xiaoyu Kong}, \bibinfo{person}{Leheng Sheng}, \bibinfo{person}{Junfei Tan}, \bibinfo{person}{Yuxin Chen}, \bibinfo{person}{Jiancan Wu}, \bibinfo{person}{An Zhang}, \bibinfo{person}{Xiang Wang}, {and} \bibinfo{person}{Xiangnan He}.} \bibinfo{year}{2025}\natexlab{}.
\newblock \showarticletitle{Minionerec: An open-source framework for scaling generative recommendation}.
\newblock \bibinfo{journal}{\emph{arXiv preprint arXiv:2510.24431}} (\bibinfo{year}{2025}).
\newblock


\bibitem[Lee et~al\mbox{.}(2022)]%
        {lee2022autoregressive}
\bibfield{author}{\bibinfo{person}{Doyup Lee}, \bibinfo{person}{Chiheon Kim}, \bibinfo{person}{Saehoon Kim}, \bibinfo{person}{Minsu Cho}, {and} \bibinfo{person}{Wook-Shin Han}.} \bibinfo{year}{2022}\natexlab{}.
\newblock \showarticletitle{Autoregressive image generation using residual quantization}. In \bibinfo{booktitle}{\emph{Proceedings of the IEEE/CVF conference on computer vision and pattern recognition}}. \bibinfo{pages}{11523--11532}.
\newblock


\bibitem[Li et~al\mbox{.}(2024)]%
        {li2024wmdp}
\bibfield{author}{\bibinfo{person}{Nathaniel Li}, \bibinfo{person}{Alexander Pan}, \bibinfo{person}{Anjali Gopal}, \bibinfo{person}{Summer Yue}, \bibinfo{person}{Daniel Berrios}, \bibinfo{person}{Alice Gatti}, \bibinfo{person}{Justin~D Li}, \bibinfo{person}{Ann-Kathrin Dombrowski}, \bibinfo{person}{Shashwat Goel}, \bibinfo{person}{Long Phan}, {et~al\mbox{.}}} \bibinfo{year}{2024}\natexlab{}.
\newblock \showarticletitle{The wmdp benchmark: Measuring and reducing malicious use with unlearning}.
\newblock \bibinfo{journal}{\emph{arXiv preprint arXiv:2403.03218}} (\bibinfo{year}{2024}).
\newblock


\bibitem[Lin et~al\mbox{.}(2026a)]%
        {lin2026volume}
\bibfield{author}{\bibinfo{person}{Luyun Lin}, \bibinfo{person}{Lixing Lin}, \bibinfo{person}{Zhen Zhang}, \bibinfo{person}{Moxuan Zheng}, {and} \bibinfo{person}{Yiqing Wang}.} \bibinfo{year}{2026}\natexlab{a}.
\newblock \showarticletitle{A Volume-Price-Adjusted MACD Trading Strategy with Sensitivity Calibration for US Equity Indices}.
\newblock \bibinfo{journal}{\emph{arXiv preprint arXiv:2604.26063}} (\bibinfo{year}{2026}).
\newblock


\bibitem[Lin and Wang(2025)]%
        {lin2025shap}
\bibfield{author}{\bibinfo{person}{Luyun Lin} {and} \bibinfo{person}{Yiqing Wang}.} \bibinfo{year}{2025}\natexlab{}.
\newblock \showarticletitle{Shap stability in credit risk management: A case study in credit card default model}.
\newblock \bibinfo{journal}{\emph{Risks}} \bibinfo{volume}{13}, \bibinfo{number}{12} (\bibinfo{year}{2025}), \bibinfo{pages}{238}.
\newblock


\bibitem[Lin et~al\mbox{.}(2026b)]%
        {lin2026reflect}
\bibfield{author}{\bibinfo{person}{Lixing Lin}, \bibinfo{person}{Juli You}, \bibinfo{person}{Yue Li}, \bibinfo{person}{Luyun Lin}, \bibinfo{person}{Yiqing Wang}, \bibinfo{person}{Zhen Zhang}, {and} \bibinfo{person}{Moxuan Zheng}.} \bibinfo{year}{2026}\natexlab{b}.
\newblock \showarticletitle{Reflect-Guard: Enhancing LLM Safeguards against Adversarial Prompts via Logical Self-Reflection}.
\newblock \bibinfo{journal}{\emph{arXiv preprint arXiv:2605.24834}} (\bibinfo{year}{2026}).
\newblock


\bibitem[Liu et~al\mbox{.}(2025b)]%
        {liu2025generative}
\bibfield{author}{\bibinfo{person}{Enze Liu}, \bibinfo{person}{Bowen Zheng}, \bibinfo{person}{Cheng Ling}, \bibinfo{person}{Lantao Hu}, \bibinfo{person}{Han Li}, {and} \bibinfo{person}{Wayne~Xin Zhao}.} \bibinfo{year}{2025}\natexlab{b}.
\newblock \showarticletitle{Generative recommender with end-to-end learnable item tokenization}. In \bibinfo{booktitle}{\emph{Proceedings of the 48th International ACM SIGIR Conference on Research and Development in Information Retrieval}}. \bibinfo{pages}{729--739}.
\newblock


\bibitem[Liu et~al\mbox{.}(2024)]%
        {liu2024rethinking}
\bibfield{author}{\bibinfo{person}{Sijia Liu}, \bibinfo{person}{Yuanshun Yao}, \bibinfo{person}{Jinghan Jia}, \bibinfo{person}{Stephen Casper}, \bibinfo{person}{Nathalie Baracaldo}, \bibinfo{person}{Peter Hase}, \bibinfo{person}{Yuguang Yao}, \bibinfo{person}{Chris~Yuhao Liu}, \bibinfo{person}{Xiaojun Xu}, \bibinfo{person}{Hang Li}, {et~al\mbox{.}}} \bibinfo{year}{2024}\natexlab{}.
\newblock \showarticletitle{Rethinking machine unlearning for large language models}.
\newblock \bibinfo{journal}{\emph{arXiv preprint arXiv:2402.08787}} (\bibinfo{year}{2024}).
\newblock


\bibitem[Liu et~al\mbox{.}(2026a)]%
        {liu2026improvingcompletenesscomparabilitysegment}
\bibfield{author}{\bibinfo{person}{Yue Liu}, \bibinfo{person}{Zhiyuan Cheng}, {and} \bibinfo{person}{Longying Lai}.} \bibinfo{year}{2026}\natexlab{a}.
\newblock \bibinfo{title}{Improving the Completeness and Comparability of Segment Disclosures: A Large Language Model Approach}.
\newblock
\showeprint[arxiv]{2605.23924}~[cs.CL]
\urldef\tempurl%
\url{https://arxiv.org/abs/2605.23924}
\showURL{%
\tempurl}


\bibitem[Liu et~al\mbox{.}(2026b)]%
        {liu2026palm}
\bibfield{author}{\bibinfo{person}{Yuanzhe Liu}, \bibinfo{person}{Jingyuan Zhu}, \bibinfo{person}{Yuchen Mo}, \bibinfo{person}{Gen Li}, \bibinfo{person}{Xu Cao}, \bibinfo{person}{Jin Jin}, \bibinfo{person}{Yifan Shen}, \bibinfo{person}{Zhengyuan Li}, \bibinfo{person}{Tianjiao Yu}, \bibinfo{person}{Wenzhen Yuan}, {et~al\mbox{.}}} \bibinfo{year}{2026}\natexlab{b}.
\newblock \showarticletitle{Palm: Progress-aware policy learning via affordance reasoning for long-horizon robotic manipulation}.
\newblock \bibinfo{journal}{\emph{arXiv preprint arXiv:2601.07060}} (\bibinfo{year}{2026}).
\newblock


\bibitem[Liu et~al\mbox{.}(2025a)]%
        {liu2025onerec}
\bibfield{author}{\bibinfo{person}{Zhanyu Liu}, \bibinfo{person}{Shiyao Wang}, \bibinfo{person}{Xingmei Wang}, \bibinfo{person}{Rongzhou Zhang}, \bibinfo{person}{Jiaxin Deng}, \bibinfo{person}{Honghui Bao}, \bibinfo{person}{Jinghao Zhang}, \bibinfo{person}{Wuchao Li}, \bibinfo{person}{Pengfei Zheng}, \bibinfo{person}{Xiangyu Wu}, {et~al\mbox{.}}} \bibinfo{year}{2025}\natexlab{a}.
\newblock \showarticletitle{Onerec-think: In-text reasoning for generative recommendation}.
\newblock \bibinfo{journal}{\emph{arXiv preprint arXiv:2510.11639}} (\bibinfo{year}{2025}).
\newblock


\bibitem[Lucic et~al\mbox{.}(2022)]%
        {lucic2022focus}
\bibfield{author}{\bibinfo{person}{Ana Lucic}, \bibinfo{person}{Harrie Oosterhuis}, \bibinfo{person}{Hinda Haned}, {and} \bibinfo{person}{Maarten De~Rijke}.} \bibinfo{year}{2022}\natexlab{}.
\newblock \showarticletitle{FOCUS: Flexible optimizable counterfactual explanations for tree ensembles}. In \bibinfo{booktitle}{\emph{Proceedings of the AAAI conference on artificial intelligence}}, Vol.~\bibinfo{volume}{36}. \bibinfo{pages}{5313--5322}.
\newblock


\bibitem[Raffel et~al\mbox{.}(2020)]%
        {raffel2020exploring}
\bibfield{author}{\bibinfo{person}{Colin Raffel}, \bibinfo{person}{Noam Shazeer}, \bibinfo{person}{Adam Roberts}, \bibinfo{person}{Katherine Lee}, \bibinfo{person}{Sharan Narang}, \bibinfo{person}{Michael Matena}, \bibinfo{person}{Yanqi Zhou}, \bibinfo{person}{Wei Li}, {and} \bibinfo{person}{Peter~J Liu}.} \bibinfo{year}{2020}\natexlab{}.
\newblock \showarticletitle{Exploring the limits of transfer learning with a unified text-to-text transformer}.
\newblock \bibinfo{journal}{\emph{Journal of machine learning research}} \bibinfo{volume}{21}, \bibinfo{number}{140} (\bibinfo{year}{2020}), \bibinfo{pages}{1--67}.
\newblock


\bibitem[Rajput et~al\mbox{.}(2023)]%
        {rajput2023recommender}
\bibfield{author}{\bibinfo{person}{Shashank Rajput}, \bibinfo{person}{Nikhil Mehta}, \bibinfo{person}{Anima Singh}, \bibinfo{person}{Raghunandan Hulikal~Keshavan}, \bibinfo{person}{Trung Vu}, \bibinfo{person}{Lukasz Heldt}, \bibinfo{person}{Lichan Hong}, \bibinfo{person}{Yi Tay}, \bibinfo{person}{Vinh Tran}, \bibinfo{person}{Jonah Samost}, {et~al\mbox{.}}} \bibinfo{year}{2023}\natexlab{}.
\newblock \showarticletitle{Recommender systems with generative retrieval}.
\newblock \bibinfo{journal}{\emph{Advances in Neural Information Processing Systems}}  \bibinfo{volume}{36} (\bibinfo{year}{2023}), \bibinfo{pages}{10299--10315}.
\newblock


\bibitem[Reisizadeh et~al\mbox{.}(2026)]%
        {reisizadeh2026blur}
\bibfield{author}{\bibinfo{person}{Hadi Reisizadeh}, \bibinfo{person}{Jinghan Jia}, \bibinfo{person}{Zhiqi Bu}, \bibinfo{person}{Bhanukiran Vinzamuri}, \bibinfo{person}{Anil Ramakrishna}, \bibinfo{person}{Kai-Wei Chang}, \bibinfo{person}{Volkan Cevher}, \bibinfo{person}{Sijia Liu}, {and} \bibinfo{person}{Mingyi Hong}.} \bibinfo{year}{2026}\natexlab{}.
\newblock \showarticletitle{Blur: A bi-level optimization approach for llm unlearning}. In \bibinfo{booktitle}{\emph{Proceedings of the 19th Conference of the European Chapter of the Association for Computational Linguistics (Volume 1: Long Papers)}}. \bibinfo{pages}{7043--7058}.
\newblock


\bibitem[Rosati et~al\mbox{.}(2024)]%
        {rosati2024representation}
\bibfield{author}{\bibinfo{person}{Domenic Rosati}, \bibinfo{person}{Jan Wehner}, \bibinfo{person}{Kai Williams}, \bibinfo{person}{{\L}ukasz Bartoszcze}, \bibinfo{person}{David Atanasov}, \bibinfo{person}{Robie Gonzales}, \bibinfo{person}{Subhabrata Majumdar}, \bibinfo{person}{Carsten Maple}, \bibinfo{person}{Hassan Sajjad}, {and} \bibinfo{person}{Frank Rudzicz}.} \bibinfo{year}{2024}\natexlab{}.
\newblock \showarticletitle{Representation noising: A defence mechanism against harmful finetuning}.
\newblock \bibinfo{journal}{\emph{Advances in Neural Information Processing Systems}}  \bibinfo{volume}{37} (\bibinfo{year}{2024}), \bibinfo{pages}{12636--12676}.
\newblock


\bibitem[Shen et~al\mbox{.}(2026)]%
        {shen2026fine}
\bibfield{author}{\bibinfo{person}{Yifan Shen}, \bibinfo{person}{Yuanzhe Liu}, \bibinfo{person}{Jingyuan Zhu}, \bibinfo{person}{Xu Cao}, \bibinfo{person}{Xiaofeng Zhang}, \bibinfo{person}{Yixiao He}, \bibinfo{person}{Wenming Ye}, \bibinfo{person}{James Rehg}, {and} \bibinfo{person}{Ismini Lourentzou}.} \bibinfo{year}{2026}\natexlab{}.
\newblock \showarticletitle{Fine-grained preference optimization improves spatial reasoning in vlms}.
\newblock \bibinfo{journal}{\emph{Advances in Neural Information Processing Systems}}  \bibinfo{volume}{38} (\bibinfo{year}{2026}), \bibinfo{pages}{17929--17960}.
\newblock


\bibitem[Shi et~al\mbox{.}(2025)]%
        {shi2025muse}
\bibfield{author}{\bibinfo{person}{Weijia Shi}, \bibinfo{person}{Jaechan Lee}, \bibinfo{person}{Yangsibo Huang}, \bibinfo{person}{Sadhika Malladi}, \bibinfo{person}{Jieyu Zhao}, \bibinfo{person}{Ari Holtzman}, \bibinfo{person}{Daogao Liu}, \bibinfo{person}{Luke Zettlemoyer}, \bibinfo{person}{Noah~A. Smith}, {and} \bibinfo{person}{Chiyuan Zhang}.} \bibinfo{year}{2025}\natexlab{}.
\newblock \showarticletitle{{MUSE}: Machine Unlearning Six-Way Evaluation for Language Models}. In \bibinfo{booktitle}{\emph{The Thirteenth International Conference on Learning Representations}}.
\newblock
\urldef\tempurl%
\url{https://openreview.net/forum?id=TArmA033BU}
\showURL{%
\tempurl}


\bibitem[Thudi et~al\mbox{.}(2022)]%
        {thudi2022unrolling}
\bibfield{author}{\bibinfo{person}{Anvith Thudi}, \bibinfo{person}{Gabriel Deza}, \bibinfo{person}{Varun Chandrasekaran}, {and} \bibinfo{person}{Nicolas Papernot}.} \bibinfo{year}{2022}\natexlab{}.
\newblock \showarticletitle{Unrolling sgd: Understanding factors influencing machine unlearning}. In \bibinfo{booktitle}{\emph{2022 IEEE 7th European Symposium on Security and Privacy (EuroS\&P)}}. IEEE, \bibinfo{pages}{303--319}.
\newblock


\bibitem[Wang et~al\mbox{.}(2024a)]%
        {wang2024learnable}
\bibfield{author}{\bibinfo{person}{Wenjie Wang}, \bibinfo{person}{Honghui Bao}, \bibinfo{person}{Xinyu Lin}, \bibinfo{person}{Jizhi Zhang}, \bibinfo{person}{Yongqi Li}, \bibinfo{person}{Fuli Feng}, \bibinfo{person}{See-Kiong Ng}, {and} \bibinfo{person}{Tat-Seng Chua}.} \bibinfo{year}{2024}\natexlab{a}.
\newblock \showarticletitle{Learnable item tokenization for generative recommendation}. In \bibinfo{booktitle}{\emph{Proceedings of the 33rd ACM International Conference on Information and Knowledge Management}}. \bibinfo{pages}{2400--2409}.
\newblock


\bibitem[Wang et~al\mbox{.}(2024b)]%
        {wang2024llm}
\bibfield{author}{\bibinfo{person}{Yaxuan Wang}, \bibinfo{person}{Jiaheng Wei}, \bibinfo{person}{Chris~Yuhao Liu}, \bibinfo{person}{Jinlong Pang}, \bibinfo{person}{Quan Liu}, \bibinfo{person}{Ankit Shah}, \bibinfo{person}{Yujia Bao}, \bibinfo{person}{Yang Liu}, {and} \bibinfo{person}{Wei Wei}.} \bibinfo{year}{2024}\natexlab{b}.
\newblock \showarticletitle{Llm unlearning via loss adjustment with only forget data}. In \bibinfo{booktitle}{\emph{The Thirteenth International Conference on Learning Representations}}.
\newblock


\bibitem[Wang et~al\mbox{.}({[n.\,d.]})]%
        {wang2410llm}
\bibfield{author}{\bibinfo{person}{Yaxuan Wang}, \bibinfo{person}{Jiaheng Wei}, \bibinfo{person}{Chris~Yuhao Liu}, \bibinfo{person}{Jinlong Pang}, \bibinfo{person}{Quan Liu}, \bibinfo{person}{Ankit~Parag Shah}, \bibinfo{person}{Yujia Bao}, \bibinfo{person}{Yang Liu}, {and} \bibinfo{person}{Wei Wei}.} \bibinfo{year}{[n.\,d.]}\natexlab{}.
\newblock \showarticletitle{Llm unlearning via loss adjustment with only forget data, 2024}.
\newblock \bibinfo{journal}{\emph{URL https://arxiv. org/abs/2410.11143}} (\bibinfo{year}{[n.\,d.]}).
\newblock


\bibitem[Xu et~al\mbox{.}(2024)]%
        {xu2024knowledge}
\bibfield{author}{\bibinfo{person}{Rongwu Xu}, \bibinfo{person}{Zehan Qi}, \bibinfo{person}{Zhijiang Guo}, \bibinfo{person}{Cunxiang Wang}, \bibinfo{person}{Hongru Wang}, \bibinfo{person}{Yue Zhang}, {and} \bibinfo{person}{Wei Xu}.} \bibinfo{year}{2024}\natexlab{}.
\newblock \showarticletitle{Knowledge conflicts for llms: A survey}.
\newblock \bibinfo{journal}{\emph{arXiv preprint arXiv:2403.08319}} (\bibinfo{year}{2024}).
\newblock


\bibitem[Yang et~al\mbox{.}(2025)]%
        {yang2025qwen3}
\bibfield{author}{\bibinfo{person}{An Yang}, \bibinfo{person}{Anfeng Li}, \bibinfo{person}{Baosong Yang}, \bibinfo{person}{Beichen Zhang}, \bibinfo{person}{Binyuan Hui}, \bibinfo{person}{Bo Zheng}, \bibinfo{person}{Bowen Yu}, \bibinfo{person}{Chang Gao}, \bibinfo{person}{Chengen Huang}, \bibinfo{person}{Chenxu Lv}, {et~al\mbox{.}}} \bibinfo{year}{2025}\natexlab{}.
\newblock \showarticletitle{Qwen3 technical report}.
\newblock \bibinfo{journal}{\emph{arXiv preprint arXiv:2505.09388}} (\bibinfo{year}{2025}).
\newblock


\bibitem[Yuan et~al\mbox{.}(2022)]%
        {yuan2022opticalflow}
\bibfield{author}{\bibinfo{person}{Shuai Yuan}, \bibinfo{person}{Xian Sun}, \bibinfo{person}{Hannah Kim}, \bibinfo{person}{Shuzhi Yu}, {and} \bibinfo{person}{Carlo Tomasi}.} \bibinfo{year}{2022}\natexlab{}.
\newblock \showarticletitle{Optical Flow Training under Limited Label Budget via Active Learning}. In \bibinfo{booktitle}{\emph{European Conference on Computer Vision (ECCV)}}. \bibinfo{publisher}{Springer}, \bibinfo{pages}{410--427}.
\newblock


\bibitem[Zhang et~al\mbox{.}(2026)]%
        {zhang2026carbon}
\bibfield{author}{\bibinfo{person}{Johnny~R. Zhang}, \bibinfo{person}{Gaoyuan Du}, \bibinfo{person}{Qianyi Sun}, \bibinfo{person}{Shiqi Wang}, \bibinfo{person}{Jiaxuan Li}, {and} \bibinfo{person}{Xian Sun}.} \bibinfo{year}{2026}\natexlab{}.
\newblock \bibinfo{title}{Carbon-Aware Compute--Power Scheduling for AI Data Centers with Microgrid Prosumer Operations}.
\newblock
\showeprint[arxiv]{2605.03751}~[cs.CE]
\urldef\tempurl%
\url{https://arxiv.org/abs/2605.03751}
\showURL{%
\tempurl}


\bibitem[Zhang and Bojja~Venkatakrishnan(2023)]%
        {zhang2023kadabra}
\bibfield{author}{\bibinfo{person}{Yunqi Zhang} {and} \bibinfo{person}{Shaileshh Bojja~Venkatakrishnan}.} \bibinfo{year}{2023}\natexlab{}.
\newblock \showarticletitle{Kadabra: Adapting kademlia for the decentralized web}. In \bibinfo{booktitle}{\emph{International Conference on Financial Cryptography and Data Security}}. Springer, \bibinfo{pages}{327--345}.
\newblock


\bibitem[Zhang and Bojja~Venkatakrishnan(2025)]%
        {zhang2025honeybee}
\bibfield{author}{\bibinfo{person}{Yunqi Zhang} {and} \bibinfo{person}{Shaileshh Bojja~Venkatakrishnan}.} \bibinfo{year}{2025}\natexlab{}.
\newblock \showarticletitle{Honeybee: Byzantine Tolerant Decentralized Peer Sampling with Verifiable Random Walks}. In \bibinfo{booktitle}{\emph{Proceedings of the Twenty-sixth International Symposium on Theory, Algorithmic Foundations, and Protocol Design for Mobile Networks and Mobile Computing}}. \bibinfo{pages}{321--330}.
\newblock


\bibitem[Zhang and Venkatakrishnan(2023)]%
        {zhang2023rethinking}
\bibfield{author}{\bibinfo{person}{Yunqi Zhang} {and} \bibinfo{person}{Shaileshh~Bojja Venkatakrishnan}.} \bibinfo{year}{2023}\natexlab{}.
\newblock \showarticletitle{Rethinking incentive in payment channel networks}. In \bibinfo{booktitle}{\emph{2023 IEEE 43rd International Conference on Distributed Computing Systems Workshops (ICDCSW)}}. IEEE, \bibinfo{pages}{61--66}.
\newblock


\bibitem[Zhou et~al\mbox{.}(2025)]%
        {zhou2025onerec}
\bibfield{author}{\bibinfo{person}{Guorui Zhou}, \bibinfo{person}{Jiaxin Deng}, \bibinfo{person}{Jinghao Zhang}, \bibinfo{person}{Kuo Cai}, \bibinfo{person}{Lejian Ren}, \bibinfo{person}{Qiang Luo}, \bibinfo{person}{Qianqian Wang}, \bibinfo{person}{Qigen Hu}, \bibinfo{person}{Rui Huang}, \bibinfo{person}{Shiyao Wang}, {et~al\mbox{.}}} \bibinfo{year}{2025}\natexlab{}.
\newblock \showarticletitle{OneRec Technical Report}.
\newblock \bibinfo{journal}{\emph{arXiv preprint arXiv:2506.13695}} (\bibinfo{year}{2025}).
\newblock


\bibitem[Zhu et~al\mbox{.}(2024)]%
        {zhu2024advancing}
\bibfield{author}{\bibinfo{person}{Chunzheng Zhu}, \bibinfo{person}{Jianxin Lin}, \bibinfo{person}{Guanghua Tan}, \bibinfo{person}{Ningbo Zhu}, \bibinfo{person}{Kenli Li}, \bibinfo{person}{Chunlian Wang}, {and} \bibinfo{person}{Shengli Li}.} \bibinfo{year}{2024}\natexlab{}.
\newblock \showarticletitle{Advancing Ultrasound Medical Continuous Learning with Task-Specific Generalization and Adaptability}. In \bibinfo{booktitle}{\emph{2024 IEEE International Conference on Bioinformatics and Biomedicine (BIBM)}}. IEEE, \bibinfo{pages}{3019--3025}.
\newblock


\bibitem[Zhu et~al\mbox{.}(2026)]%
        {zhu2026medeyes}
\bibfield{author}{\bibinfo{person}{Chunzheng Zhu}, \bibinfo{person}{Yangfang Lin}, \bibinfo{person}{Shen Chen}, \bibinfo{person}{Yijun Wang}, {and} \bibinfo{person}{Jianxin Lin}.} \bibinfo{year}{2026}\natexlab{}.
\newblock \showarticletitle{Medeyes: Learning dynamic visual focus for medical progressive diagnosis}. In \bibinfo{booktitle}{\emph{Proceedings of the AAAI Conference on Artificial Intelligence}}, Vol.~\bibinfo{volume}{40}. \bibinfo{pages}{13916--13924}.
\newblock


\bibitem[Zhu et~al\mbox{.}(2025)]%
        {zhu2025pathology}
\bibfield{author}{\bibinfo{person}{Chunzheng Zhu}, \bibinfo{person}{Yangfang Lin}, \bibinfo{person}{Jialin Shao}, \bibinfo{person}{Jianxin Lin}, {and} \bibinfo{person}{Yijun Wang}.} \bibinfo{year}{2025}\natexlab{}.
\newblock \showarticletitle{Pathology-Aware Prototype Evolution via LLM-Driven Semantic Disambiguation for Multicenter Diabetic Retinopathy Diagnosis}. In \bibinfo{booktitle}{\emph{Proceedings of the 33rd ACM International Conference on Multimedia}}. \bibinfo{pages}{9196--9205}.
\newblock


\end{thebibliography}

\appendix
\newpage
\section{Broader Impact}
\label{sec:broader_impact}

This work addresses the growing need for \emph{machine unlearning} in recommender systems, enabling platforms to comply with privacy regulations such as GDPR's right to be forgotten and CCPA by selectively removing the influence of specific items or brands from a trained model without full retraining.
Our framework empowers content providers and users to exercise meaningful control over how their data is represented in recommendation models, reducing the risk of unwanted brand association or continued promotion of recalled, delisted, or sensitive products.
We emphasize that unlearning should be viewed as one component of a broader responsible-AI strategy rather than a standalone guarantee.
Our experiments are conducted on English-language Amazon product review datasets; the effectiveness of SID-based unlearning in multilingual or cross-domain settings remains an interesting direction for future work~\cite{yuan2022opticalflow,gupta2025dameflame,zhang2026carbon}.

\section{Theoretical Analysis}
\label{apx:theory}
\subsection{Theoretical results}

We first define $\mathcal F$ be the forget set, $\mathcal{R}$ be the retain set and $\Omega$ as the average token overlap between a forget item
$i\in\mathcal F$ and a retain item $j\in\mathcal R$, averaged over all
codebook levels. Specifically,
\begin{equation}
\Omega
=
\mathbb{E}_{i\sim\mathcal F,\;j\sim\mathcal R}
\left[
\omega(i,j)
\right],
\qquad
\omega(i,j)
=
\frac{1}{L}
\sum_{\ell=1}^{L}
\sum_{k\in\mathcal T^\ell}
\mathbf{1}[s_i^\ell=k]\mathbf{1}[s_j^\ell=k].
\end{equation}
Here, $\mathcal T^\ell$ denotes the token set at codebook level $\ell$,
and $t_i^\ell$ is the token assigned to item $i$ at level $\ell$.
We define the loss of generating a token $t$ conditioned on a sequence
$\mathcal T$ as
\[
\ell(\theta;t\mid\mathcal T)
=
-\log p_\theta(t\mid \mathcal T).
\]

\begin{assumption}
\label{asm:loss}
We assume that the token-level loss satisfies the following properties.
\begin{itemize}
    \item[(a)] \emph{Lipschitzness.} For any token $t$ and conditioning
    sequence $\mathcal T$ in the codebook,
    \[
    \|\nabla_\theta \ell(\theta;t\mid\mathcal T)\|_2
    \le \xi,
    \qquad
    \forall \theta.
    \]
    \item[(b)] \emph{Smoothness.} For any token $t$ and conditioning
    sequence $\mathcal T$ in the codebook,
    \[
    \|\nabla^2_\theta \ell(\theta;t\mid\mathcal T)\|_{\mathrm{op}}
    \le M,
    \qquad
    \forall \theta.
    \]
\end{itemize}
\end{assumption}

The above assumptions are standard in the optimization literature; see,
for example, \citep[Assumption~1]{reisizadeh2026blur}.

\begin{proposition}
\label{prop:ulb-retain}
Suppose Assumption~\ref{asm:loss} holds. Let
\[
\theta_+
:=
\theta
-
\eta\nabla_\theta \mathcal L_{\mathrm{unlearn}}(\theta)
\]
be the model parameters after one gradient descent step on the unlearning
objective. Then the retain-loss change satisfies
\begin{equation}
\label{eq:retain-change-upper-simplified}
\mathcal L_R(\theta_+) - \mathcal L_R(\theta)
\le
\eta\alpha\frac{\xi^2}{L}\Omega
+
C_U .
\end{equation}
Furthermore, if for any fixed token $t$ and any conditioning sequences
$\mathcal T,\mathcal T'$, the token-level gradients satisfy
\[
\left\langle
\nabla_\theta \ell(\theta;t\mid\mathcal T),
\nabla_\theta \ell(\theta;t\mid\mathcal T')
\right\rangle
\ge
\gamma
>
0,
\]
then
\begin{equation}
\label{eq:retain-change-lower-simplified}
\mathcal L_R(\theta_+) - \mathcal L_R(\theta)
\ge
\eta\alpha\frac{\gamma}{L}\Omega
+
C_L .
\end{equation}
Here, $C_U$ and $C_L$ collect residual gradient-alignment terms and
second-order optimization remainders; they are not assumed to be invariant
under token reassignment.
\end{proposition}

Proposition~\ref{prop:ulb-retain} identifies average token overlap $\Omega$ as a
directly controllable source of forget-retain interference: the shared-token
component of the forget-retain gradient alignment scales linearly with
$\Omega$ and enters the retain-loss change with positive coefficient
$\eta\alpha$ during unlearning. Thus, reducing $\Omega$ suppresses an explicit
overlap-mediated source of retain-loss increase. We do not claim that $\Omega$
fully determines retain degradation, since the residual term $C_U, C_L$ may also vary under reassignment; rather,
the point is to isolate this shared-token mechanism and use it to
motivate token reassignment, which we next show reduces the overlap $\Omega.$ In the lemma below, we consider fixed the model $\theta,$ and run one step of gradient descent on $\phi$ reduce the overlap.
 
We then made the following Assumption on the structure of the history and the .

\begin{assumption}
    \label{asm:overlap} We assume one of the following holds on the history of the items and the retain set: \begin{itemize}
        \item[(A1)] All the items in the history is not reassigned.
        \item[(A2)] \begin{itemize}
            \item[(1)] The tokenizations of the items in the retain set and forget set are disjoint.
            \item[(2)] For any $\mathcal{H}_u$ in the retain set, there exist a small enough $\varepsilon_{\mathrm{hist}} > 0$ such that: \begin{equation*}
                \frac{1}{|\mathcal{H}_u|} \sum_{i \in \mathcal H_u} \mathbf{1}[i \in \mathcal{F}] \leq \varepsilon_{\mathrm{hist}}.
            \end{equation*}
        \end{itemize} 
    \end{itemize}

    Further, we assume that $\Omega(\phi)$ is $\beta_{\phi, \Omega}$-smooth in $\phi.$

\end{assumption}

Assumption~\ref{asm:overlap} formalizes the idea that retain examples are
only weakly coupled with the reassignment parameters of forget items.
Condition (A1) corresponds to the simplified setting where reassigned
tokens do not appear in histories, so the history component does not
contribute to the \(\phi\)-gradient. Condition (A2) covers a more general
case: retain and forget items are token-separated, and forget items occur
only sparsely in retain histories. Under this condition, the gradient of
the retain loss with respect to forget-item reassignment parameters is
small, of order \(O(\varepsilon_{\mathrm{hist}})\). The smoothness
assumption is used only to convert first-order descent directions into
one-step decreases for sufficiently small step size.

\begin{lemma}
    \label{lem:reduce_con}
    Under Assumption \ref{asm:overlap}, for any fixed model $\theta,$ let $\phi_+ = \phi - \eta_\phi M \odot \nabla_{\phi} \mathcal L_{\mathrm{unlearn}}(\phi),$ where $M$ is define in \ref{eq:mask} and the $\odot$ denote element-wise product. For small enough $\eta_\phi,$ we have: \begin{equation*}
        \begin{split}
            \Omega(\phi_+) - \Omega(\phi) \leq 0, \quad \mathcal{L}_F(\phi_+) - \mathcal{L}_F(\phi) \leq 0.
        \end{split}
    \end{equation*}
\end{lemma}

Lemma~\ref{lem:reduce_con} shows that the masked reassignment update
simultaneously reduces the forget log-likelihood and the average
forget-retain token overlap, provided that the retain examples are only
weakly coupled with forget-item reassignment. The mask \(M\) selects
coordinates for which increasing the reassignment score would both
increase the forget objective gradient and correspond to above-average
overlap with retain tokens. Consequently, moving in the negative masked
unlearning-gradient direction decreases the forget objective while also
moving against the overlap gradient.

Combined with Proposition~\ref{prop:ulb-retain}, this lemma provides a
mechanistic justification for token reassignment. Proposition~\ref{prop:ulb-retain}
identifies \(\Omega\) as an explicit overlap-mediated component of the
retain-loss increase during unlearning, while Lemma~\ref{lem:reduce_con}
shows that optimizing \(\phi\) can directly reduce this component.
Therefore, token reassignment can mitigate retain degradation by
suppressing shared-token interference between forget and retain items.
We emphasize that this does not imply that \(\Omega\) fully determines
the retain loss: the residual terms \(C_U\) and \(C_L\) in
Proposition~\ref{prop:ulb-retain} may also change under reassignment.
Rather, the result isolates one controllable mechanism through which
reassignment can improve the retain-loss behavior.

\subsection{Proof of Proposition~\ref{prop:ulb-retain}: Token Overlap Affects Unlearning}

\label{apx:pf-prop}

{\color{black}
We first write the forget and retain losses as averages of item-level
generation losses:
\begin{equation}
\mathcal L_F(\theta)
=
\frac{1}{|\mathcal F|}
\sum_{i\in\mathcal F}
\ell_i(\theta),
\qquad
\mathcal L_R(\theta)
=
\frac{1}{|\mathcal R|}
\sum_{j\in\mathcal R}
\ell_j(\theta).
\end{equation}
For an item $i$, we define the item-level loss as the average token-level
generation loss across codebook levels:
\begin{equation}
\ell_i(\theta)
=
\frac{1}{L}
\sum_{\ell=1}^{L}
\ell(\theta;t_i^\ell\mid \mathcal T(\mathcal H_i), t_i^{<\ell}),
\end{equation}
where
\begin{equation}
\ell(\theta;t_i^\ell\mid \mathcal T(\mathcal H_i),t_i^{<\ell})
=
-
\log
p_\theta
\left(
t_i^\ell
\mid
\mathcal T(\mathcal H_i), t_i^{<\ell}
\right).
\end{equation}
Thus,
\begin{equation}
\nabla_\theta \mathcal L_F
=
\frac{1}{|\mathcal F|}
\sum_{i\in\mathcal F}
\nabla_\theta \ell_i,
\qquad
\nabla_\theta \mathcal L_R
=
\frac{1}{|\mathcal R|}
\sum_{j\in\mathcal R}
\nabla_\theta \ell_j.
\end{equation}

The forget-retain gradient alignment can be expanded as
\begin{equation}
\begin{aligned}
\left\langle
\nabla_\theta \mathcal L_F,
\nabla_\theta \mathcal L_R
\right\rangle
&=
\left\langle
\frac{1}{|\mathcal F|}
\sum_{i\in\mathcal F}
\nabla_\theta \ell_i,
\frac{1}{|\mathcal R|}
\sum_{j\in\mathcal R}
\nabla_\theta \ell_j
\right\rangle \\
&=
\frac{1}{|\mathcal F||\mathcal R|}
\sum_{i\in\mathcal F}
\sum_{j\in\mathcal R}
\left\langle
\nabla_\theta \ell_i,
\nabla_\theta \ell_j
\right\rangle \\
&=
\mathbb E_{i\sim\mathcal F,\;j\sim\mathcal R}
\left[
\left\langle
\nabla_\theta \ell_i,
\nabla_\theta \ell_j
\right\rangle
\right].
\end{aligned}
\end{equation}
Therefore, the global forget-retain gradient alignment is the expected
pairwise gradient alignment between forget and retain items.

For each item $i$ and level $\ell$, define the token-level gradient component
\begin{equation}
g_{i,\ell}
:=
\nabla_\theta
\ell(\theta;t_i^\ell\mid \mathcal T(\mathcal H_i),t_i^{<\ell}).
\end{equation}
For notational convenience, we also write $g_{i,\ell,k}$ for this gradient
when $t_i^\ell=k$. Since
\[
\ell_i(\theta)
=
\frac{1}{L}
\sum_{\ell=1}^{L}
\ell(\theta;t_i^\ell\mid \mathcal T(\mathcal H_i), t_i^{<\ell}),
\]
we have
\[
\nabla_\theta \ell_i(\theta)
=
\frac{1}{L}
\sum_{\ell=1}^{L}
g_{i,\ell}.
\]
Therefore, the pairwise gradient alignment can be decomposed as
\begin{equation}
\begin{aligned}
\left\langle
\nabla_\theta \ell_i,
\nabla_\theta \ell_j
\right\rangle
&=
\frac{1}{L^2}
\sum_{\ell=1}^{L}
\sum_{m=1}^{L}
\left\langle
g_{i,\ell},
g_{j,m}
\right\rangle \\
&=
\frac{1}{L^2}
\sum_{\ell=1}^{L}
\sum_{k\in\mathcal T^\ell}
\mathbf 1[t_i^\ell=k]
\mathbf 1[t_j^\ell=k]
\left\langle
g_{i,\ell,k},
g_{j,\ell,k}
\right\rangle
+
C_{ij},
\end{aligned}
\end{equation}
where the residual term $C_{ij}$ is given explicitly by
\begin{equation}
\begin{aligned}
C_{ij}
:=
&
\frac{1}{L^2}
\sum_{\ell=1}^{L}
\sum_{\substack{k,k'\in\mathcal T^\ell\\ k\neq k'}}
\mathbf 1[t_i^\ell=k]
\mathbf 1[t_j^\ell=k']
\left\langle
g_{i,\ell,k},
g_{j,\ell,k'}
\right\rangle \\
&+
\frac{1}{L^2}
\sum_{\substack{\ell,m=1\\ \ell\neq m}}^{L}
\sum_{k\in\mathcal T^\ell}
\sum_{k'\in\mathcal T^m}
\mathbf 1[t_i^\ell=k]
\mathbf 1[t_j^m=k']
\left\langle
g_{i,\ell,k},
g_{j,m,k'}
\right\rangle .
\end{aligned}
\end{equation}
The first term in $C_{ij}$ collects same-level but non-shared-token
contributions, while the second term collects cross-level interactions.

Using the positive alignment assumption for shared tokens and the uniform
gradient bound in Assumption~\ref{asm:loss}, we have
\begin{equation}
\gamma
\le
\left\langle
g_{i,\ell,k},
g_{j,\ell,k}
\right\rangle
\le
\|g_{i,\ell,k}\|_2
\|g_{j,\ell,k}\|_2
\le
\xi^2 .
\end{equation}
Therefore, from the above decomposition,
\begin{equation}
\begin{aligned}
&
\frac{\gamma}{L^2}
\sum_{\ell=1}^{L}
\sum_{k\in\mathcal T^\ell}
\mathbf 1[t_i^\ell=k]
\mathbf 1[t_j^\ell=k]
+
C_{ij} \\
&\qquad\le
\left\langle
\nabla_\theta \ell_i,
\nabla_\theta \ell_j
\right\rangle \\
&\qquad\le
\frac{\xi^2}{L^2}
\sum_{\ell=1}^{L}
\sum_{k\in\mathcal T^\ell}
\mathbf 1[t_i^\ell=k]
\mathbf 1[t_j^\ell=k]
+
C_{ij}.
\end{aligned}
\end{equation}
By the definition of token overlap,
\[
\omega(i,j)
=
\frac{1}{L}
\sum_{\ell=1}^{L}
\sum_{k\in\mathcal T^\ell}
\mathbf 1[t_i^\ell=k]
\mathbf 1[t_j^\ell=k],
\]
we have
\[
\sum_{\ell=1}^{L}
\sum_{k\in\mathcal T^\ell}
\mathbf 1[t_i^\ell=k]
\mathbf 1[t_j^\ell=k]
=
L\omega(i,j).
\]
Hence,
\begin{equation}
\label{eq:pairwise-alignment-overlap-bound}
\frac{\gamma}{L}\omega(i,j)+C_{ij}
\le
\left\langle
\nabla_\theta \ell_i,
\nabla_\theta \ell_j
\right\rangle
\le
\frac{\xi^2}{L}\omega(i,j)+C_{ij}.
\end{equation}

Taking expectation over $i\sim\mathcal F$ and $j\sim\mathcal R$ gives
\begin{equation}
\begin{aligned}
\left\langle
\nabla_\theta \mathcal L_F,
\nabla_\theta \mathcal L_R
\right\rangle
&=
\mathbb E_{i\sim\mathcal F,\;j\sim\mathcal R}
\left[
\left\langle
\nabla_\theta \ell_i,
\nabla_\theta \ell_j
\right\rangle
\right] \\
&\ge
\mathbb E_{i\sim\mathcal F,\;j\sim\mathcal R}
\left[
\frac{\gamma}{L}\omega(i,j)+C_{ij}
\right] \\
&=
\frac{\gamma}{L}
\mathbb E_{i\sim\mathcal F,\;j\sim\mathcal R}
\left[
\omega(i,j)
\right]
+
\mathbb E_{i\sim\mathcal F,\;j\sim\mathcal R}
\left[
C_{ij}
\right].
\end{aligned}
\end{equation}
Similarly, the upper bound in
\eqref{eq:pairwise-alignment-overlap-bound} yields
\begin{equation}
\begin{aligned}
\left\langle
\nabla_\theta \mathcal L_F,
\nabla_\theta \mathcal L_R
\right\rangle
&\le
\frac{\xi^2}{L}
\mathbb E_{i\sim\mathcal F,\;j\sim\mathcal R}
\left[
\omega(i,j)
\right]
+
\mathbb E_{i\sim\mathcal F,\;j\sim\mathcal R}
\left[
C_{ij}
\right].
\end{aligned}
\end{equation}
Since
\[
\Omega
=
\mathbb E_{i\sim\mathcal F,\;j\sim\mathcal R}
\left[
\omega(i,j)
\right],
\]
and defining
\[
C_0
:=
\mathbb E_{i\sim\mathcal F,\;j\sim\mathcal R}
\left[
C_{ij}
\right],
\]
we obtain the two-sided gradient-alignment bound
\begin{equation}
\label{eq:global-alignment-overlap-bound}
\frac{\gamma}{L}\Omega+C_0
\le
\left\langle
\nabla_\theta \mathcal L_F,
\nabla_\theta \mathcal L_R
\right\rangle
\le
\frac{\xi^2}{L}\Omega+C_0.
\end{equation}

We now translate this gradient-alignment bound into a bound on the retain-loss
change after one unlearning step. Recall that
\[
\mathcal L_{\mathrm{unlearn}}(\theta)
=
\mathcal L_R(\theta)
-
\alpha \mathcal L_F(\theta),
\qquad
\alpha>0.
\]
One gradient descent step gives
\[
\theta_+
=
\theta
-
\eta\nabla_\theta \mathcal L_{\mathrm{unlearn}}(\theta)
=
\theta
-
\eta\nabla_\theta \mathcal L_R(\theta)
+
\eta\alpha\nabla_\theta \mathcal L_F(\theta).
\]
For brevity, write
\[
G_R:=\nabla_\theta \mathcal L_R(\theta),
\qquad
G_F:=\nabla_\theta \mathcal L_F(\theta),
\]
and
\[
\Delta
:=
\theta_+-\theta
=
-\eta G_R+\eta\alpha G_F
=
-\eta(G_R-\alpha G_F).
\]

By Assumption~\ref{asm:loss}, the token-level loss is $M$-smooth. Since
the item-level and retain losses are averages of token-level losses,
$\mathcal L_R$ is also $M$-smooth. Hence, Taylor's theorem with remainder
gives
\begin{equation}
\label{eq:retain-smooth-two-sided}
\left\langle
G_R,\Delta
\right\rangle
-
\frac{M}{2}
\|\Delta\|^2
\le
\mathcal L_R(\theta_+)-\mathcal L_R(\theta)
\le
\left\langle
G_R,\Delta
\right\rangle
+
\frac{M}{2}
\|\Delta\|^2 .
\end{equation}
Moreover,
\[
\left\langle
G_R,\Delta
\right\rangle
=
-\eta\|G_R\|^2
+
\eta\alpha
\left\langle
G_R,G_F
\right\rangle,
\]
and
\[
\|\Delta\|^2
=
\eta^2
\|G_R-\alpha G_F\|^2.
\]
Substituting these identities into
\eqref{eq:retain-smooth-two-sided}, we obtain
\begin{equation}
\label{eq:retain-change-gradient-alignment}
\begin{aligned}
&-\eta\|G_R\|^2
+
\eta\alpha
\left\langle
G_R,G_F
\right\rangle
-
\frac{M\eta^2}{2}
\|G_R-\alpha G_F\|^2 \\
&\qquad\le
\mathcal L_R(\theta_+)-\mathcal L_R(\theta) \\
&\qquad\le
-\eta\|G_R\|^2
+
\eta\alpha
\left\langle
G_R,G_F
\right\rangle
+
\frac{M\eta^2}{2}
\|G_R-\alpha G_F\|^2 .
\end{aligned}
\end{equation}

Combining
\eqref{eq:retain-change-gradient-alignment}
with the lower bound in
\eqref{eq:global-alignment-overlap-bound}, we get
\begin{equation}
\label{eq:retain-change-lower-overlap}
\begin{aligned}
\mathcal L_R(\theta_+)-\mathcal L_R(\theta)
&\ge
-\eta\|G_R\|^2
+
\eta\alpha
\left(
\frac{\gamma}{L}\Omega+C_0
\right)
-
\frac{M\eta^2}{2}
\|G_R-\alpha G_F\|^2 .
\end{aligned}
\end{equation}
Similarly, combining
\eqref{eq:retain-change-gradient-alignment}
with the upper bound in
\eqref{eq:global-alignment-overlap-bound}, we get
\begin{equation}
\label{eq:retain-change-upper-overlap}
\begin{aligned}
\mathcal L_R(\theta_+)-\mathcal L_R(\theta)
&\le
-\eta\|G_R\|^2
+
\eta\alpha
\left(
\frac{\xi^2}{L}\Omega+C_0
\right)
+
\frac{M\eta^2}{2}
\|G_R-\alpha G_F\|^2 .
\end{aligned}
\end{equation}

Finally, all terms in
\eqref{eq:retain-change-lower-overlap}
and
\eqref{eq:retain-change-upper-overlap}
that do not explicitly depend on $\Omega$ can be absorbed into constants.
Thus, there exist constants $C_L$ and $C_U$, independent of the explicit
shared-token overlap term, such that
\begin{equation}
\label{eq:retain-change-final-overlap}
\eta\alpha\frac{\gamma}{L}\Omega + C_L
\le
\mathcal L_R(\theta_+)-\mathcal L_R(\theta)
\le
\eta\alpha\frac{\xi^2}{L}\Omega + C_U.
\end{equation}
This proves the desired two-sided control of the retain-loss change in terms
of the average token overlap $\Omega$. }

\subsection{Proof of Lemma \ref{lem:reduce_con}: Optimizing the Unlearning loss Reduces Token Overlap}

In this section, we show that optimizing the token reassignment parameters
can reduce the overlap measure \(\Omega\), thereby tightening the
overlap-dependent component in the bound on
\(\mathcal L_R(\theta_+) - \mathcal L_R(\theta)\) from
Proposition~\ref{prop:ulb-retain}. We first introduce the token
reassignment mechanism and formulate item generation probabilistically.
We then prove the overlap-reduction result under the two cases in
Assumption~\ref{asm:overlap}. Section~\ref{sec:independent} treats
Assumption~\ref{asm:overlap}(A1), where histories do not contain
reassigned items and hence the history tokenization does not depend on
\(\phi\). Section~\ref{sec:dependent} treats
Assumption~\ref{asm:overlap}(A2), where histories may contain reassigned
items, but retain and forget items are token-separated and forget items
appear only sparsely in retain histories.

\subsubsection{Unlearning loss as a function of the reassignment parameter $\phi$}

Given the reassignment parameter $\phi,$ we recall that the probability of tokenize user $i$ into token $k$ at level $\ell$ is: \begin{equation*}
    \begin{split}
        q_{\phi}(s_i^\ell = k | i) = \frac{
\exp\left((-\|\boldsymbol{r}_i^{l}-\boldsymbol{c}^{l}_{k}\|^{2}+\phi_{i,k}^{l})/\tau\right)
}{
\sum_{j=1}^{K}
\exp\left((-\|\boldsymbol{r}_i^{l}-\boldsymbol{c}^{l}_{j}\|^{2}+\phi_{i,j}^{l})/\tau\right)
}
    \end{split}
\end{equation*}

Thus, given the history $\mathcal H_u,$ and the target item $i,$ the probability of generating the target item $i$ given the history $\mathcal H_u$ is \begin{equation*}
    \begin{split}
        \Pr_{\phi, \theta}[ i| \mathcal H_u  ] = & \sum_{\mathcal T(\mathcal H_u)} \sum_{\mathcal T } \Pr[ i | \mathcal{T} ] p_{\theta}(\mathcal T \mid \mathcal T(\mathcal H_u)) \Pr[\mathcal{T}(\mathcal{H}_u) | \mathcal{H}_u ] \\
        = & \sum_{\mathcal T(\mathcal H_u)} \sum_{\mathcal T }  \frac{q_{\phi}(\mathcal{T}|i) \Pr(i)}{\sum_j q_{\phi}(\mathcal{T}|j) \Pr(j) } p_{\theta}(\mathcal T \mid \mathcal T (\mathcal H_u)) \Pr[\mathcal{T}(\mathcal{H}_u) | \mathcal{H}_u ] \\
        = &  \sum_{\mathcal T }  \frac{q_{\phi}(\mathcal{T}|i) \Pr(i)}{\sum_j q_{\phi}(\mathcal{T}|j) \Pr(j) } \sum_{\mathcal T(\mathcal H_u)} p_{\theta}(\mathcal T \mid \mathcal T (\mathcal H_u))  \prod_{j \in \mathcal{H}_u} \frac{q_{\phi}(\mathcal{T}(j)|j) \Pr(j)}{\sum_k q_{\phi}(\mathcal{T}(j)|k) \Pr(k) }
    \end{split}
\end{equation*}  where $\mathcal T$ denote a sequence of tokens of length $L,$ which is a possible tokenization of the target item $i,$ $\mathcal T(\mathcal{H}_u) = \{\mathcal T(j)\}_{j \in \mathcal{H}_u}$ which is the tokenization of the history of user $u.$ $\Pr[i]$ denote the frequency of item $i$ in the dataset. The first step is by the law of total probability, and the second and third step is by the Bayes rule, as well as the fact that each item in the history is tokenized independently.

Now for fixed model $\theta$ the forget and retain loss can be equivalently written as: \begin{equation*}
    \begin{split}
    &\mathcal{L}_F(\phi;\theta) = \sum_{(\mathcal H_u, i ) \in \mathcal \mathcal F} \log \Pr_{\phi, \theta}[ i| \mathcal H_u  ] \\
    &\mathcal{L}_R(\phi;\theta) = - \sum_{(\mathcal H_r, i_r ) \in \mathcal R} \log \Pr_{\phi, \theta}[ i_r| \mathcal H_r  ] \\
    & \mathcal{L}_{\mathrm{unlearn}}(\phi; \theta) = \mathcal{L}_F((\phi; \theta) - \alpha  \mathcal{L}_R((\phi; \theta)
    \end{split}
\end{equation*} and we drop the dependency on $\theta$ when there's no confusion. Next, we study how one step of gradient descent on $\phi$ effect the forget loss and retain loss. In particular, recall that the token overlap between a forget item $i$ and a
retain item $j$ is defined as
\begin{equation}
\omega(i,j)
=\frac{1}{L}
\sum_{\ell=1}^{L}
\sum_{k\in\mathcal T^\ell}
\mathbf{1}[s_i^\ell=k]\mathbf{1}[s_j^\ell=k].
\end{equation}  Since the token of a forget item is reassigned according to
$q_\phi(\cdot\mid i)$, while the retain item tokens are fixed, the
expected overlap can be written as
\begin{equation}
 \mathbb{E}[\omega(i,j)]
=
\frac{1}{L}
\sum_{\ell=1}^{L}
\sum_{k\in\mathcal{T}^{\ell}}
q_\phi(s_i^\ell = k\mid i) q_\phi(s_j^\ell = k\mid j) 
\end{equation}
Therefore, the average expected sensitive-token overlap is
\begin{equation}
\Omega(\phi)
=
\frac{1}{|\mathcal{F}||\mathcal{R}|L}
\sum_{i\in\mathcal{F}}
\sum_{j\in\mathcal{R}}
\sum_{\ell=1}^{L}
\sum_{k\in\mathcal{T}^{\ell}}
q_\phi(s_i^\ell = k\mid i)
q_\phi(s_j^\ell = k\mid j).
\end{equation}

The update of the reassignment parameters on the unlearning loss $\mathcal L_{\mathrm{unlearn}}$ is: \begin{equation*}
    \begin{split}
        (\phi_{i,b}^{\ell})_+ = \phi_{i,b}^{\ell} - \eta_\phi M_{i,b}^{\ell} \cdot \nabla_{\phi_{i,b}^{\ell} }\mathcal{L}_{\mathrm{unlearn}}(\phi)
    \end{split}
\end{equation*} with the selective mask $M$ define as \begin{equation*}
    \begin{split}
           M_{i,b}^{\ell} &=
\mathbf{1}
\left[ \rho_\ell(b)
 > \bar \rho_{i, \ell}
\right]
\cdot
\mathbf{1}
\left[
\nabla_{\phi_{i,b}^{\ell}} \mathcal{L}_F > 0
 \right],
    \end{split}
\end{equation*}  where \begin{equation}
\bar{\rho}_{i,\ell}
=
\sum_{k\in\mathcal{V}^{\ell}}
q_\phi(k\mid i)\rho_\ell(k),
\qquad
\rho_\ell(k)
=
\frac{1}{|\mathcal{R}|}
\sum_{j\in\mathcal{R}}
\mathbf{1}\left[t_j^\ell=k\right].
\end{equation}

We aim to show that: \begin{equation*}
    \begin{split}
        &\Omega(\phi_+) - \Omega(\phi) \leq 0, \\
        & \mathcal L_{F}(\phi_+) - \mathcal L_F(\phi_+) \leq 0.
    \end{split}
\end{equation*} 

\subsubsection{Proof under Assumption \ref{asm:overlap}(A1)}
\label{sec:independent}

We first consider a simple case with the following assumptions: 
\begin{itemize}
    \item[(a)] The history of all items does not contain items in the forget set.
\end{itemize}

Note that the token reassignment is only for items in the forget set, which implies that $\mathcal{L}_R$ does not depend on $\phi$ in this case and thus  $\nabla_\phi \mathcal{L}_{\mathrm{unlearn}} = \nabla_\phi \mathcal{L}_F.$ Thus, we immediately have: \begin{equation*}
    \begin{split}
        \mathcal{L}_F (\phi_+) \leq \mathcal{L}_F (\phi)
    \end{split}
\end{equation*} for small enough $\eta_\phi.$

Next, recall the definition of $\Omega,$ \begin{equation*}
    \begin{split}
        \Omega(\phi)
=
\frac{1}{|\mathcal{F}||\mathcal{R}|L}
\sum_{i\in\mathcal{F}}
\sum_{j\in\mathcal{R}}
\sum_{\ell=1}^{L}
\sum_{k\in\mathcal{T}^{\ell}}
q_\phi(s_i^\ell = k\mid i)
q_\phi(s_j^\ell = k\mid j).
    \end{split}
\end{equation*} and directly compute the gradient, we obtained that: \begin{equation}
\begin{aligned}
\frac{\partial \Omega}
{\partial \phi_{i,b}^{\ell}}
&=
\frac{1}{|\mathcal{F}|L}
\sum_{k\in\mathcal{V}^{\ell}}
\rho_\ell(k)
\frac{\partial q_\phi(k\mid i)}
{\partial \phi_{i,b}^{\ell}}
\\
&=
\frac{1}{\tau|\mathcal{F}|L}
\sum_{k\in\mathcal{V}^{\ell}}
\rho_\ell(k)
q_\phi(k\mid i)
\left(
\mathbf{1}[k=b]-q_\phi(b\mid i)
\right)
\\
&=
\frac{1}{\tau|\mathcal{F}|L}
q_\phi(b\mid i)
\left(
\rho_\ell(b)-\bar{\rho}_{i,\ell}
\right),
\end{aligned}
\end{equation} where \begin{equation}
\bar{\rho}_{i,\ell}
=
\sum_{k\in\mathcal{V}^{\ell}}
q_\phi(k\mid i)\rho_\ell(k),
\qquad
\rho_\ell(k)
=
\frac{1}{|\mathcal{R}|}
\sum_{j\in\mathcal{R}}
\mathbf{1}\left[t_j^\ell=k\right].
\end{equation}

It is sufficient to show \begin{equation*}
    \langle \nabla_\phi \Omega, M \cdot \nabla_{\phi}  \mathcal L_{\mathrm{unlearn}}  \rangle \geq 0.
\end{equation*}

Indeed we have: \begin{equation}
\begin{split}
&
\left\langle
    \nabla_\phi \Omega(\phi),
    M\odot \nabla_\phi \mathcal L_{\mathrm{unlearn}}(\phi)
\right\rangle
\\
&=
\sum_{i\in\mathcal F}
\sum_{\ell=1}^{L}
\sum_{b\in\mathcal V^\ell}
\frac{\partial \Omega(\phi)}
{\partial \phi_{i,b}^{\ell}}
M_{i,b}^{\ell}
\frac{\partial \mathcal L_{\mathrm{unlearn}}(\phi)}
{\partial \phi_{i,b}^{\ell}}
\\
&=
\frac{1}{\tau|\mathcal F|L}
\sum_{i\in\mathcal F}
\sum_{\ell=1}^{L}
\sum_{b\in\mathcal V^\ell}
q_\phi(b\mid i)
\left(
    \rho_\ell(b)-\bar\rho_{i,\ell}
\right)
M_{i,b}^{\ell}
\frac{\partial \mathcal L_F(\phi)}
{\partial \phi_{i,b}^{\ell}}
\\
&\ge 0 .
\end{split}
\end{equation}

\subsubsection{Proof under Assumption \ref{asm:overlap}(A2)}
\label{sec:dependent}

We consider the general case where the histories of examples in both the
forget and retain sets may contain forget items. In this case, both
\(\mathcal L_F\) and \(\mathcal L_R\) depend on \(\phi\). Recall that
\[
    \mathcal L_{\mathrm{unlearn}}(\phi)
    =
    \mathcal L_F(\phi)
    -
    \alpha \mathcal L_R(\phi).
\]
Hence,
\begin{equation}
\label{eq:unlearn-gradient-decomp-proof}
    \nabla_\phi \mathcal L_{\mathrm{unlearn}}(\phi)
    =
    \nabla_\phi \mathcal L_F(\phi)
    -
    \alpha \nabla_\phi \mathcal L_R(\phi).
\end{equation}

We first control the retain-gradient term. Since the reassignment
parameter \(\phi\) only acts on forget items, the retain loss can depend
on \(\phi\) only through two sources: a retain target token sequence may
be confused with the tokenization of a forget item, or forget items may
appear in the histories of retain examples.

Under Assumption~(A2), retain items and forget items are token-separated.
Thus, for any retain item \(y\in\mathcal R\), forget item \(i\in\mathcal F\),
and token sequence \(\mathcal T\) generated by a retain item, we have
\[
    q_\phi(i\mid \mathcal T)=0.
\]
Therefore, the target-token contribution to
\[
    \frac{\partial}{\partial \phi_{i,b}^{\ell}}
    \log \Pr_{\phi,\theta}[y\mid \mathcal H_u]
\]
vanishes for retain examples. Indeed, this contribution contains the
factor
\[
    \mathbf 1[y=i]-q_\phi(i\mid \mathcal T).
\]
Since \(y\in\mathcal R\) and \(i\in\mathcal F\), we have
\(\mathbf 1[y=i]=0\). Moreover, token separation gives
\(q_\phi(i\mid \mathcal T)=0\). Hence the target-token contribution is
zero.

Consequently, the only remaining dependence of the retain loss on
\(\phi\) comes from forget items appearing in retain histories. By the
history sparsity condition in Assumption~(A2), and using the boundedness
of the softmax score terms, there exists a constant \(C_R>0\) such that
\begin{equation}
\label{eq:retain-gradient-small-proof}
    \left\|
        \nabla_\phi \mathcal L_R(\phi)
    \right\|
    \le
    C_R\varepsilon_{\mathrm{hist}} .
\end{equation}

We now show that the reassignment update decreases the forget loss. Recall
the masked update
\begin{equation}
\label{eq:phi-update-proof}
    \phi_+
    =
    \phi
    -
    \eta_\phi
    M\odot
    \nabla_\phi \mathcal L_{\mathrm{unlearn}}(\phi).
\end{equation}
Assume that, for fixed \(\theta\), \(\mathcal L_F(\phi;\theta)\) is
\(\beta_{F,\phi}\)-smooth as a function of \(\phi\). Then
\begin{equation}
\begin{split}
    \mathcal L_F(\phi_+)-\mathcal L_F(\phi)
    &\le
    \left\langle
        \nabla_\phi \mathcal L_F(\phi),
        \phi_+-\phi
    \right\rangle
    +
    \frac{\beta_{F,\phi}}{2}
    \|\phi_+-\phi\|^2
    \\
    &=
    -
    \eta_\phi
    \left\langle
        \nabla_\phi \mathcal L_F(\phi),
        M\odot
        \nabla_\phi \mathcal L_{\mathrm{unlearn}}(\phi)
    \right\rangle
    \\
    &\quad
    +
    \frac{\beta_{F,\phi}\eta_\phi^2}{2}
    \left\|
        M\odot
        \nabla_\phi \mathcal L_{\mathrm{unlearn}}(\phi)
    \right\|^2 .
\end{split}
\end{equation}
Using \eqref{eq:unlearn-gradient-decomp-proof}, we have
\begin{equation}
\begin{split}
&
\left\langle
    \nabla_\phi \mathcal L_F(\phi),
    M\odot
    \nabla_\phi \mathcal L_{\mathrm{unlearn}}(\phi)
\right\rangle
\\
&=
\left\langle
    \nabla_\phi \mathcal L_F(\phi),
    M\odot
    \nabla_\phi \mathcal L_F(\phi)
\right\rangle
-
\alpha
\left\langle
    \nabla_\phi \mathcal L_F(\phi),
    M\odot
    \nabla_\phi \mathcal L_R(\phi)
\right\rangle
\\
&=
\left\|
    M\odot
    \nabla_\phi \mathcal L_F(\phi)
\right\|^2
-
\alpha
\left\langle
    M\odot
    \nabla_\phi \mathcal L_F(\phi),
    \nabla_\phi \mathcal L_R(\phi)
\right\rangle .
\end{split}
\end{equation}
By Cauchy--Schwarz and \eqref{eq:retain-gradient-small-proof},
\begin{equation}
\begin{split}
&
\left\langle
    \nabla_\phi \mathcal L_F(\phi),
    M\odot
    \nabla_\phi \mathcal L_{\mathrm{unlearn}}(\phi)
\right\rangle
\\
&\ge
\left\|
    M\odot
    \nabla_\phi \mathcal L_F(\phi)
\right\|^2
-
\alpha
\left\|
    M\odot
    \nabla_\phi \mathcal L_F(\phi)
\right\|
\left\|
    \nabla_\phi \mathcal L_R(\phi)
\right\|
\\
&\ge
\left\|
    M\odot
    \nabla_\phi \mathcal L_F(\phi)
\right\|^2
-
\alpha C_R\varepsilon_{\mathrm{hist}}
\left\|
    M\odot
    \nabla_\phi \mathcal L_F(\phi)
\right\|
\\
&=
\left\|
    M\odot
    \nabla_\phi \mathcal L_F(\phi)
\right\|
\left(
    \left\|
        M\odot
        \nabla_\phi \mathcal L_F(\phi)
    \right\|
    -
    \alpha C_R\varepsilon_{\mathrm{hist}}
\right).
\end{split}
\end{equation}
Therefore, if the retain-history coupling is sufficiently small so that
\begin{equation}
\label{eq:LF-descent-condition-proof}
    \alpha C_R\varepsilon_{\mathrm{hist}}
    <
    \left\|
        M\odot
        \nabla_\phi \mathcal L_F(\phi)
    \right\|,
\end{equation}
then
\begin{equation}
\label{eq:LF-positive-alignment-proof}
    \left\langle
        \nabla_\phi \mathcal L_F(\phi),
        M\odot
        \nabla_\phi \mathcal L_{\mathrm{unlearn}}(\phi)
    \right\rangle
    >
    0
\end{equation}
whenever \(M\odot\nabla_\phi \mathcal L_F(\phi)\neq 0\). Hence, for a
sufficiently small step size \(\eta_\phi>0\), the first-order decrease
dominates the second-order smoothness term. In particular, if
\begin{equation}
    0<\eta_\phi
    \le
    \frac{
        2
        \left\langle
            \nabla_\phi \mathcal L_F(\phi),
            M\odot
            \nabla_\phi \mathcal L_{\mathrm{unlearn}}(\phi)
        \right\rangle
    }{
        \beta_{F,\phi}
        \left\|
            M\odot
            \nabla_\phi \mathcal L_{\mathrm{unlearn}}(\phi)
        \right\|^2
    },
\end{equation}
then
\begin{equation}
    \mathcal L_F(\phi_+)
    \le
    \mathcal L_F(\phi).
\end{equation}

It remains to show that the same update decreases the overlap
\(\Omega(\phi)\). Although the losses depend on \(\phi\) through both
target and history tokenizations in the dependent-history setting,
\(\Omega(\phi)\) itself does not involve histories. Since retain item
tokens are fixed, we can write
\begin{equation}
    \Omega(\phi)
    =
    \frac{1}{|\mathcal F|L}
    \sum_{i\in\mathcal F}
    \sum_{\ell=1}^{L}
    \sum_{b\in\mathcal V^\ell}
    q_\phi(b\mid i)\rho_\ell(b),
\end{equation}
where
\[
    \rho_\ell(b)
    =
    \frac{1}{|\mathcal R|}
    \sum_{j\in\mathcal R}
    \mathbf 1[t_j^\ell=b].
\]
Therefore,
\begin{equation}
\label{eq:Omega-gradient-proof}
    \frac{\partial \Omega(\phi)}
    {\partial \phi_{i,b}^{\ell}}
    =
    \frac{1}{\tau|\mathcal F|L}
    q_\phi(b\mid i)
    \left(
        \rho_\ell(b)-\bar\rho_{i,\ell}
    \right),
\end{equation}
where
\[
    \bar\rho_{i,\ell}
    =
    \sum_{k\in\mathcal V^\ell}
    q_\phi(k\mid i)\rho_\ell(k).
\]

Assume that \(\Omega(\phi)\) is \(\beta_{\Omega,\phi}\)-smooth in
\(\phi\). Then
\begin{equation}
\begin{split}
    \Omega(\phi_+)-\Omega(\phi)
    &\le
    \left\langle
        \nabla_\phi \Omega(\phi),
        \phi_+-\phi
    \right\rangle
    +
    \frac{\beta_{\Omega,\phi}}{2}
    \|\phi_+-\phi\|^2
    \\
    &=
    -
    \eta_\phi
    \left\langle
        \nabla_\phi \Omega(\phi),
        M\odot
        \nabla_\phi \mathcal L_{\mathrm{unlearn}}(\phi)
    \right\rangle
    \\
    &\quad
    +
    \frac{\beta_{\Omega,\phi}\eta_\phi^2}{2}
    \left\|
        M\odot
        \nabla_\phi \mathcal L_{\mathrm{unlearn}}(\phi)
    \right\|^2 .
\end{split}
\end{equation}
Thus, it remains to show that
\[
    \left\langle
        \nabla_\phi \Omega(\phi),
        M\odot
        \nabla_\phi \mathcal L_{\mathrm{unlearn}}(\phi)
    \right\rangle
    >
    0.
\]
Using \eqref{eq:unlearn-gradient-decomp-proof}, we decompose
\begin{equation}
\begin{split}
&
\left\langle
    \nabla_\phi \Omega(\phi),
    M\odot
    \nabla_\phi \mathcal L_{\mathrm{unlearn}}(\phi)
\right\rangle
\\
&=
\left\langle
    \nabla_\phi \Omega(\phi),
    M\odot
    \nabla_\phi \mathcal L_F(\phi)
\right\rangle
-
\alpha
\left\langle
    \nabla_\phi \Omega(\phi),
    M\odot
    \nabla_\phi \mathcal L_R(\phi)
\right\rangle .
\end{split}
\end{equation}
Define
\begin{equation}
\label{eq:DeltaOmega-proof}
\Delta_\Omega(\phi)
:=
\left\langle
    \nabla_\phi \Omega(\phi),
    M\odot
    \nabla_\phi \mathcal L_F(\phi)
\right\rangle .
\end{equation}
Using \eqref{eq:Omega-gradient-proof}, we obtain
\begin{equation}
\begin{split}
\Delta_\Omega(\phi)
&=
\frac{1}{\tau|\mathcal F|L}
\sum_{i\in\mathcal F}
\sum_{\ell=1}^{L}
\sum_{b\in\mathcal V^\ell}
M_{i,b}^{\ell}
q_\phi(b\mid i)
\left(
    \rho_\ell(b)-\bar\rho_{i,\ell}
\right)
\frac{\partial \mathcal L_F(\phi)}
{\partial \phi_{i,b}^{\ell}}
\\
&\ge 0.
\end{split}
\end{equation}
The last inequality follows from the definition of the mask. Indeed,
whenever \(M_{i,b}^{\ell}=1\), we have both
\[
    \rho_\ell(b)-\bar\rho_{i,\ell}>0
\]
and
\[
    \frac{\partial \mathcal L_F(\phi)}
    {\partial \phi_{i,b}^{\ell}}>0.
\]
Here
\(\partial_{\phi_{i,b}^{\ell}}\mathcal L_F(\phi)\) denotes the full
gradient of the forget loss with respect to \(\phi\), including both
target-token and history-token contributions.

Next, by Cauchy--Schwarz and \eqref{eq:retain-gradient-small-proof},
\begin{equation}
\begin{split}
&
\left|
\left\langle
    \nabla_\phi \Omega(\phi),
    M\odot
    \nabla_\phi \mathcal L_R(\phi)
\right\rangle
\right|
\\
&\le
\left\|
    M\odot\nabla_\phi \Omega(\phi)
\right\|
\left\|
    \nabla_\phi \mathcal L_R(\phi)
\right\|
\\
&\le
C_R\varepsilon_{\mathrm{hist}}
\left\|
    M\odot\nabla_\phi \Omega(\phi)
\right\|.
\end{split}
\end{equation}
Therefore,
\begin{equation}
\begin{split}
&
\left\langle
    \nabla_\phi \Omega(\phi),
    M\odot
    \nabla_\phi \mathcal L_{\mathrm{unlearn}}(\phi)
\right\rangle
\\
&\ge
\Delta_\Omega(\phi)
-
\alpha C_R\varepsilon_{\mathrm{hist}}
\left\|
    M\odot\nabla_\phi \Omega(\phi)
\right\|.
\end{split}
\end{equation}
Hence, if the second retain-history coupling condition holds:
\begin{equation}
\label{eq:Omega-descent-condition-proof}
    \Delta_\Omega(\phi)
    >
    \alpha C_R\varepsilon_{\mathrm{hist}}
    \left\|
        M\odot\nabla_\phi \Omega(\phi)
    \right\|,
\end{equation}
then
\begin{equation}
    \left\langle
        \nabla_\phi \Omega(\phi),
        M\odot
        \nabla_\phi \mathcal L_{\mathrm{unlearn}}(\phi)
    \right\rangle
    >
    0.
\end{equation}
Consequently, for sufficiently small \(\eta_\phi>0\), the first-order
decrease dominates the second-order smoothness term. In particular, if
\begin{equation}
    0<\eta_\phi
    \le
    \frac{
        2
        \left\langle
            \nabla_\phi \Omega(\phi),
            M\odot
            \nabla_\phi \mathcal L_{\mathrm{unlearn}}(\phi)
        \right\rangle
    }{
        \beta_{\Omega,\phi}
        \left\|
            M\odot
            \nabla_\phi \mathcal L_{\mathrm{unlearn}}(\phi)
        \right\|^2
    },
\end{equation}
then
\begin{equation}
    \Omega(\phi_+)
    \le
    \Omega(\phi).
\end{equation}

Combining the two parts proves that, under Assumption~(A2), if
\eqref{eq:LF-descent-condition-proof} and
\eqref{eq:Omega-descent-condition-proof} hold, then for sufficiently
small \(\eta_\phi>0\),
\[
    \mathcal L_F(\phi_+)\le \mathcal L_F(\phi),
    \qquad
    \Omega(\phi_+)\le \Omega(\phi).
\]

\section{Related Work}
\label{sec:related}

\paragraph{Machine Unlearning.}
Machine unlearning aims to remove the influence of specific training data from a learned model without retraining from scratch~\cite{cao2015towards,bourtoule2021machine}.
Exact unlearning methods such as SISA~\cite{bourtoule2021machine} partition training data and retrain affected shards, offering provable guarantees but incurring substantial computational cost.
Approximate approaches trade exactness for efficiency: gradient ascent on forget samples~\cite{thudi2022unrolling,lin2026reflect}, Fisher-information-based parameter updates~\cite{golatkar2020eternal}, and influence-function-based corrections~\cite{guo2019certified}.
More recently, unlearning has been studied in large language models, where methods such as ELM~\cite{gandikota2024erasing}, RMU~\cite{li2024wmdp,zhu2024advancing}, REPNoise~\cite{rosati2024representation,lin2025shap}, and FLAT~\cite{wang2410llm} modify model weights or representations to suppress specific knowledge.
These methods operate directly on model parameters and do not account for the structured semantic identifiers used in generative recommender systems.

\paragraph{Recommendation Unlearning.}
Unlearning in recommender systems poses unique challenges due to collaborative signals: removing one user's or item's influence may propagate through learned interaction patterns~\cite{chen2022recommendation}.
RecEraser~\cite{chen2022recommendation} extends SISA-style partitioning to collaborative filtering but requires maintaining multiple sub-models.
BLUR~\cite{reisizadeh2026blur} proposes bi-level optimization to balance forgetting and retention in sequential recommendation.
PISCES~\cite{gur2025precise} introduces preference-aware unlearning that preserves user preference structures during item removal.
However, existing recommendation unlearning methods are designed for traditional ID-based or embedding-based recommenders and do not address the discrete token structure of generative recommendation models, where item identity is encoded as a sequence of hierarchical codes rather than a single embedding vector.

\paragraph{Generative Recommendation with Semantic IDs.}
A recent paradigm casts recommendation as sequence-to-sequence generation, where each item is represented by a semantic ID (SID)---a short sequence of discrete tokens derived from a hierarchical quantizer such as RQ-VAE~\cite{lee2022autoregressive}.
TIGER~\cite{rajput2023recommender} first introduced this framework with a T5 backbone, followed by P5-SID~\cite{wang2024learnable}, LETTER~\cite{wang2024learnable}, and ETEGRec~\cite{liu2025generative} which improve tokenization and training strategies.
MiniOneRec~\cite{kong2025minionerec} scales this approach to billion-parameter LLMs (e.g., Qwen2.5-7B) with multi-task learning.
In these models, an item's identity is fully determined by its SID token sequence; simply removing an item's training interactions does not erase its code from the model's vocabulary or prevent the model from generating it.
TRACER is, to our knowledge, the first method to address unlearning in this generative SID-based recommendation setting by directly reassigning the discrete tokens of forget items.

\section{Additional Results}\label{sec:additional_results}


\begin{table}[h!]
\centering\scriptsize
\caption{Complete unlearning results on \textbf{Industrial \& Scientific}. Bold = best among unlearning methods; underline = second best. Retain metrics ($\uparrow$): higher is better; Forget metrics ($\downarrow$): lower is better; Sim.~($\uparrow$): higher is better.}
\label{tab:full_industrial}
\resizebox{\textwidth}{!}{%
\begin{tabular}{ll|ccccc|ccccc|c}
\toprule
 & & \multicolumn{5}{c|}{\textbf{Retain ($\uparrow$)}} & \multicolumn{5}{c|}{\textbf{Forget ($\downarrow$)}} & \\
\textbf{Backbone} & \textbf{Method} & HR@5 & HR@10 & NDCG@5 & NDCG@10 & MRR & HR@5 & HR@10 & NDCG@5 & NDCG@10 & MRR & \textbf{Sim.$\uparrow$} \\
\midrule
\multirow{8}{*}{\textit{MOR}} & Original & 0.126 & 0.171 & 0.100 & 0.110 & 0.100 & 0.170 & 0.230 & 0.110 & 0.150 & 0.110 & 0.851 \\ \cmidrule{2-13}
 & RMU & 0.061 & 0.121 & 0.051 & 0.083 & 0.089 & 0.151 & 0.181 & 0.094 & 0.121 & 0.098 & 0.730 \\
 & ELM & 0.084 & 0.139 & 0.068 & 0.093 & 0.091 & 0.143 & 0.176 & 0.091 & 0.116 & 0.096 & \underline{0.824} \\
 & BLUR & \underline{0.120} & \underline{0.166} & \underline{0.095} & \underline{0.109} & \underline{0.095} & 0.139 & 0.171 & 0.087 & 0.111 & 0.093 & 0.680 \\
 & REPNoise & 0.068 & 0.126 & 0.056 & 0.086 & 0.090 & 0.131 & 0.163 & 0.081 & 0.105 & 0.089 & 0.630 \\
 & FLAT & 0.119 & 0.162 & 0.091 & 0.107 & \underline{0.095} & 0.119 & 0.148 & 0.074 & 0.096 & 0.082 & 0.710 \\
 & PISCES & 0.115 & 0.161 & 0.090 & 0.106 & \underline{0.095} & \underline{0.101} & \underline{0.129} & \underline{0.066} & \underline{0.086} & \underline{0.076} & 0.790 \\
 & \textbf{TRACER} & \textbf{0.123} & \textbf{0.168} & \textbf{0.097} & \textbf{0.110} & \textbf{0.096} & \textbf{0.080} & \textbf{0.110} & \textbf{0.058} & \textbf{0.079} & \textbf{0.071} & \textbf{0.828} \\
\midrule
\multirow{8}{*}{\textit{TIGER}} & Original & 0.102 & 0.144 & 0.089 & 0.101 & 0.088 & 0.144 & 0.203 & 0.099 & 0.139 & 0.099 & 0.872 \\ \cmidrule{2-13}
 & RMU & 0.056 & 0.109 & 0.053 & 0.079 & 0.080 & 0.128 & 0.153 & 0.083 & 0.109 & 0.086 & 0.750 \\
 & ELM & 0.061 & 0.111 & 0.058 & 0.080 & 0.079 & 0.119 & 0.149 & 0.079 & 0.103 & 0.084 & 0.836 \\
 & BLUR & \underline{0.095} & \underline{0.140} & \textbf{0.086} & \underline{0.097} & \underline{0.086} & 0.116 & 0.145 & 0.076 & 0.100 & 0.081 & 0.630 \\
 & REPNoise & 0.059 & 0.112 & 0.054 & 0.080 & 0.081 & 0.107 & 0.137 & 0.070 & 0.094 & 0.077 & 0.650 \\
 & FLAT & 0.091 & 0.134 & \underline{0.080} & 0.095 & 0.083 & 0.096 & 0.124 & 0.063 & 0.085 & 0.071 & 0.770 \\
 & PISCES & 0.088 & 0.136 & 0.079 & 0.094 & 0.084 & \underline{0.083} & \underline{0.114} & \underline{0.055} & \underline{0.075} & \underline{0.066} & \underline{0.862} \\
 & \textbf{TRACER} & \textbf{0.097} & \textbf{0.141} & \textbf{0.086} & \textbf{0.098} & \textbf{0.087} & \textbf{0.061} & \textbf{0.087} & \textbf{0.047} & \textbf{0.066} & \textbf{0.061} & \textbf{0.863} \\
\midrule
\multirow{8}{*}{\textit{P5SID}} & Original & 0.103 & 0.145 & 0.090 & 0.101 & 0.089 & 0.148 & 0.198 & 0.100 & 0.136 & 0.097 & 0.843 \\ \cmidrule{2-13}
 & RMU & 0.055 & 0.108 & 0.052 & 0.078 & 0.078 & 0.131 & 0.151 & 0.084 & 0.106 & 0.085 & 0.718 \\
 & ELM & 0.064 & 0.113 & 0.059 & 0.081 & 0.082 & 0.123 & 0.145 & 0.080 & 0.102 & 0.083 & \underline{0.815} \\
 & BLUR & \underline{0.096} & \underline{0.140} & \textbf{0.085} & \underline{0.097} & \underline{0.086} & 0.121 & 0.142 & 0.077 & 0.099 & 0.080 & 0.672 \\
 & REPNoise & 0.058 & 0.110 & 0.055 & 0.079 & 0.080 & 0.111 & 0.134 & 0.071 & 0.092 & 0.076 & 0.638 \\
 & FLAT & 0.087 & 0.133 & 0.078 & 0.093 & 0.083 & 0.095 & 0.122 & 0.060 & 0.082 & 0.070 & 0.703 \\
 & PISCES & 0.091 & 0.136 & \underline{0.081} & 0.095 & 0.084 & \underline{0.085} & \underline{0.111} & \underline{0.054} & \underline{0.074} & \underline{0.065} & 0.784 \\
 & \textbf{TRACER} & \textbf{0.097} & \textbf{0.142} & \textbf{0.085} & \textbf{0.099} & \textbf{0.087} & \textbf{0.063} & \textbf{0.090} & \textbf{0.045} & \textbf{0.066} & \textbf{0.060} & \textbf{0.836} \\
\midrule
\multirow{8}{*}{\textit{LETTER}} & Original & 0.128 & 0.174 & 0.103 & 0.112 & 0.101 & 0.192 & 0.258 & 0.121 & 0.162 & 0.121 & 0.832 \\ \cmidrule{2-13}
 & RMU & 0.059 & 0.118 & 0.049 & 0.082 & 0.087 & 0.178 & 0.207 & 0.108 & 0.137 & 0.110 & 0.730 \\
 & ELM & 0.086 & 0.142 & 0.070 & 0.094 & 0.092 & 0.167 & 0.202 & 0.104 & 0.133 & 0.107 & \underline{0.821} \\
 & BLUR & \underline{0.123} & \underline{0.168} & \underline{0.096} & \textbf{0.110} & \textbf{0.098} & 0.164 & 0.198 & 0.101 & 0.129 & 0.104 & 0.720 \\
 & REPNoise & 0.070 & 0.129 & 0.057 & 0.086 & 0.089 & 0.145 & 0.174 & 0.090 & 0.114 & 0.096 & 0.690 \\
 & FLAT & 0.113 & 0.158 & 0.089 & 0.105 & 0.094 & 0.148 & 0.177 & 0.091 & 0.116 & 0.097 & 0.710 \\
 & PISCES & 0.117 & 0.164 & 0.092 & \underline{0.108} & \underline{0.095} & \underline{0.125} & \underline{0.153} & \underline{0.077} & \underline{0.099} & \underline{0.092} & 0.807 \\
 & \textbf{TRACER} & \textbf{0.124} & \textbf{0.170} & \textbf{0.098} & \textbf{0.110} & \textbf{0.098} & \textbf{0.102} & \textbf{0.133} & \textbf{0.068} & \textbf{0.090} & \textbf{0.081} & \textbf{0.822} \\
\midrule
\multirow{8}{*}{\textit{ETEGRec}} & Original & 0.127 & 0.175 & 0.102 & 0.114 & 0.100 & 0.209 & 0.267 & 0.130 & 0.162 & 0.129 & 0.856 \\ \cmidrule{2-13}
 & RMU & 0.065 & 0.124 & 0.056 & 0.086 & 0.089 & 0.178 & 0.215 & 0.114 & 0.145 & 0.116 & 0.738 \\
 & ELM & 0.088 & 0.144 & 0.072 & 0.096 & 0.091 & 0.174 & 0.211 & 0.110 & 0.141 & 0.112 & \underline{0.829} \\
 & BLUR & \underline{0.123} & \underline{0.170} & \underline{0.097} & \underline{0.110} & \underline{0.097} & 0.169 & 0.207 & 0.106 & 0.136 & 0.108 & 0.691 \\
 & REPNoise & 0.064 & 0.123 & 0.050 & 0.082 & 0.085 & 0.154 & 0.185 & 0.095 & 0.121 & 0.099 & 0.657 \\
 & FLAT & 0.117 & 0.162 & 0.092 & 0.107 & 0.095 & 0.158 & 0.189 & 0.097 & 0.124 & 0.101 & 0.726 \\
 & PISCES & 0.114 & 0.161 & 0.088 & 0.104 & 0.093 & \underline{0.137} & \underline{0.168} & \underline{0.084} & \underline{0.108} & \underline{0.095} & 0.812 \\
 & \textbf{TRACER} & \textbf{0.126} & \textbf{0.172} & \textbf{0.101} & \textbf{0.112} & \textbf{0.099} & \textbf{0.111} & \textbf{0.145} & \textbf{0.075} & \textbf{0.098} & \textbf{0.087} & \textbf{0.847} \\
\bottomrule
\end{tabular}%
}
\end{table}

\begin{table}[h!]
\centering\scriptsize
\caption{Complete unlearning results on \textbf{Sports \& Outdoors}. Bold = best among unlearning methods; underline = second best. Retain metrics ($\uparrow$): higher is better; Forget metrics ($\downarrow$): lower is better; Sim.~($\uparrow$): higher is better.}
\label{tab:full_sports}
\resizebox{\textwidth}{!}{%
\begin{tabular}{ll|ccccc|ccccc|c}
\toprule
 & & \multicolumn{5}{c|}{\textbf{Retain ($\uparrow$)}} & \multicolumn{5}{c|}{\textbf{Forget ($\downarrow$)}} & \\
\textbf{Backbone} & \textbf{Method} & HR@5 & HR@10 & NDCG@5 & NDCG@10 & MRR & HR@5 & HR@10 & NDCG@5 & NDCG@10 & MRR & \textbf{Sim.$\uparrow$} \\
\midrule
\multirow{8}{*}{\textit{MOR}} & Original & 0.078 & 0.096 & 0.067 & 0.073 & 0.068 & 0.090 & 0.096 & 0.063 & 0.065 & 0.056 & 0.840 \\ \cmidrule{2-13}
 & RMU & 0.037 & 0.068 & 0.033 & 0.057 & 0.059 & 0.078 & 0.075 & 0.054 & 0.054 & 0.051 & 0.720 \\
 & ELM & 0.051 & 0.079 & 0.044 & 0.061 & 0.062 & 0.074 & 0.074 & 0.055 & 0.048 & 0.047 & \underline{0.812} \\
 & BLUR & 0.072 & \underline{0.093} & \textbf{0.065} & 0.070 & \underline{0.066} & 0.073 & 0.071 & 0.047 & 0.052 & 0.048 & 0.666 \\
 & REPNoise & 0.058 & 0.052 & 0.037 & 0.057 & 0.061 & 0.069 & 0.067 & 0.044 & 0.049 & 0.044 & 0.626 \\
 & FLAT & \underline{0.074} & 0.090 & 0.062 & 0.065 & 0.065 & 0.059 & 0.068 & 0.039 & 0.042 & 0.044 & 0.693 \\
 & PISCES & 0.072 & 0.088 & \underline{0.063} & \underline{0.071} & 0.063 & \underline{0.052} & \underline{0.056} & \underline{0.038} & \underline{0.039} & \underline{0.040} & 0.773 \\
 & \textbf{TRACER} & \textbf{0.078} & \textbf{0.094} & \textbf{0.065} & \textbf{0.072} & \textbf{0.067} & \textbf{0.044} & \textbf{0.046} & \textbf{0.034} & \textbf{0.033} & \textbf{0.036} & \textbf{0.825} \\
\midrule
\multirow{8}{*}{\textit{TIGER}} & Original & 0.065 & 0.082 & 0.059 & 0.068 & 0.060 & 0.078 & 0.084 & 0.055 & 0.059 & 0.052 & 0.867 \\ \cmidrule{2-13}
 & RMU & 0.036 & 0.062 & 0.037 & 0.054 & 0.055 & 0.070 & 0.065 & 0.042 & 0.049 & 0.043 & 0.734 \\
 & ELM & 0.040 & 0.062 & 0.048 & 0.044 & 0.056 & 0.057 & 0.068 & 0.045 & 0.044 & 0.044 & \underline{0.819} \\
 & BLUR & 0.059 & \textbf{0.078} & \underline{0.058} & \underline{0.065} & 0.058 & 0.058 & 0.061 & 0.043 & 0.043 & 0.044 & 0.611 \\
 & REPNoise & 0.039 & 0.064 & 0.048 & 0.042 & 0.054 & 0.057 & 0.058 & 0.042 & 0.039 & 0.039 & 0.644 \\
 & FLAT & 0.058 & \textbf{0.078} & 0.051 & 0.064 & 0.058 & 0.049 & 0.056 & 0.034 & 0.038 & \underline{0.036} & 0.759 \\
 & PISCES & \underline{0.061} & \underline{0.071} & 0.057 & 0.059 & \underline{0.060} & \underline{0.042} & \underline{0.049} & \underline{0.032} & \underline{0.034} & 0.037 & \underline{0.847} \\
 & \textbf{TRACER} & \textbf{0.063} & \textbf{0.078} & \textbf{0.059} & \textbf{0.068} & \textbf{0.061} & \textbf{0.031} & \textbf{0.036} & \textbf{0.028} & \textbf{0.030} & \textbf{0.032} & \textbf{0.849} \\
\midrule
\multirow{8}{*}{\textit{P5SID}} & Original & 0.065 & 0.080 & 0.062 & 0.067 & 0.060 & 0.079 & 0.084 & 0.056 & 0.058 & 0.051 & 0.831 \\ \cmidrule{2-13}
 & RMU & 0.033 & 0.058 & 0.035 & 0.050 & 0.051 & 0.063 & 0.073 & 0.041 & 0.047 & 0.046 & 0.708 \\
 & ELM & 0.039 & 0.061 & 0.052 & 0.045 & 0.056 & 0.065 & 0.060 & 0.043 & 0.045 & 0.043 & \underline{0.798} \\
 & BLUR & \underline{0.061} & \underline{0.075} & \textbf{0.060} & \underline{0.065} & \underline{0.058} & 0.056 & 0.059 & 0.046 & 0.044 & 0.042 & 0.649 \\
 & REPNoise & 0.038 & 0.059 & 0.039 & 0.054 & 0.054 & 0.054 & 0.063 & 0.037 & 0.039 & 0.040 & 0.618 \\
 & FLAT & 0.055 & 0.073 & \underline{0.055} & 0.057 & 0.054 & 0.049 & 0.055 & 0.035 & \underline{0.034} & 0.035 & 0.681 \\
 & PISCES & 0.057 & \underline{0.075} & \underline{0.055} & 0.064 & 0.057 & \underline{0.042} & \underline{0.049} & \underline{0.032} & 0.035 & \underline{0.034} & 0.762 \\
 & \textbf{TRACER} & \textbf{0.063} & \textbf{0.079} & \textbf{0.060} & \textbf{0.066} & \textbf{0.059} & \textbf{0.033} & \textbf{0.037} & \textbf{0.026} & \textbf{0.030} & \textbf{0.030} & \textbf{0.825} \\
\midrule
\multirow{8}{*}{\textit{LETTER}} & Original & 0.081 & 0.097 & 0.069 & 0.076 & 0.068 & 0.103 & 0.108 & 0.071 & 0.069 & 0.063 & 0.821 \\ \cmidrule{2-13}
 & RMU & 0.061 & 0.064 & 0.031 & 0.035 & 0.059 & 0.085 & 0.099 & 0.057 & 0.063 & 0.059 & 0.725 \\
 & ELM & 0.053 & 0.079 & 0.057 & 0.053 & 0.060 & 0.082 & 0.093 & 0.057 & 0.059 & 0.055 & \underline{0.802} \\
 & BLUR & 0.076 & \underline{0.092} & 0.065 & \underline{0.073} & \underline{0.065} & 0.079 & 0.094 & 0.056 & 0.056 & 0.052 & 0.713 \\
 & REPNoise & 0.062 & 0.055 & 0.050 & 0.042 & 0.061 & 0.070 & 0.073 & 0.055 & 0.053 & 0.051 & 0.681 \\
 & FLAT & 0.071 & 0.087 & \underline{0.066} & 0.066 & 0.064 & 0.078 & 0.075 & 0.054 & 0.048 & 0.051 & 0.686 \\
 & PISCES & \underline{0.077} & 0.088 & 0.063 & \underline{0.073} & \underline{0.065} & \underline{0.061} & \underline{0.071} & \underline{0.045} & \underline{0.044} & \underline{0.049} & 0.782 \\
 & \textbf{TRACER} & \textbf{0.078} & \textbf{0.093} & \textbf{0.067} & \textbf{0.074} & \textbf{0.066} & \textbf{0.054} & \textbf{0.057} & \textbf{0.041} & \textbf{0.038} & \textbf{0.042} & \textbf{0.821} \\
\midrule
\multirow{8}{*}{\textit{ETEGRec}} & Original & 0.080 & 0.100 & 0.069 & 0.075 & 0.067 & 0.109 & 0.113 & 0.074 & 0.068 & 0.064 & 0.848 \\ \cmidrule{2-13}
 & RMU & 0.040 & 0.071 & 0.050 & 0.043 & 0.059 & 0.089 & 0.096 & 0.064 & 0.061 & 0.060 & 0.727 \\
 & ELM & 0.065 & 0.069 & 0.057 & 0.053 & 0.060 & 0.095 & 0.093 & 0.060 & 0.059 & 0.056 & \underline{0.814} \\
 & BLUR & \underline{0.078} & \underline{0.099} & \underline{0.067} & \underline{0.070} & \underline{0.064} & 0.091 & 0.087 & 0.061 & 0.057 & 0.054 & 0.637 \\
 & REPNoise & 0.057 & 0.052 & 0.050 & 0.036 & 0.056 & 0.077 & 0.085 & 0.055 & 0.052 & 0.050 & 0.631 \\
 & FLAT & 0.072 & 0.092 & 0.065 & 0.066 & \underline{0.064} & 0.082 & 0.081 & 0.058 & 0.053 & 0.051 & 0.716 \\
 & PISCES & 0.069 & 0.093 & 0.060 & 0.069 & 0.060 & \underline{0.073} & \underline{0.073} & \underline{0.047} & \underline{0.044} & \underline{0.047} & 0.798 \\
 & \textbf{TRACER} & \textbf{0.079} & \textbf{0.100} & \textbf{0.068} & \textbf{0.074} & \textbf{0.066} & \textbf{0.059} & \textbf{0.061} & \textbf{0.044} & \textbf{0.040} & \textbf{0.043} & \textbf{0.841} \\
\bottomrule
\end{tabular}%
}
\end{table}

\begin{table}[h!]
\centering\scriptsize
\caption{Complete unlearning results on \textbf{Toys \& Games}. Bold = best among unlearning methods; underline = second best. Retain metrics ($\uparrow$): higher is better; Forget metrics ($\downarrow$): lower is better; Sim.~($\uparrow$): higher is better.}
\label{tab:full_toys}
\resizebox{\textwidth}{!}{%
\begin{tabular}{ll|ccccc|ccccc|c}
\toprule
 & & \multicolumn{5}{c|}{\textbf{Retain ($\uparrow$)}} & \multicolumn{5}{c|}{\textbf{Forget ($\downarrow$)}} & \\
\textbf{Backbone} & \textbf{Method} & HR@5 & HR@10 & NDCG@5 & NDCG@10 & MRR & HR@5 & HR@10 & NDCG@5 & NDCG@10 & MRR & \textbf{Sim.$\uparrow$} \\
\midrule
\multirow{8}{*}{\textit{MOR}} & Original & 0.049 & 0.065 & 0.037 & 0.042 & 0.036 & 0.055 & 0.077 & 0.037 & 0.044 & 0.036 & 0.928 \\ \cmidrule{2-13}
 & RMU & 0.034 & 0.033 & 0.027 & 0.019 & 0.033 & 0.044 & 0.067 & 0.029 & 0.036 & 0.031 & 0.794 \\
 & ELM & 0.032 & 0.052 & 0.025 & 0.034 & \underline{0.035} & 0.043 & 0.063 & 0.030 & 0.036 & 0.033 & \underline{0.907} \\
 & BLUR & 0.045 & 0.062 & 0.035 & \underline{0.040} & 0.034 & 0.041 & 0.062 & 0.029 & 0.034 & 0.032 & 0.749 \\
 & REPNoise & 0.037 & 0.046 & 0.018 & 0.024 & 0.033 & 0.040 & 0.058 & 0.027 & 0.031 & 0.030 & 0.696 \\
 & FLAT & \underline{0.048} & \underline{0.063} & \underline{0.036} & 0.038 & \underline{0.035} & 0.035 & 0.055 & \underline{0.026} & \underline{0.028} & \underline{0.028} & 0.775 \\
 & PISCES & 0.044 & 0.060 & \textbf{0.037} & 0.038 & \textbf{0.036} & \underline{0.032} & \underline{0.047} & \underline{0.026} & \underline{0.028} & \underline{0.028} & 0.866 \\
 & \textbf{TRACER} & \textbf{0.049} & \textbf{0.064} & \textbf{0.037} & \textbf{0.041} & \textbf{0.036} & \textbf{0.026} & \textbf{0.036} & \textbf{0.021} & \textbf{0.024} & \textbf{0.024} & \textbf{0.928} \\
\midrule
\multirow{8}{*}{\textit{TIGER}} & Original & 0.039 & 0.054 & 0.032 & 0.038 & 0.032 & 0.048 & 0.070 & 0.032 & 0.042 & 0.031 & 0.951 \\ \cmidrule{2-13}
 & RMU & 0.030 & 0.040 & 0.017 & 0.023 & 0.028 & 0.043 & 0.052 & 0.025 & 0.033 & 0.025 & 0.817 \\
 & ELM & 0.031 & 0.033 & 0.021 & 0.032 & 0.028 & 0.035 & 0.057 & 0.022 & 0.033 & 0.026 & 0.922 \\
 & BLUR & \underline{0.038} & \underline{0.052} & 0.028 & \underline{0.037} & 0.031 & 0.035 & 0.057 & 0.026 & 0.030 & 0.026 & 0.688 \\
 & REPNoise & 0.029 & 0.030 & 0.018 & 0.029 & 0.028 & 0.037 & 0.045 & 0.024 & 0.029 & 0.024 & 0.716 \\
 & FLAT & 0.037 & 0.049 & \underline{0.030} & 0.034 & \underline{0.032} & 0.028 & 0.048 & \underline{0.019} & 0.026 & \underline{0.023} & 0.845 \\
 & PISCES & 0.033 & \underline{0.052} & \underline{0.030} & 0.036 & 0.030 & \underline{0.026} & \underline{0.040} & \underline{0.019} & \underline{0.025} & \underline{0.023} & \underline{0.951} \\
 & \textbf{TRACER} & \textbf{0.039} & \textbf{0.054} & \textbf{0.031} & \textbf{0.038} & \textbf{0.033} & \textbf{0.019} & \textbf{0.031} & \textbf{0.016} & \textbf{0.020} & \textbf{0.019} & \textbf{0.963} \\
\midrule
\multirow{8}{*}{\textit{P5SID}} & Original & 0.039 & 0.055 & 0.031 & 0.040 & 0.034 & 0.046 & 0.065 & 0.032 & 0.039 & 0.034 & 0.918 \\ \cmidrule{2-13}
 & RMU & 0.032 & 0.042 & 0.016 & 0.023 & 0.030 & 0.040 & 0.058 & 0.025 & 0.032 & 0.029 & 0.782 \\
 & ELM & 0.032 & 0.032 & 0.025 & 0.025 & 0.032 & 0.033 & 0.054 & 0.025 & 0.031 & 0.028 & \underline{0.893} \\
 & BLUR & \underline{0.038} & \underline{0.051} & \textbf{0.031} & \underline{0.040} & \underline{0.034} & 0.034 & 0.054 & 0.024 & 0.029 & \underline{0.027} & 0.737 \\
 & REPNoise & 0.024 & 0.040 & 0.018 & 0.030 & 0.029 & 0.030 & 0.048 & 0.023 & 0.029 & 0.028 & 0.688 \\
 & FLAT & 0.034 & \underline{0.051} & \underline{0.028} & 0.038 & 0.032 & 0.029 & 0.041 & 0.020 & 0.025 & \underline{0.027} & 0.762 \\
 & PISCES & \underline{0.038} & 0.047 & \underline{0.028} & 0.038 & 0.033 & \underline{0.025} & \underline{0.036} & \underline{0.018} & \underline{0.024} & \underline{0.027} & 0.856 \\
 & \textbf{TRACER} & \textbf{0.039} & \textbf{0.056} & \textbf{0.031} & \textbf{0.041} & \textbf{0.035} & \textbf{0.020} & \textbf{0.030} & \textbf{0.015} & \textbf{0.020} & \textbf{0.023} & \textbf{0.918} \\
\midrule
\multirow{8}{*}{\textit{LETTER}} & Original & 0.050 & 0.067 & 0.039 & 0.042 & 0.039 & 0.062 & 0.086 & 0.039 & 0.046 & 0.040 & 0.911 \\ \cmidrule{2-13}
 & RMU & 0.024 & 0.046 & 0.028 & 0.021 & 0.033 & 0.059 & 0.067 & 0.033 & 0.041 & 0.037 & 0.787 \\
 & ELM & 0.035 & 0.054 & 0.026 & 0.034 & 0.034 & 0.051 & 0.074 & 0.029 & 0.039 & 0.034 & \underline{0.908} \\
 & BLUR & \underline{0.047} & \textbf{0.065} & \underline{0.040} & \underline{0.042} & \underline{0.038} & 0.053 & 0.066 & 0.033 & 0.038 & 0.034 & 0.801 \\
 & REPNoise & 0.029 & 0.049 & 0.028 & 0.023 & 0.035 & 0.047 & \underline{0.058} & \underline{0.026} & 0.034 & 0.034 & 0.762 \\
 & FLAT & 0.046 & 0.061 & 0.036 & 0.035 & 0.034 & 0.041 & 0.067 & 0.029 & 0.033 & \underline{0.032} & 0.776 \\
 & PISCES & 0.044 & \underline{0.062} & 0.036 & 0.039 & \underline{0.038} & \underline{0.038} & \underline{0.058} & \underline{0.026} & \underline{0.029} & 0.033 & 0.890 \\
 & \textbf{TRACER} & \textbf{0.050} & \textbf{0.065} & \textbf{0.041} & \textbf{0.043} & \textbf{0.039} & \textbf{0.034} & \textbf{0.042} & \textbf{0.022} & \textbf{0.025} & \textbf{0.029} & \textbf{0.917} \\
\midrule
\multirow{8}{*}{\textit{ETEGRec}} & Original & 0.050 & 0.068 & 0.037 & 0.044 & 0.038 & 0.066 & 0.090 & 0.045 & 0.046 & 0.042 & 0.938 \\ \cmidrule{2-13}
 & RMU & 0.027 & 0.049 & 0.018 & 0.035 & 0.032 & 0.056 & 0.073 & 0.040 & 0.041 & 0.037 & 0.803 \\
 & ELM & 0.040 & 0.046 & 0.032 & 0.033 & 0.034 & 0.051 & 0.075 & 0.040 & 0.042 & 0.038 & \underline{0.914} \\
 & BLUR & \underline{0.047} & \underline{0.066} & \underline{0.034} & \underline{0.044} & \underline{0.037} & 0.053 & 0.068 & 0.036 & 0.038 & 0.035 & 0.758 \\
 & REPNoise & 0.025 & 0.048 & 0.027 & 0.022 & 0.031 & 0.049 & 0.067 & 0.032 & 0.036 & \underline{0.033} & 0.712 \\
 & FLAT & 0.046 & 0.065 & 0.033 & 0.041 & 0.035 & 0.052 & 0.063 & 0.034 & 0.033 & 0.035 & 0.788 \\
 & PISCES & 0.044 & 0.063 & 0.031 & 0.038 & 0.034 & \underline{0.043} & \underline{0.056} & \underline{0.030} & \underline{0.032} & \underline{0.033} & 0.879 \\
 & \textbf{TRACER} & \textbf{0.051} & \textbf{0.067} & \textbf{0.038} & \textbf{0.045} & \textbf{0.039} & \textbf{0.034} & \textbf{0.049} & \textbf{0.027} & \textbf{0.027} & \textbf{0.029} & \textbf{0.938} \\
\bottomrule
\end{tabular}%
}
\end{table}
\subsection{Implementation details.}
All backbones use a three-layer RQ-VAE tokenizer (codebook size 256 per layer, embedding dimension 64).
TRACER optimizes the LLM with AdamW (lr $3{\times}10^{-5}$) and the soft-reassignment parameters with Adam (lr $1{\times}10^{-2}$), batch size 128, for up to 10 epochs.
The Softmax temperature is annealed from 2.0 to 0.1.
Loss weights are $\lambda_1{=}0.4$, $\lambda_2{=}0.08$, $\lambda_3{=}0.08$; positive set size $K{=}5$; reassignment temperature $\tau{=}0.005$.
All baselines start from the same SFT checkpoint with tuned learning rates in $\{10^{-3}, 10^{-4}, 10^{-5}\}$.
Evaluation uses beam search (width 10).
All results are averaged over 5 runs on 4$\times$ A100 GPUs.

\subsection{More results on diverse datasets}
\label{sec:additional_results}

We report the complete set of unlearning results across all five backbones (MOR, TIGER, P5SID, LETTER, ETEGRec) and three datasets.
The main paper presents results for two backbone–dataset combinations (MOR and LETTER on Industrial \& Scientific and Sports \& Outdoors); here we provide the full results including all five backbones~\cite{lin2026volume}.
Table~\ref{tab:full_industrial} reports results on \textbf{Industrial \& Scientific},
Table~\ref{tab:full_sports} on \textbf{Sports \& Outdoors}, and
Table~\ref{tab:full_toys} on \textbf{Toys \& Games}.
Each table includes Retain metrics (HR@5, HR@10, NDCG@5, NDCG@10, MRR; higher is better), Forget metrics (the same five; lower is better), and Semantic Similarity (higher is better).

Across all 15 backbone–dataset combinations, TRACER consistently achieves the best performance on every metric group.
On the Forget side, TRACER attains the lowest leakage in all settings, often by a substantial margin over the second-best method (PISCES).
On the Retain side, TRACER matches or exceeds the Original model's utility in most configurations, confirming that targeted token reassignment effectively preserves recommendation quality for non-forget users.
For Semantic Similarity, TRACER ranks first in the majority of settings, demonstrating that the coherence-aware design maintains meaningful item representations after unlearning~\cite{liu2026improvingcompletenesscomparabilitysegment,cheng2026toward}.
\subsection{Efficiency Comparison}
\label{sec:efficiency}

Table~\ref{tab:efficiency} reports the wall-clock unlearning time (in minutes) for each method across all four backbone–dataset combinations used in the main experiments.
All timings are averaged over 5 runs on 4$\times$ NVIDIA A100 GPUs.
TRACER is consistently the fastest method, requiring only 19--27 minutes depending on the setting—roughly 2$\times$ faster than the next-fastest baseline (ELM) and up to 3.5$\times$ faster than the slowest (RMU)~\cite{zhang2023kadabra,zhang2023rethinking}.
Notably, LETTER-based configurations are faster than MOR across all methods, as the joint-training backbone involves fewer sequential decoding steps during unlearning.

\begin{table}[t]
\centering\scriptsize
\caption{Unlearning time (minutes, mean $\pm$ std over 5 runs) on 4$\times$ A100 GPUs.
\textbf{Bold}: fastest; \underline{underline}: second fastest.}
\label{tab:efficiency}
\renewcommand{\arraystretch}{1.10}
\setlength{\tabcolsep}{3.5pt}
\begin{tabular}{lcccc}
\toprule
\multirow{2}{*}{\textbf{Method}}
  & \multicolumn{2}{c}{\textbf{Ind.\ \& Sci.}}
  & \multicolumn{2}{c}{\textbf{Spo.\ \& Out.}} \\
\cmidrule(lr){2-3}\cmidrule(l){4-5}
  & MOR & LETTER & MOR & LETTER \\
\midrule
RMU      & \ms{67.5}{1.8} & \ms{51.3}{1.4} & \ms{63.8}{1.7} & \ms{48.6}{1.3} \\
ELM      & \msu{35.2}{0.9} & \msu{26.8}{0.7} & \msu{33.1}{0.8} & \msu{25.4}{0.6} \\
REPNoise & \ms{43.5}{1.1} & \ms{33.1}{0.8} & \ms{41.2}{1.0} & \ms{31.5}{0.7} \\
FLAT     & \ms{51.2}{1.3} & \ms{39.4}{1.0} & \ms{48.6}{1.2} & \ms{37.2}{0.9} \\
BLUR     & \ms{47.5}{1.2} & \ms{36.2}{0.9} & \ms{44.7}{1.1} & \ms{34.1}{0.8} \\
PISCES   & \ms{41.6}{1.0} & \ms{31.7}{0.8} & \ms{39.8}{0.9} & \ms{30.2}{0.7} \\
\midrule
\textbf{TRACER} & \msb{26.7}{0.5} & \msb{20.3}{0.4} & \msb{24.9}{0.5} & \msb{19.1}{0.4} \\
\bottomrule
\end{tabular}
\end{table}

\subsection{Generalization to Item-Level Unlearning}
\label{app:item_level}

While our main experiments focus on brand-level unlearning (removing all items associated with a target brand), 
we further evaluate whether TRACER generalizes to \emph{item-level} forgetting, where individual items are designated for removal.
Table~\ref{tab:item_industrial} reports results on the Industrial \& Scientific dataset.
TRACER consistently achieves the lowest forget leakage across all five backbones while maintaining competitive retain performance and high semantic coherence,
confirming that our SID-code reassignment strategy is not limited to brand-level forgetting but extends naturally to arbitrary item-level unlearning requests~\cite{zhang2025honeybee}.

\begin{table}[h!]
\centering\scriptsize
\caption{Item-level unlearning results on \textbf{Industrial \& Scientific}. Bold = best among unlearning methods; underline = second best. Retain ($\uparrow$); Forget ($\downarrow$); Sim.~($\uparrow$).}
\label{tab:item_industrial}
\resizebox{\textwidth}{!}{%
\begin{tabular}{ll|ccccc|ccccc|c}
\toprule
 & & \multicolumn{5}{c|}{\textbf{Retain ($\uparrow$)}} & \multicolumn{5}{c|}{\textbf{Forget ($\downarrow$)}} & \\
\textbf{Backbone} & \textbf{Method} & HR@5 & HR@10 & NDCG@5 & NDCG@10 & MRR & HR@5 & HR@10 & NDCG@5 & NDCG@10 & MRR & \textbf{Sim.$\uparrow$} \\
\midrule
\multirow{8}{*}{\textit{MOR}} & Original & 0.115 & 0.162 & 0.083 & 0.098 & 0.085 & 0.206 & 0.256 & 0.167 & 0.187 & 0.167 & 0.866 \\ \cmidrule{2-13}
 & RMU & 0.080 & 0.080 & 0.061 & 0.049 & 0.075 & 0.164 & 0.226 & 0.142 & 0.155 & 0.148 & 0.735 \\
 & ELM & 0.077 & 0.131 & 0.056 & 0.081 & 0.078 & 0.159 & 0.215 & 0.130 & 0.151 & 0.144 & \underline{0.817} \\
 & BLUR & 0.110 & \underline{0.156} & \textbf{0.081} & \underline{0.096} & \textbf{0.081} & 0.156 & 0.211 & 0.125 & 0.146 & 0.141 & 0.663 \\
 & REPNoise & 0.083 & 0.086 & 0.046 & 0.075 & 0.076 & 0.152 & 0.183 & 0.124 & 0.131 & 0.134 & 0.622 \\
 & FLAT & \underline{0.111} & 0.155 & \textbf{0.081} & 0.087 & \underline{0.079} & 0.145 & 0.166 & 0.109 & 0.124 & 0.124 & 0.703 \\
 & PISCES & 0.108 & 0.150 & \underline{0.074} & 0.093 & \underline{0.079} & \underline{0.121} & \underline{0.143} & \underline{0.100} & \underline{0.108} & \underline{0.117} & 0.799 \\
 & \textbf{TRACER} & \textbf{0.114} & \textbf{0.157} & \textbf{0.081} & \textbf{0.098} & \textbf{0.081} & \textbf{0.097} & \textbf{0.122} & \textbf{0.089} & \textbf{0.097} & \textbf{0.106} & \textbf{0.834} \\
\midrule
\multirow{8}{*}{\textit{TIGER}} & Original & 0.095 & 0.134 & 0.076 & 0.092 & 0.074 & 0.173 & 0.227 & 0.152 & 0.172 & 0.149 & 0.833 \\ \cmidrule{2-13}
 & RMU & 0.066 & 0.063 & 0.057 & 0.049 & 0.066 & 0.134 & 0.202 & 0.130 & 0.144 & 0.134 & 0.734 \\
 & ELM & 0.062 & 0.108 & 0.054 & 0.078 & 0.069 & 0.133 & 0.189 & 0.119 & 0.138 & 0.129 & \underline{0.831} \\
 & BLUR & \textbf{0.092} & \underline{0.132} & \underline{0.072} & 0.090 & \underline{0.072} & 0.130 & 0.187 & 0.111 & 0.136 & 0.128 & 0.661 \\
 & REPNoise & 0.068 & 0.072 & 0.041 & 0.074 & 0.067 & 0.128 & 0.160 & 0.111 & 0.120 & 0.122 & 0.630 \\
 & FLAT & \underline{0.090} & 0.129 & \textbf{0.074} & 0.083 & 0.071 & 0.123 & 0.145 & 0.096 & 0.114 & 0.111 & 0.699 \\
 & PISCES & 0.088 & 0.124 & 0.067 & \underline{0.091} & 0.071 & \underline{0.102} & \underline{0.126} & \underline{0.091} & \underline{0.097} & \underline{0.104} & 0.796 \\
 & \textbf{TRACER} & \textbf{0.092} & \textbf{0.133} & \textbf{0.074} & \textbf{0.093} & \textbf{0.073} & \textbf{0.083} & \textbf{0.107} & \textbf{0.080} & \textbf{0.092} & \textbf{0.095} & \textbf{0.833} \\
\midrule
\multirow{8}{*}{\textit{P5SID}} & Original & 0.095 & 0.139 & 0.075 & 0.093 & 0.074 & 0.177 & 0.219 & 0.152 & 0.170 & 0.147 & 0.868 \\ \cmidrule{2-13}
 & RMU & 0.069 & 0.067 & 0.055 & 0.049 & 0.065 & 0.140 & 0.193 & 0.131 & 0.142 & 0.131 & 0.738 \\
 & ELM & 0.064 & 0.114 & 0.052 & 0.077 & 0.068 & 0.138 & 0.185 & 0.116 & 0.137 & 0.127 & \underline{0.827} \\
 & BLUR & 0.089 & \underline{0.133} & 0.072 & \underline{0.093} & \underline{0.071} & 0.137 & 0.178 & 0.113 & 0.133 & 0.123 & 0.664 \\
 & REPNoise & 0.070 & 0.075 & 0.043 & 0.071 & 0.066 & 0.132 & 0.156 & 0.112 & 0.118 & 0.118 & 0.635 \\
 & FLAT & \underline{0.092} & 0.132 & \underline{0.075} & 0.085 & 0.069 & 0.124 & 0.141 & 0.097 & 0.114 & 0.111 & 0.730 \\
 & PISCES & 0.089 & 0.127 & 0.070 & 0.091 & 0.068 & \underline{0.103} & \underline{0.121} & \underline{0.093} & \underline{0.097} & \underline{0.103} & 0.776 \\
 & \textbf{TRACER} & \textbf{0.093} & \textbf{0.138} & \textbf{0.076} & \textbf{0.094} & \textbf{0.072} & \textbf{0.087} & \textbf{0.104} & \textbf{0.079} & \textbf{0.089} & \textbf{0.095} & \textbf{0.829} \\
\midrule
\multirow{8}{*}{\textit{LETTER}} & Original & 0.119 & 0.165 & 0.086 & 0.102 & 0.087 & 0.231 & 0.288 & 0.185 & 0.201 & 0.184 & 0.847 \\ \cmidrule{2-13}
 & RMU & 0.084 & 0.081 & 0.066 & 0.054 & 0.077 & 0.182 & 0.256 & 0.159 & 0.165 & 0.163 & 0.735 \\
 & ELM & 0.080 & 0.134 & 0.058 & 0.085 & 0.080 & 0.180 & 0.242 & 0.143 & 0.162 & 0.160 & \underline{0.829} \\
 & BLUR & \underline{0.112} & \underline{0.161} & \underline{0.083} & \underline{0.100} & \underline{0.083} & 0.175 & 0.234 & 0.135 & 0.159 & 0.156 & 0.662 \\
 & REPNoise & 0.089 & 0.089 & 0.050 & 0.080 & 0.076 & 0.172 & 0.203 & 0.135 & 0.139 & 0.151 & 0.625 \\
 & FLAT & \textbf{0.114} & 0.156 & \underline{0.083} & 0.093 & \underline{0.083} & 0.163 & 0.184 & 0.117 & 0.136 & 0.137 & 0.715 \\
 & PISCES & \textbf{0.114} & 0.151 & 0.077 & 0.098 & 0.080 & \underline{0.138} & \underline{0.162} & \underline{0.110} & \underline{0.116} & \underline{0.126} & 0.790 \\
 & \textbf{TRACER} & \textbf{0.114} & \textbf{0.164} & \textbf{0.084} & \textbf{0.101} & \textbf{0.084} & \textbf{0.109} & \textbf{0.136} & \textbf{0.096} & \textbf{0.106} & \textbf{0.119} & \textbf{0.842} \\
\midrule
\multirow{8}{*}{\textit{ETEGRec}} & Original & 0.117 & 0.163 & 0.084 & 0.105 & 0.084 & 0.249 & 0.297 & 0.200 & 0.200 & 0.195 & 0.856 \\ \cmidrule{2-13}
 & RMU & 0.083 & 0.081 & 0.065 & 0.052 & 0.073 & 0.196 & 0.264 & 0.172 & 0.165 & 0.172 & 0.714 \\
 & ELM & 0.077 & 0.131 & 0.059 & 0.090 & 0.075 & 0.193 & 0.249 & 0.152 & 0.160 & 0.171 & \underline{0.807} \\
 & BLUR & \underline{0.112} & \underline{0.160} & 0.079 & \underline{0.105} & \underline{0.082} & 0.191 & 0.245 & 0.147 & 0.160 & 0.165 & 0.688 \\
 & REPNoise & 0.086 & 0.088 & 0.049 & 0.083 & 0.074 & 0.187 & 0.213 & 0.146 & 0.141 & 0.160 & 0.613 \\
 & FLAT & 0.111 & 0.154 & \underline{0.082} & 0.097 & 0.079 & 0.176 & 0.193 & 0.126 & 0.133 & 0.147 & 0.723 \\
 & PISCES & \underline{0.112} & 0.150 & 0.075 & 0.102 & 0.081 & \underline{0.148} & \underline{0.167} & \underline{0.119} & \underline{0.116} & \underline{0.136} & 0.798 \\
 & \textbf{TRACER} & \textbf{0.113} & \textbf{0.161} & \textbf{0.083} & \textbf{0.106} & \textbf{0.083} & \textbf{0.119} & \textbf{0.140} & \textbf{0.107} & \textbf{0.104} & \textbf{0.127} & \textbf{0.811} \\
\bottomrule
\end{tabular}%
}
\end{table}

\subsection{Effect of Positive Set Size $K$ and Temperature $\tau$}
\label{sec:ablation_K_tau}

\begin{figure}[t]
    \centering
    \includegraphics[width=0.48\linewidth]{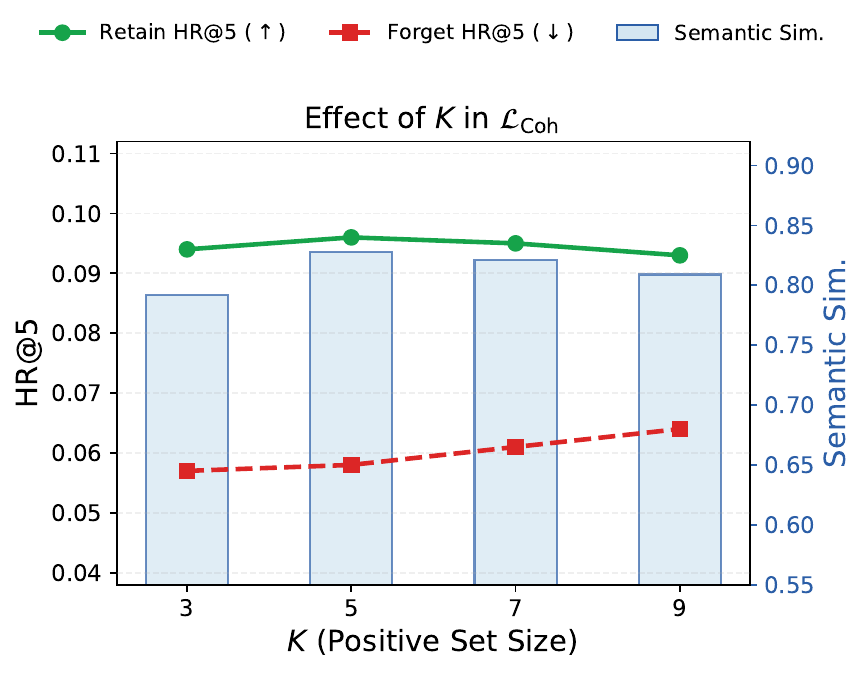}
    \hfill
    \includegraphics[width=0.48\linewidth]{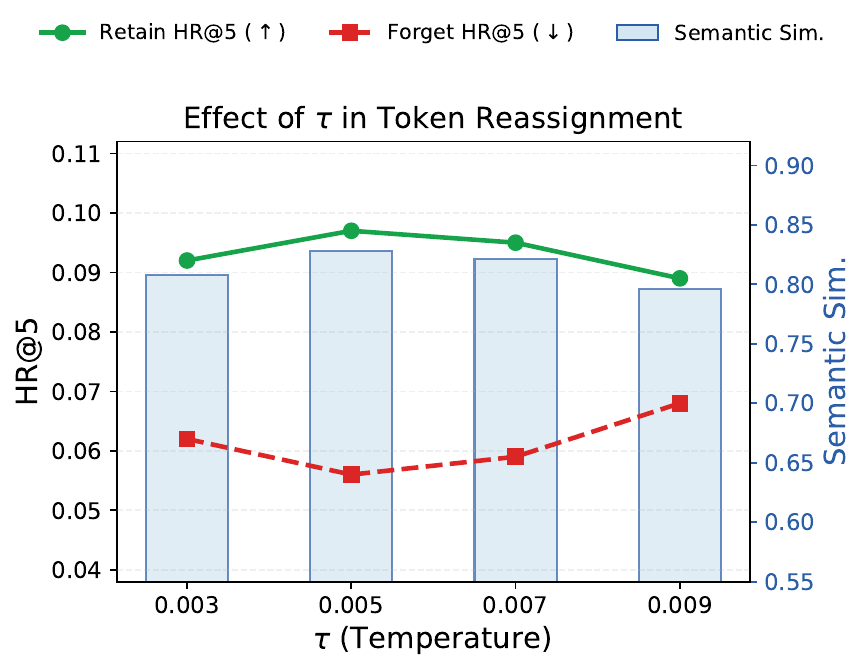}
    \caption{Ablation studies on $K$ (left) and $\tau$ (right).
    \textbf{Left:} $K{=}5$ achieves the best trade-off between semantic coherence and unlearning effectiveness.
    \textbf{Right:} $\tau{=}0.005$ yields the strongest performance; larger $\tau$ gradually degrades all metrics as the token-reassignment distribution becomes too diffuse.}
    \label{fig:ablation_K_tau}
\end{figure}

\begin{figure*}[t]
    \centering
    \includegraphics[width=\textwidth]{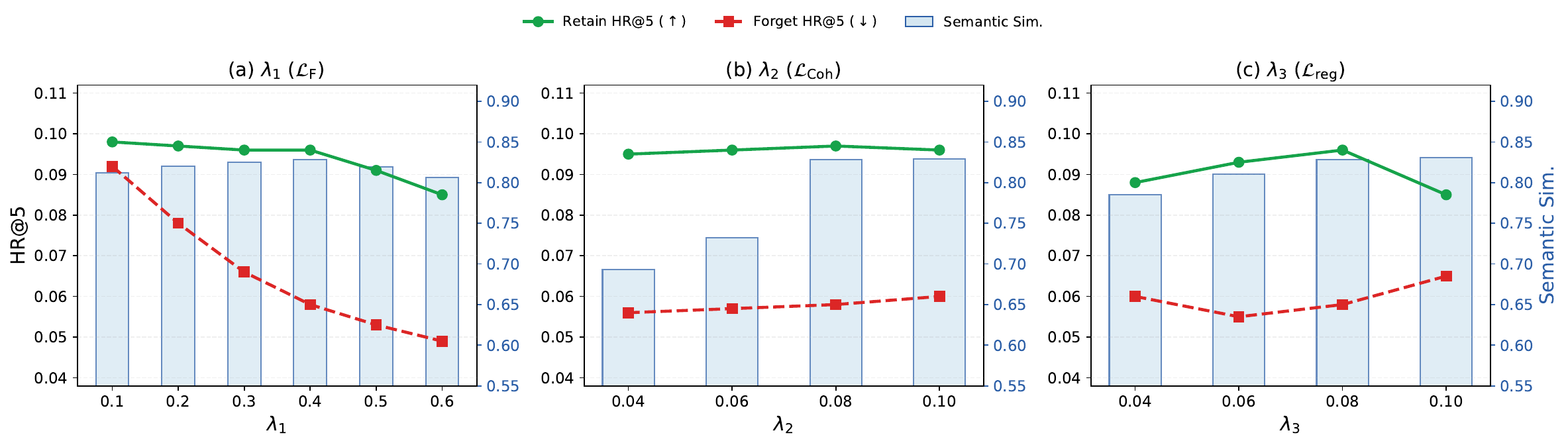}
    \caption{Ablation study on loss weights. 
    (a)~$\lambda_1$ controls the forget loss $\mathcal{L}_{\mathrm{F}}$: increasing $\lambda_1$ improves forgetting but degrades retain performance when too large.
    (b)~$\lambda_2$ controls the coherence loss $\mathcal{L}_{\mathrm{Coh}}$: higher values improve semantic similarity with diminishing returns.
    (c)~$\lambda_3$ controls the regularization loss $\mathcal{L}_{\mathrm{reg}}$: stronger regularization preserves retain quality but weakens forgetting.
    }
    \label{fig:ablation_lambda}
\end{figure*}

\paragraph{Effect of $K$.}
The hyperparameter $K$ controls the number of semantically similar retain items used to construct the positive set $\mathcal{P}(i)$ in the coherence loss $\mathcal{L}_{\mathrm{Coh}}$ (Eq.~\ref{eq:coh}).
A smaller $K$ limits the diversity of the coherence signal, resulting in lower semantic similarity after reassignment.
Conversely, a larger $K$ introduces less relevant items into the positive set, diluting the coherence objective and slightly degrading both unlearning effectiveness and similarity.
As shown in Figure~\ref{fig:ablation_K_tau} (left), $K{=}5$ achieves the best trade-off, yielding the highest semantic similarity (0.828) while maintaining strong forget performance (Forget HR@5\,=\,0.058) and retain quality (Retain HR@5\,=\,0.096).

\paragraph{Effect of $\tau$.}
The temperature $\tau$ in Eq.~\ref{eq:reassign} governs the sharpness of the soft codeword-assignment distribution during token reassignment.
A very small $\tau$ ($e.g.$, 0.003) makes the distribution excessively peaked, reducing the flexibility of the reassignment and limiting the model's ability to find alternative semantic codes for the forget item.
A very large $\tau$ ($e.g.$, 0.009) makes the distribution nearly uniform, causing the model to scatter probability mass across irrelevant codewords and degrading semantic coherence.
As shown in Figure~\ref{fig:ablation_K_tau} (right), $\tau{=}0.005$ achieves the best balance, yielding the lowest forget leakage (Forget HR@5\,=\,0.056), the highest retain quality (Retain HR@5\,=\,0.097), and the strongest semantic coherence (Sim\,=\,0.828).
$\tau{=}0.007$ is a close second, while $\tau{=}0.009$ is the worst-performing setting across all metrics, confirming that over-softening the assignment distribution is more harmful than over-sharpening it.

\newpage
\subsection{Concrete Examples}
\textit{}
\label{app:concrete_examples}

To provide qualitative insight into how TRACER modifies recommendation behavior,
we present top-10 recommendation lists produced by the base model and by TRACER
for three representative forget-brand queries, one from each dataset.
Items belonging to the forget set are highlighted with \colorbox{forgetbg}{\textcolor{forgetred}{red shading}};
the ``Forget Leaked'' row counts how many of the 10 retrieved items originate from the target brand's forget set.
As shown in Table~\ref{tab:qualitative_1}, the base model frequently surfaces memorized forget-set items
(up to 9 out of 10), whereas TRACER consistently achieves 0/10 leakage while
returning semantically relevant alternatives from retain brands.



\begin{table*}[t]
\centering\footnotesize
\setlength{\tabcolsep}{4pt}
\caption{Qualitative Example 1 (Industrial \& Scientific).
  Target: \textcolor{forgetred}{\textbf{Anderson Metals}} ---
  ``Anderson Metals Brass Hose Fitting, Tee, 5/16'' x 5/16'' x 5/16'' Barb''.
  Original model leaks 5/10 forget items (rank\,=\,1); \method{}: 0/10.}
\label{tab:qualitative_1}
\begin{tabular}{@{} c L{12.4cm} @{}}
\toprule
\multicolumn{2}{l}{\textbf{User Interaction History} (most recent 5 items)} \\
\midrule
1 & \cellcolor{forgetbg}\textcolor{forgetred}{J-B Weld 8276 KwikWeld Quick Setting Steel Reinforced Epoxy - 2\,oz.} \\
2 & \cellcolor{forgetbg}\textcolor{forgetred}{J-B Weld 8265S Original Cold-Weld Steel Reinforced Epoxy - 2\,oz.} \\
3 & Accessbuy 180-Piece Rubber Grommet Shop Assortment \\
4 & Plymor 6'' x 9'', 4\,Mil (Pack of 100) Heavy Duty Plastic Reclosable Zipper Bags \\
5 & \cellcolor{forgetbg}\textcolor{forgetred}{Loctite Liquid Professional Super Glue 20-Gram Bottle (1365882)} \\
\midrule
\multicolumn{2}{@{}l@{}}{%
\begin{tabular}{c L{5.8cm} L{5.8cm}}
\multicolumn{1}{c}{\textbf{Rank}}
  & \multicolumn{1}{c}{\textbf{Original (Base Model)}}
  & \multicolumn{1}{c}{\textbf{\method{} (Ours)}} \\
\midrule
1 & \cellcolor{forgetbg}\textcolor{forgetred}{\textbf{Anderson Metals}: Anderson Metals Brass Hose Fitting, Tee, 5/16''} & Dremel: Dremel 3D Printing BT20-01 Build Sheets (Pack of 3) \\
2 & Neiko: Neiko 01412A Electronic Digital Caliper & Pro Tapes: ProTapes Pro Duct 110 PE-Coated Cloth General Purpose \\
3 & Merit Brass: Stainless Steel 304 Cast Pipe Fitting, Hex Locknut & Anycubic: Anycubic All Metal Right Hand MK8 Extruder \\
4 & Pomona: Pomona 1296 BNC (M) Binding Posts & Scotch: Scotch 1110-C Multi Use Duct Tape, 10-Yards \\
5 & Dremel: Dremel 3D Printing BT20-01 Build Sheets (Pack of 3) & AcuRite: AcuRite 00613 Humidity Monitor with Indoor Thermometer \\
6 & \cellcolor{forgetbg}\textcolor{forgetred}{\textbf{Loctite}: Loctite 40140 Clear 401 Prism Surface Insensitive} & Camco: Camco 4ft TastePURE Drinking Water Hose \\
7 & \cellcolor{forgetbg}\textcolor{forgetred}{\textbf{Gardner Bender}: Gardner Bender GHG-1538 5PK 3/8'' Grommet, Black} & Crest: Pro-Health Clean Mint Toothpaste 6\,oz. (Pack of 4) \\
8 & Pro Tapes: ProTapes Pro Duct 110 PE-Coated Cloth General Purpose & Extech: Extech TG20 Wire Tracer and Tone Generator \\
9 & \cellcolor{forgetbg}\textcolor{forgetred}{\textbf{DEWALT}: DEWALT DCT418S1 12V Li-Ion Hand Held Radar Scanner} & Forney: Forney 72391 Replacement Cutters for Bench Grinding Wheel \\
10 & \cellcolor{forgetbg}\textcolor{forgetred}{\textbf{DAP}: Dap 18172 10.1\,Oz Alex Plus Acrylic Latex Caulk} & Cambridge: Cambridge Hose Clamps SAE Size 16, 10\,Pcs. \\
\midrule
\textbf{Forget} & \cellcolor{forgetbg}\textcolor{forgetred}{\textbf{5/10}} & \cellcolor{retainbg}\textbf{0/10} \\
\bottomrule
\end{tabular}} \\
\end{tabular}
\end{table*}


\begin{table*}[t]
\centering\footnotesize
\setlength{\tabcolsep}{4pt}
\caption{Qualitative Example 2 (Sports \& Outdoors).
  Target: \textcolor{forgetred}{\textbf{Speedo}} ---
  ``Speedo Vanquisher 2.0 Mirrored Swim Goggle''.
  Original model leaks 9/10 forget items (rank\,=\,3); \method{}: 0/10.}
\label{tab:qualitative_2}
\begin{tabular}{@{} c L{12.4cm} @{}}
\toprule
\multicolumn{2}{l}{\textbf{User Interaction History} (most recent 5 items)} \\
\midrule
1 & WOTOW 16 in 1 Multi-Function Bike Bicycle Cycling Mechanic Repair Tool \\
2 & AmazonBasics 1/4-Inch Yoga and Exercise Mat with Carrying Strap \\
3 & \cellcolor{forgetbg}\textcolor{forgetred}{Speedo Vanquisher 2.0 Mirrored Swim Goggle} \\
4 & \cellcolor{forgetbg}\textcolor{forgetred}{Speedo Vanquisher 2.0 Mirrored Swim Goggle} \\
5 & \cellcolor{forgetbg}\textcolor{forgetred}{Speedo Vanquisher 2.0 Mirrored Swim Goggle} \\
\midrule
\multicolumn{2}{@{}l@{}}{%
\begin{tabular}{c L{5.8cm} L{5.8cm}}
\multicolumn{1}{c}{\textbf{Rank}}
  & \multicolumn{1}{c}{\textbf{Original (Base Model)}}
  & \multicolumn{1}{c}{\textbf{\method{} (Ours)}} \\
\midrule
1 & \cellcolor{forgetbg}\textcolor{forgetred}{\textbf{Speedo}: Speedo Short Blade Swim Training Fins} & Duduma: Duduma Polarized Sports Sunglasses for Men \\
2 & \cellcolor{forgetbg}\textcolor{forgetred}{\textbf{Speedo}: Speedo Silicone Long Hair Swim Cap} & HODGSON: HODGSON Polarized Sports Sunglasses with 5 Lenses \\
3 & \cellcolor{forgetbg}\textcolor{forgetred}{\textbf{Speedo}: Speedo Vanquisher 2.0 Mirrored Swim Goggle} & Duduma: Duduma Polarized Designer Fashion Sports Sunglasses \\
4 & \cellcolor{forgetbg}\textcolor{forgetred}{\textbf{Speedo}: Speedo Deluxe Ventilator Mesh Bag} & Zionor: Zionor X4 Ski Snowboard Snow Goggles Magnet Dual Layers \\
5 & \cellcolor{forgetbg}\textcolor{forgetred}{\textbf{Speedo}: Speedo Ventilator Mesh Equipment Bag} & DRSKIN: DRSKIN Compression Cool Dry Sports Tights Shirt \\
6 & \cellcolor{forgetbg}\textcolor{forgetred}{\textbf{Speedo}: Speedo Silicone Solid Swim Cap} & COSVER: COSVER Men's Polarized Sunglasses for Men Sports \\
7 & \cellcolor{forgetbg}\textcolor{forgetred}{\textbf{Speedo}: Speedo Kids' Skoogles Swim Goggle} & Baleaf: Baleaf Men's Running Fitness Workout Compression Pants \\
8 & \cellcolor{forgetbg}\textcolor{forgetred}{\textbf{Speedo}: Speedo Men's Endurance+ Launch Splice Jammer} & Tesla: Tesla Men's Cool Dry Compression Muscle Tank \\
9 & \cellcolor{forgetbg}\textcolor{forgetred}{\textbf{Speedo}: Speedo Women's Pro LT Super Pro Swimsuit} & LUENX: LUENX Men Aviator Sunglasses Polarized Women \\
10 & adidas: adidas Men's Tiro 15 Training Pants & DRSKIN: DRSKIN Men's 3/4 Compression Tight Pants \\
\midrule
\textbf{Forget} & \cellcolor{forgetbg}\textcolor{forgetred}{\textbf{9/10}} & \cellcolor{retainbg}\textbf{0/10} \\
\end{tabular}} \\
\bottomrule
\end{tabular}
\end{table*}


\begin{table*}[t]
\centering\footnotesize
\setlength{\tabcolsep}{4pt}
\caption{Qualitative Example 3 (Toys \& Games).
  Target: \textcolor{forgetred}{\textbf{Learning Resources}} ---
  ``Learning Resources New Sprouts Reel it''.
  Original model leaks 8/10 forget items (not in top-20); \method{}: 0/10.}
\label{tab:qualitative_3}
\begin{tabular}{@{} c L{12.4cm} @{}}
\toprule
\multicolumn{2}{l}{\textbf{User Interaction History} (most recent 5 items)} \\
\midrule
1 & \cellcolor{forgetbg}\textcolor{forgetred}{Green Toys Airplane Vehicle Toy, Purple, 8.5'' x 9'' x 4.5''} \\
2 & Fisher-Price Go Baby Go Poppity-Pop Musical Dino \\
3 & \cellcolor{forgetbg}\textcolor{forgetred}{Learning Resources New Sprouts Stir Fry Set} \\
4 & VTech Lil' Critters Roll and Discover Ball \\
5 & \cellcolor{forgetbg}\textcolor{forgetred}{Learning Resources Pretend \& Play Doctor Kit for Kids, 19 Piece Set} \\
\midrule
\multicolumn{2}{@{}l@{}}{%
\begin{tabular}{c L{5.8cm} L{5.8cm}}
\multicolumn{1}{c}{\textbf{Rank}}
  & \multicolumn{1}{c}{\textbf{Original (Base Model)}}
  & \multicolumn{1}{c}{\textbf{\method{} (Ours)}} \\
\midrule
1 & Peppa Pig: Peppa Pig 92602 Fancy Dress Party Toy Figure & Peppa Pig: Peppa Pig 92602 Fancy Dress Party Toy Figure \\
2 & Peppa Pig: Peppa Pig Character Options Classroom Playset & Peppa Pig: Peppa Pig Character Options Classroom Playset \\
3 & \cellcolor{forgetbg}\textcolor{forgetred}{\textbf{Paw Patrol}: Paw Patrol, Lights and Sounds Air Patroller Plane} & Fisher-Price: Fisher-Price Think \& Learn Code-a-Pillar Toy \\
4 & \cellcolor{forgetbg}\textcolor{forgetred}{\textbf{Paw Patrol}: Paw Patrol Super Pup Rubble's Crane, Vehicle and Figure} & Fisher-Price: Fisher-Price DoodlePro, Trip, (Red) \\
5 & \cellcolor{forgetbg}\textcolor{forgetred}{\textbf{Paw Patrol}: Paw Patrol - Paw Terrain Vehicle (Amazon Exclusive)} & Fisher-Price: Fisher-Price Train Inflatable Ball Pit \\
6 & \cellcolor{forgetbg}\textcolor{forgetred}{\textbf{Paw Patrol}: Paw Patrol Action Pack Pups 3pk Figure Set} & Fisher-Price: Fisher-Price Bright Beats Dance \& Move BeatBo \\
7 & \cellcolor{forgetbg}\textcolor{forgetred}{\textbf{Paw Patrol}: Paw Patrol, Adventure Bay Railway Track Set} & Fisher-Price: Fisher-Price Bright Beats Smart Touch Play Space \\
8 & \cellcolor{forgetbg}\textcolor{forgetred}{\textbf{Paw Patrol}: Paw Patrol Chase's Cruiser, Vehicle and Figure} & Fisher-Price: Fisher-Price Laugh \& Learn Smart Stages Piggy Bank \\
9 & \cellcolor{forgetbg}\textcolor{forgetred}{\textbf{Paw Patrol}: Paw Patrol Chase's Spy Cruiser, Vehicle and Figure} & VTech: VTech Swim \& Spray Musical Dolphin \\
10 & \cellcolor{forgetbg}\textcolor{forgetred}{\textbf{Paw Patrol}: Paw Patrol Racers 3-Pack Vehicle Set, Rubble/Rocky/Skye} & Fisher-Price: Fisher-Price Laugh \& Learn Let's Get Ready Sink \\
\midrule
\textbf{Forget} & \cellcolor{forgetbg}\textcolor{forgetred}{\textbf{8/10}} & \cellcolor{retainbg}\textbf{0/10} \\
\end{tabular}} \\
\bottomrule
\end{tabular}
\end{table*}



\end{document}